\documentclass[aps,nofootinbib,preprintnumbers,showpacs,prd,
twocolumn,superscriptaddress]{revtex4-2}
\usepackage{array}
\usepackage{booktabs}
\usepackage{subfigure}
\usepackage{capt-of}
\usepackage{graphicx}
\usepackage{lipsum}
\usepackage{epstopdf}
\usepackage{amsmath}
\usepackage{bm}
\usepackage{amssymb}
\usepackage{color,xcolor}
\usepackage[bookmarks=false]{hyperref}
\hypersetup{colorlinks=true,
	citecolor=green,
	linkcolor=blue,
	urlcolor=blue,
	pdfstartview=FitH,
	bookmarksopen=true}

\begin{document}
	
	\title{Twist-4 GTMDs of sea quarks at zero skewness in the light-cone quark model}
	
	\author{Xiaoyan Luan} \email[]{xiaoyanluan@jsnu.edu.cn}
\affiliation{Department of Physics, Jiangsu Normal University, Xuzhou 221116, China}	
\author{Zhun Lu}\email[]{zhunlu@seu.edu.cn}
	\affiliation{School of Physics, Southeast University, Nanjing 211189, China}
	
	\begin{abstract}
We investigate the twist-4 generalized transverse momentum dependent parton distributions (GTMDs) of $\bar{u}$ and $\bar{d}$ quarks inside the proton at zero skewness by adopting the overlap representation within the light-cone formalism. Using the light-cone wave functions of the proton derived from the baryon-meson fluctuation model in terms of the $|q\bar{q}B\rangle$ Fock states, we calculate the twist-4 GTMDs of $\bar{u}$ and $\bar{d}$ quarks. Additionally, the intricate relations between the GTMDs and the generalized parton distributions are also explored in our research. Numerical results for these twist-4 GTMDs and twist-4 GPDs at zero skewness are analyzed and presented. 
	\end{abstract}
	\maketitle
	
	\section{Introduction}\label{Sec1}
Unraveling the internal structures of hadrons in terms of quarks and gluons is one of the main goals of QCD and hadronic physics. Deep inelastic scattering (DIS) provides a key experimental tool for revealing hadronic structure via the extraction of parton distribution functions~\cite{Collins:1981uw,Martin:1998sq,Gluck:1994uf,Gluck:1998xa}. 
The PDF $f_{i/h}(x)$, first introduced by Feynman~\cite{Feynman:1969ej}, represents the probability density of finding a parton $i$ carrying a longitudinal momentum fraction $x$ in a hadron $h$. 
Despite their great success, PDFs only encode one-dimensional longitudinal information.
A more comprehensive picture of the nucleon in momentum space can be revealed by the transverse momentum dependent parton distributions (TMDs) $f(x,\bm k_\perp)$~\cite{Goeke:2005hb,Bacchetta:2006tn}, which depend on both the longitudinal momentum fraction $x$ and the parton transverse momentum $\bm k_\perp$ relative to the parent hadron, thus allowing a three-dimensional description of parton structure in momentum space. 
TMDs can be extracted from the processes involving two hadrons such as the semi-inclusive deep inelastic scattering (SIDIS) and the Drell-Yan process~\cite{Mulders:1995dh,Boer:1997nt,Barone:2001sp,Brodsky:2002cx,Bacchetta:2006tn,Arnold:2008kf,Bacchetta:2017gcc}. 

As the extension of PDFs from the forward-scattering region to the off-forward scattering region, generalized parton distributions (GPDs) provide three-dimensional structural information of the nucleon~\cite{Ji:1996nm,Radyushkin:1996nd,Belitsky:2001ns} complementary to the TMDs. 
GPDs appear in exclusive processes such as deeply virtual Compton scattering (DVCS)~\cite{Ji:1996nm,Radyushkin:1996nd,Belitsky:2001ns} and deeply virtual meson production (DVMP), where the target acquires a recoil momentum $\Delta$~\cite{Goloskokov:2006hr,Goloskokov:2007nt,Goloskokov:2009ia,Goloskokov:2011rd}. Besides $x$, GPDs depend on the momentum transfer squared $t=\Delta^2$ and the skewness parameter $\xi=\Delta^+ /(2P^+)$. At zero skewness, the Fourier transform of GPDs with respect to the transverse momentum transfer yields impact-parameter dependent parton distributions (IPDs), which describe parton transverse positions and enable nucleon tomography~\cite{Burkardt:2000za,Burkardt:2002hr,Diehl:2002he}.

A more comprehensive understanding of the partonic structure of the nucleon can be achieved by combining information in momentum and position spaces. To this end, the Wigner distributions of quarks and gluons in the nucleon~\cite{Ji:2003ak,Belitsky:2003nz} have been proposed and extensively studied. 
The Wigner distributions are six-dimensional phase-space distributions that encode the joint dependences of partons on their transverse position and transverse momentum in the nucleon.
Moreover, the orbital angular momentum (OAM) of partons and their spin-orbit correlations can be extracted from the Wigner distributions through phase-space averages~\cite{Lorce:2011kd,Chakrabarti:2016yuw,Chakrabarti:2017teq}. 

Through Fourier transformation, the Wigner distributions are related to generalized transverse-momentum-dependent parton distributions (GTMDs), which are often regarded as the ``mother distributions'' of TMDs and GPDs~\cite{Belitsky:2005qn,Ji:2003ak,Belitsky:2003nz}. GTMDs depend on the light-cone momentum fraction, the transverse momentum of the parton, and the transverse momentum transfer to the nucleon. They are obtained from the generalized parton correlation functions (GPCFs)~\cite{Meissner:2008ay,Meissner:2009ww} by integrating over the minus component of the parton momentum. Quark GTMDs may be accessible through the exclusive double Drell-Yan process~\cite{Bhattacharya:2017bvs}, while the feasibility of measuring gluon GTMDs in diffractive dijet production has also been investigated~\cite{Hagiwara:2017fye,Ji:2016jgn,Hatta:2016aoc,Bhattacharya:2022vvo}. In certain kinematic limits, GTMDs reduce to TMDs and GPDs. A comprehensive classification of the various parton distributions and their interrelations was presented in Refs.~\cite{Meissner:2008ay,Meissner:2009ww}.

Parton distribution functions are classified according to their twist, which characterizes the order in the $1/Q$ expansion at which they contribute, where $Q$ denotes the hard scale of the process. Most studies of parton distributions have focused on leading-twist distributions. However, at the energy scales relevant to current experiments, higher-twist contributions may not be negligible. Despite their importance, higher-twist distributions are difficult to determine experimentally because of the challenge of disentangling them from leading-twist contributions. In general, higher-twist contributions are suppressed relative to leading-twist contributions by powers of the hard scale. In recent years, higher-twist distributions have also been investigated in a variety of theoretical models. For example, higher-twist TMDs have been studied within the MIT bag model~\cite{Signal:1996ct,Avakian:2010br,Lorce:2014hxa}, diquark spectator models~\cite{Lu:2012gu,Mao:2013waa,Mao:2014aoa,Liu:2021ype}, and light-front approaches~\cite{Burkardt:2001iy,Kundu:2001pk,Mukherjee:2009uy,Lorce:2016ugb,Pasquini:2018oyz,Sharma:2023azu,Sharma:2023wha,Puhan:2023ekt,Zhu:2023lst}. Other approaches, including chiral quark soliton models~\cite{Schweitzer:2003uy,Wakamatsu:2003uu,Wakamatsu:2007nc,Ohnishi:2003mf,Cebulla:2007ej} and the instanton model~\cite{Balla:1997hf,Dressler:1999hc}, have also been employed to study higher-twist distributions. 
Meanwhile, twist-3 GPDs have been investigated in the quark target model~\cite{Mukherjee:2002pq,Mukherjee:2002xi,Aslan:2018tff} and the scalar diquark model~\cite{Aslan:2018tff} for the nucleon. 
The chiral-even twist-3 GPDs of the proton have also been studied using lattice QCD~\cite{Bhattacharya:2023nmv}. 
More recently, twist-3 GPDs and twist-4 chiral-even GPDs of the proton have been calculated within basis light-front quantization~\cite{Zhang:2023xfe} and the light-front quark-diquark model (LFQDM)~\cite{Sharma:2023ibp}, respectively, using overlaps of light-front wave functions.

For GTMDs, the parameterization of quark GTMDs up to twist-4 for spin-0 and spin-$1/2$ targets has been established in Refs.~\cite{Meissner:2008ay} and~\cite{Meissner:2009ww}, respectively. Various model calculations have been performed to evaluate twist-2 GTMDs of the nucleon~\cite{Lorce:2011kd,Lorce:2011ni,Lorce:2011dv,Mukherjee:2014nya,Mukherjee:2015aja,
More:2017zqq,Liu:2015eqa,Chakrabarti:2016yuw,Chakrabarti:2017teq,Chakrabarti:2019wjx,
Gutsche:2016gcd,Kaur:2019lox,Kumar:2017xcm,Lorce:2011kd,Lorce:2011ni,Kanazawa:2014nha}. 
Meson GTMDs have also been investigated using theoretical approaches~\cite{Ma:2018ysi,Kaur:2019jow,Kaur:2019kpi,Zhang:2021tnr}. 
Moreover, the twist-2 GTMDs $F_{1,4}$ and $G_{1,1}$ have been used to investigate the canonical OAM~\cite{Lorce:2011kd,Hatta:2011ku,Ji:2012sj,Lorce:2012ce} and spin-orbit correlations~\cite{Lorce:2014mxa,Tan:2021osk} of partons, respectively. Compared with leading-twist GTMDs and higher-twist TMDs and GPDs, relatively few model calculations of higher-twist GTMDs are available. Recently, twist-3 and twist-4 proton GTMDs have been studied within the LFQDM framework~\cite{Sharma:2023tre,Sharma:2024arf}. Notably, a complete overlap representation of GTMDs has been developed within the light-cone formalism, in which the Fock-state expansion of a hadron is expressed in terms of $N$-parton Fock states, with the corresponding coefficients given by the light-cone wave functions (LCWFs) of the $N$ partons~\cite{Brodsky:2000xy,Muller:2014tqa,Pasquini:2006dv}. Within this framework, GTMDs can be interpreted in terms of overlaps of LCWFs.
	
Since GTMDs reduce to different parton distributions in different kinematic limits, and higher-twist effects cannot always be neglected, a detailed understanding of higher-twist GTMDs is important for advancing our knowledge of the proton structure. 
In this paper, we employ the light-cone quark model (LCQM) to calculate, for the first time, the twist-4 GTMDs of $\bar{u}$ and $\bar{d}$ quarks at zero skewness using the overlap representation. As emphasized in Refs.~\cite{Brodsky:1996hc,Pasquini:2006dv}, sea-quark degrees of freedom can be generated through the meson-baryon fluctuation mechanism, in which the proton fluctuates into a composite state consisting of a meson $M$ and a baryon $B$, with the meson containing a $q\bar{q}$ component. Accordingly, the LCWFs of the proton can be constructed from the $|q\bar{q}B\rangle$ Fock states, as demonstrated in Ref.~\cite{Luan:2022fjc}. Based on these LCWFs and the overlap representation, we derive the twist-4 GTMDs of $\bar{u}$ and $\bar{d}$ quarks. We present numerical results for these twist-4 GTMDs using appropriate values of the model parameters. Furthermore, we investigate the relations between GTMDs and GPDs and, based on these relations, present numerical results for the corresponding twist-4 GPDs at $\xi=0$.
	
The remaining part of this paper is organized as follows. In Sec.~\ref{Sec2}, we introduce the definition of twist-4 GTMDs. In Sec.~\ref{Sec3}, we derive the overlap representation and obtain the analytic expressions for the twist-4 GTMDs and GPDs of sea quarks in terms of the LCWFs. In Sec.~\ref{Sec4}, we present the numerical results for these distributions. Finally, we summarize our results in Sec.~\ref{Sec5}.

\section{Definition of twist-4 GTMDs}\label{Sec2}
A complete set of GTMDs for the nucleon has been presented in Refs.~\cite{Meissner:2008ay,Meissner:2009ww}. For a spin-$1/2$ hadron, the GTMDs are defined through the fully unintegrated quark-quark correlator at fixed light-cone time $z^{+}=0$:
\begin{widetext}
	\begin{align} \label{correlator}
		\notag
		&W_{\Lambda^{\prime} \Lambda }^{[\Gamma]}\left(P, x, \boldsymbol{k}_T, \xi,\boldsymbol{\Delta}_T \right) = \int d k^{-} W_{ \Lambda^{\prime} \Lambda}^{[\Gamma]}(P, k, \Delta) \\
		& = \left.\frac{1}{2} \int \frac{d z^{-} d^{2} \vec{z}_{T}}{(2 \pi)^{3}} e^{i k \cdot z}\left\langle p^{\prime}, \Lambda^{\prime}\left|\bar{\psi}\left(-\frac{1}{2} z\right) \Gamma \mathcal{W}\left(-\frac{1}{2} z, \left.\frac{1}{2} z \right\rvert\, n\right) \psi\left(\frac{1}{2} z\right)\right| p, \Lambda\right\rangle\right|_{z^{+} = 0},
	\end{align}
    \end{widetext}
where $\mathcal{W}$ denotes the gauge link connecting the quark fields at positions $-z/2$ and $z/2$, thereby ensuring color gauge invariance. Here, $p=P-\Delta/2$ ($p^{\prime}=P+\Delta/2$) and $\Lambda$ ($\Lambda^{\prime}$) represent the momenta and helicities of the initial and final nucleons, respectively. The average nucleon momentum and the momentum transfer are defined as $P=(p+p^{\prime})/2$ and $\Delta=p^{\prime}-p$, respectively, with $t=\Delta^{2}=-\boldsymbol{\Delta}_T^2$ denoting the squared momentum transfer. In addition, $x=k^+/P^+$ is the average light-cone momentum fraction carried by the active quark, while the skewness parameter $\xi=-\Delta^{+}/(2P^{+})$ characterizes the longitudinal momentum transfer to the nucleon. Finally, $\Gamma$ specifies the Dirac structure of the correlator. In this work, we consider $\Gamma=\gamma^{-}$, $\gamma^{-}\gamma^{5}$, and $i\sigma^{i-}\gamma_{5}$.
    
The generalized correlator in Eq.~(\ref{correlator}) can be parameterized in terms of GTMDs~\cite{Meissner:2009ww}. At twist-4, there are sixteen independent GTMDs, which are defined as follows:
    \begin{widetext}
    \begin{align}
    	\notag
    	W_{\Lambda^{\prime} \Lambda}^{\left[\gamma^{-}\right]} & = \frac{M}{2\left(P^{+}\right)^{2}} \bar{u}\left(p^{\prime}, \Lambda^{\prime}\right)\left[F_{3,1}+\frac{i \sigma^{i+} k_{i}}{P^{+}} F_{3,2}+\frac{i \sigma^{i+} \Delta_{i}}{P^{+}} F_{3,3}+\frac{i \sigma^{i j} k_{i} \Delta_{j}}{M^{2}} F_{3,4}\right] u(p, \Lambda)\label{W1}
    	\\
    	&=\frac{M^{2}}{\left(P^{+}\right)^{2}}\left[F_{3,1}+\frac{i \Lambda \epsilon^{i j}_{T} k_{i} \Delta_{j}}{M^{2}} F_{3,4}\right] \delta_{\Lambda^{\prime} \Lambda}+\left[\frac{\Lambda \Delta_{1}+i \Delta_{2}}{2 M}\left(2 F_{3,3}-F_{3,1}\right)+\frac{\Lambda k_{1}+i k_{2}}{M} F_{3,2}\right] \delta_{-\Lambda^{\prime} \Lambda},
    	\\\notag\label{W2}
    	W_{\Lambda^{\prime} \Lambda }^{\left[\gamma^{-} \gamma_{5}\right]} & = \frac{M}{2\left(P^{+}\right)^{2}} \bar{u}\left(p^{\prime}, \Lambda^{\prime}\right)\left[-\frac{i \varepsilon_{T}^{i j} k_{i} \Delta_{j}}{M^{2}} G_{3,1}+\frac{i \sigma^{i+} \gamma_{5} k_{i}}{P^{+}} G_{3,2}+\frac{i \sigma^{i+} \gamma_{5} \Delta_{i}}{P^{+}} G_{3,3}+i \sigma^{+-} \gamma_{5}G_{3,4}\right] u(p, \Lambda)
    	\\\notag&=\frac{M^{2}}{\left(P^{+}\right)^{2}}\left[-\frac{i\left(\boldsymbol{k}_T \times \boldsymbol{\Delta}_T\right)}{M^{2}} G_{3,1}+\Lambda G_{3,4}\right] \delta_{\Lambda^{\prime} \Lambda}+\left[\frac{\Delta_{1}+i \Lambda \Delta_{2}}{M}\left(G_{3,3}+\frac{i \Lambda\left(\boldsymbol{k}_T \times \boldsymbol{\Delta}_T\right)}{2 M^{2}} G_{3,1}\right)
    	\right.\\&\left.
    	+\frac{k_{1}+i \Lambda k_{2}}{M} G_{3,2}\right] \delta_{-\Lambda^{\prime} \Lambda},
    	\\\notag\label{W3}
    	W_{\Lambda^{\prime} \Lambda }^{\left[i \sigma^{i-} \gamma_{5}\right]} & = \frac{M}{2\left(P^{+}\right)^{2}} \bar{u}\left(p^{\prime}, \Lambda^{\prime}\right)\left[\frac{i \varepsilon_{T}^{i j} k_{j}}{M} H_{3,1}+\frac{i \varepsilon_{T}^{i j} \Delta_{j}}{M} H_{3,2}+\frac{M i \sigma^{i+} \gamma_{5}}{P^{+}} H_{3,3}+\frac{k_{i} i \sigma^{k+} \gamma_{5} k_{k}}{M P^{+}}H_{3,4}\right. \\\notag
    	& \left. +\frac{\Delta_{i} i \sigma^{k+} \gamma_{5} k_{k}}{M P^{+}} H_{3,5}+\frac{\Delta_{i} i \sigma^{k+} \gamma_{5} \Delta_{k}}{M P^{+}} H_{3,6} +\frac{k_{i} i \sigma^{+-} \gamma_{5}}{M} H_{3,7}+\frac{\Delta_{i} i \sigma^{+-} \gamma_{5}}{M} H_{3,8}\right] u(p, \Lambda)
    	\\\notag& =\frac{M^{2}}{\left(P^{+}\right)^{2}}  \left[i \epsilon^{i j}_{T}\left(\frac{k_{j}}{M} H_{3,1}+\frac{\Delta_{j}}{M} H_{3,2}\right)+\Lambda\left(\frac{k_{i}}{M} H_{3,7}+\frac{\Delta_{i}}{M} H_{3,8}\right)\right] \delta_{\Lambda^{\prime} \Lambda}
    	\\\notag&
    	+\left[-i \epsilon^{i j}_{T} \frac{\Lambda \Delta_{1}+i \Delta_{2}}{2 M}\left(\frac{k_{j}}{M} H_{3,1}+\frac{\Delta_{j}}{M} H_{3,2}\right)+\left(\delta_{i 1}+i \Lambda \delta_{i 2}\right) H_{3,3}+\frac{k_{1}+i \Lambda k_{2}}{M}\left(\frac{k_{i}}{M} H_{3,4}+\frac{\Delta_{i}}{M} H_{3,5}\right)
    	\right. \\	& \left.
    	+\frac{\left(\Delta_{1}+i \Lambda \Delta_{2}\right) \Delta_{i}}{M^{2}} H_{3,6}\right] \delta_{-\Lambda^{\prime} \Lambda} .
    \end{align}
Here $\varepsilon^{ij}_{T}=\varepsilon^{-+ij}$ and $\varepsilon^{0123}=1$. The four F-type GTMDs describe unpolarized quarks, while the four G-type GTMDs describe quark helicity distributions. For transversely polarized quarks, an additional set of eight H-type GTMDs is required. The GTMDs considered here are complex functions of the kinematic variables $x$, $\xi$, $\boldsymbol{\Delta}_{T}^{2}$, $\boldsymbol{k}_{T}\cdot\boldsymbol{\Delta}_{T}$, and $\boldsymbol{k}_{T}^{2}$. The twist-4 GTMDs can be expressed in terms of the correlators contracted with appropriate tensors as follows:   
    \begin{align} 
    	{F}_{3,1}&=\frac{(P^{+})^{2} }{2M^{2}} \left(W^{[\gamma^{-}]}_{++}+W^{[\gamma^{-}]}_{--}\right), 
    	\\
    	\frac{i\left(\boldsymbol{k}_{T} \times \Delta_{T}\right)}{M^{2}}{F}_{3,4}&=\frac{(P^{+})^{2} }{2M^{2}} \left(W^{[\gamma^{-}]}_{++}	-W^{[\gamma^{-}]}_{--}\right),
    	\\
    	-\frac{i\left(\boldsymbol{k}_{T} \times \Delta_{T}\right)}{M} F_{3,2}&=\frac{(P^{+})^{2} }{2M^{2}}\left(\left(\Delta_{1}-i \Delta_{2}\right) W_{-+}^{[\gamma^{-}]}+\left(\Delta_{1}+i \Delta_{2}\right) W_{+-}^{[\gamma^{-}]}\right),
    	\\
    	\frac{i\left(\boldsymbol{k}_{T} \times \Delta_{T}\right)}{2M} \left(2F_{3,3}-F_{3,1}\right)&=\frac{(P^{+})^{2} }{2M^{2}}\left(\left(k_{1}-i k_{2}\right) W_{-+}^{[\gamma^{-}]}+\left(k_{1}+i k_{2}\right) W_{+-}^{[\gamma^{-}]}\right),
    \end{align}
    for F-type GTMDs,
    \begin{align} 
    	{G}_{3,4}&=\frac{(P^{+})^{2} }{2M^{2}} \left(W^{[\gamma^{-}\gamma_{5}]}_{++}-W^{[\gamma^{-}\gamma_{5}]}_{--}\right), 
    	\\
    	-\frac{i\left(\boldsymbol{k}_{T} \times \Delta_{T}\right)}{M^{2}}{G}_{3,1}&=\frac{(P^{+})^{2} }{2M^{2}} \left(W^{[\gamma^{-}\gamma_{5}]}_{++}	+W^{[\gamma^{-}\gamma_{5}]}_{--}\right),
    	\\
    	\frac{ \boldsymbol{\Delta}_{T}^{2}}{M} G_{3,3}+\frac{\boldsymbol{k}_{T} \cdot \boldsymbol{\Delta}_{T}}{M} G_{3,2}&=\frac{(P^{+})^{2} }{2M^{2}}\left(\left(\Delta_{1}-i \Delta_{2}\right) W_{-+}^{[\gamma^{-}\gamma_{5}]}+\left(\Delta_{1}+i \Delta_{2}\right) W_{+-}^{[\gamma^{-}\gamma_{5}]}\right),
    	\\
    	-\frac{i\left(\boldsymbol{k}_{T} \times \Delta_{T}\right)}{M} \left(G_{3,2}-\frac{ \boldsymbol{\Delta}_{T}^{2}}{4M^{2}}G_{3,1}\right)&=\frac{(P^{+})^{2} }{2M^{2}}\left(\left(\Delta_{1}-i \Delta_{2}\right) W_{-+}^{[\gamma^{-}\gamma_{5}]}-\left(\Delta_{1}+i \Delta_{2}\right) W_{+-}^{[\gamma^{-}\gamma_{5}]}\right),
    \end{align}
     for G-type GTMDs, and 
    \begin{align} 
    	-\frac{i\left(\boldsymbol{k}_{T} \times \Delta_{T}\right)}{M}{H}_{3,1}&=\frac{(P^{+})^{2} }{2M^{2}}\left[\Delta_{1}\left(W_{++}^{[i \sigma^{1-} \gamma_{5}]}+W_{--}^{[i \sigma^{1-} \gamma_{5}]}\right) +\Delta_{2}\left(W_{++}^{[i \sigma^{2-} \gamma_{5}]}+W_{--}^{[i \sigma^{2-} \gamma_{5}]}\right)\right],
    	\\
    	\frac{i\left(\boldsymbol{k}_{T} \times \Delta_{T}\right)}{M}{H}_{3,2}&=\frac{(P^{+})^{2} }{2M^{2}}\left[k_{1}\left(W_{++}^{[i \sigma^{1-} \gamma_{5}]}+W_{--}^{[i \sigma^{1-} \gamma_{5}]}\right) +k_{2}\left(W_{++}^{[i \sigma^{2-} \gamma_{5}]}+W_{--}^{[i \sigma^{2-} \gamma_{5}]}\right)\right],
    	\\
    	\frac{\boldsymbol{k}_{T} \times \Delta_{T}}{M}{H}_{3,7}&=\frac{(P^{+})^{2} }{2M^{2}}\left[\Delta_{2}\left(W_{++}^{[i \sigma^{1-} \gamma_{5}]}-W_{--}^{[i \sigma^{1-} \gamma_{5}]}\right) -\Delta_{1}\left(W_{++}^{[i \sigma^{2-} \gamma_{5}]}-W_{--}^{[i \sigma^{2-} \gamma_{5}]}\right)\right],
    	\\
    	-\frac{\boldsymbol{k}_{T} \times \Delta_{T}}{M}{H}_{3,8}&=\frac{(P^{+})^{2} }{2M^{2}}\left[k_{2}\left(W_{++}^{[i \sigma^{1-} \gamma_{5}]}-W_{--}^{[i \sigma^{1-} \gamma_{5}]}\right) -k_{1}\left(W_{++}^{[i \sigma^{2-} \gamma_{5}]}-W_{--}^{[i \sigma^{2-} \gamma_{5}]}\right)\right],
    	\\
    	\Delta_{i} H_{3,3}+\frac{\boldsymbol{k}_{T} \cdot \boldsymbol{\Delta}_{T}}{M^{2} }\left(k_{i} H_{3,4}\right.&\left.+\Delta_{i} H_{3,5}\right)+\frac{\Delta_{i} \boldsymbol{\Delta}_{T}^{2}}{M^{2}} H_{3,6}  =\frac{(P^{+})^{2} }{2M^{2}} \left(\left(\Delta_{1}-i \Delta_{2}\right) W_{-+}^{[i \sigma^{i-} \gamma_{5}]}+\left(\Delta_{1}+i \Delta_{2}\right) W_{+-}^{[i \sigma^{i-} \gamma_{5}]}\right),
    	\\\notag
    	-\frac{i \epsilon^{i j}_{T} \boldsymbol{\Delta}_{T}^{2}}{2 M ^{2
    	}}\left(k_{j} H_{3,1}+\Delta_{j} H_{3,2}\right)-&\frac{i \epsilon^{i j}_{T} \Delta_{j}}{M} H_{3,3}-\frac{i\left(\boldsymbol{k}_{T} \times \boldsymbol{\Delta}_{T}\right)}{M^{2}}\left(k_{i} H_{3,4}+\Delta_{i} H_{3,5}\right) 
    	\\& =\frac{(P^{+})^{2} }{2M^{2}}\left(\left(\Delta_{1}-i \Delta_{2}\right) W_{-+}^{[i \sigma^{i-} \gamma_{5}]}-\left(\Delta_{1}+i \Delta_{2}\right) W_{+-}^{[i \sigma^{i-} \gamma_{5}]}\right),
    \end{align}
for H-type GTMDs, where $+$ ($-$) denotes the positive (negative) helicity of the proton.
    
In the overlap representation, the generalized correlator in Eq.~(\ref{correlator}) can be expressed as~\cite{Sharma:2023ibp}   
    \begin{align}\label{correlation}
    	\notag W^{[\Gamma]}_{\Lambda^{\prime}\Lambda}\left(x,\boldsymbol{k}_{T}, \boldsymbol{\Delta}_{T}\right)  &=\sum_{\lambda_{i}} \int [d x][d^{2} k_{T}] \psi^{\Lambda^{\prime}*}_{\lambda_{1}^{\prime} \lambda_{i}^{\prime}}\left(x_{i}, \boldsymbol{k}_{T}^{i\prime}\right) \psi^{\Lambda}_{\lambda_{1} \lambda_{i}}\left(x_{i}, \boldsymbol{k}_{T}^{i},\right)[\delta_{\lambda_i^\prime,\lambda_i}(i=2...n)]
    	\\
    	&\times\frac{u^{\dagger}\left(x_{1} P^{+}, \boldsymbol{k}_{T}^{1}-\frac{\boldsymbol{\Delta}_{T}}{2},\lambda_{1}^{\prime}\right) \gamma^{0} \Gamma u\left(x_{1} P^{+}, \boldsymbol{k}_{T}^
    		{1}+\frac{\boldsymbol{\Delta}_{T}}{2},\lambda_{1}\right)}{2 x P^{+}} ,
    \end{align}
where $\lambda_1(\lambda_1^\prime)$ represents the helicity of the initial (final) struck quark, and $\lambda_i(\lambda_i^\prime) (i=2\dots n)$ denotes the helicity of the initial(final) spectators. The spinor product $u^{\dagger}\left(x_{1} P^{+}, \boldsymbol{k}_{T}^{1}-\frac{\boldsymbol{\Delta}_{T}}{2},\lambda_{1}^{\prime}\right) \gamma^{0} \Gamma u\left(x_{1} P^{+}, \boldsymbol{k}_{T}^{1}+\frac{\boldsymbol{\Delta}_{T}}{2},\lambda_{1}\right)$ corresponds to the higher-twist Dirac matrices and encodes struck quark helicity combinations. And
\begin{align}
    	{[d x][d^{2} k_{T}] } = \prod \frac{d x_{i} \boldsymbol{k}_{T}^{i}}{16 \pi^{3}} 16 \pi^{3} \delta\left(1-\sum x_{i}\right) \delta^{2}_{T}\left(\sum \boldsymbol{k}_{T}^{i}\right) \delta\left(x-x_{1}\right).
    \end{align}
\end{widetext}

\section{Twist-4 GTMDs and GPDs of sea quarks in the overlap representation}\label{Sec3}
    
\subsection{Twist-3 GTMDs of sea quarks}
In this section, we present the calculations for the twist-4 GTMDs of the  $\bar{u}$ and $\bar{d}$ quarks in the proton at zero skewness in the LCQM using the overlap representation. The light-cone formalism has been widely applied in the calculation of PDFs of nucleons and mesons~\cite{Lepage:1980fj}. Within the light-cone approach, a hadronic composite state can be expanded in terms of LCWFs on the Fock-state basis. The overlap representation has also been used to study various FFs of the hadrons~\cite{Brodsky:2000ii,Xiao:2003wf}, anomalous magnetic moment of the nucleon~\cite{Brodsky:2000ii}, TMDs~\cite{Bacchetta:2008af,Lu:2006kt,Luan:2022fjc}, GPDs~\cite{Muller:2014tqa,Brodsky:2000xy,Burkardt:2003je,Luan:2023lmt}, as well as the quark Wigner distributions~\cite{Ma:2018ysi,Kaur:2020par,Luan:2024nwc}. Here we extend the light-cone formalism to calculate the twist-4 GTMDs of the sea quarks. 
    
In the light-cone approach, the wave functions of the hadron that describe a composite state at a specific light-cone time are expanded in accordance with LCWFs in the Fock state basis. In order to generate the sea quark degrees of freedom, we apply the baryon-meson fluctuation model~\cite{Brodsky:1996hc,Pasquini:2006dv}, in which the proton can fluctuate to a composite system formed by a meson $M$ and a baryon $B$, where the meson is composed in terms of $q\bar{q}$:
\begin{align}\label{fock state}
    	|p\rangle\to| M B\rangle\to|q\bar{q}B\rangle.
    \end{align}
    In our work, we consider the fluctuation $|p\rangle \to |\pi^+ n\rangle $ and $|p\rangle \to |\pi^- \Delta^{++}\rangle $. 
    The corresponding LCWFs have the form which have been derived in Ref.~\cite{Luan:2022fjc}
    \begin{align}\label{LCWFs} \psi^{\lambda_N}_{{\lambda_B}{\lambda_q}{\lambda_{\bar{q}}}}(x,y,\boldsymbol{k}_T,\boldsymbol{r}_T)		=&\psi^{\lambda_N}_{\lambda_B}(y,\boldsymbol{r}_T)\psi_{{\lambda_q}{\lambda_{\bar{q}}}}
    	(x,y,\boldsymbol{k}_T,\boldsymbol{r}_T),
    \end{align}
where $\psi^{\lambda_N}_{\lambda_B}(y,r_T)$ can be viewed as the wave function of the nucleon in terms of $\pi B$ components, and $\psi_{{\lambda_q}{\lambda_{\bar{q}}}}(x,y,\boldsymbol{k}_T,\boldsymbol{r}_T)$ is the pion wave function in terms of $q \bar{q}$ components. The indices $\lambda_N$, $\lambda_B$, $\lambda_q$, $\lambda_{\bar{q}}$ denote the helicities of the proton, the baryon, the quark and the sea quark, respectively;  
$x$ and $y$ represent their light-cone momentum fractions, $\boldsymbol{k}_T$ and $\boldsymbol{r}_T$ denote the transverse momenta of the antiquark and the meson. 

For the former one of Eq.~(\ref{LCWFs}), they have the expression:
    \begin{align}\label{former}
    	\notag\psi^+_+(y,\boldsymbol{r}_T)&=\frac{M_B-(1-y)M}{\sqrt{1-y}}\phi_1, \\
    	\notag\psi^+_-(y,\boldsymbol{r}_T)&=\frac{r_1+ir_2}{\sqrt{1-y}}\phi_1, \\
    	\notag\psi^-_+(y,\boldsymbol{r}_T)&=\frac{r_1-ir_2}{\sqrt{1-y}}\phi_1 , \\
    	\psi^-_-(y,\boldsymbol{r}_T)&=\frac{(1-y)M-M_B}{\sqrt{1-y}}\phi_1.
    \end{align}	
Here, $M$ and $M_B$ are the masses of proton and baryon, respectively. $\phi_1$ is the momentum space wave function of the bayron-meson Fock state 
\begin{align}		
\phi_1(y,\boldsymbol{r}_T)=-\frac{g(r^2)\sqrt{y(1-y)}}{\boldsymbol{r}_T^2+L_1^2(m_\pi^2)},
\end{align}
where $m_\pi$ is the mass of $\pi$ meson, $g(r^2)$ is the form factor for the coupling of the nucleon-pion meson-baryon vertex, and
\begin{align}
L_1^2({m_\pi^2})=yM_B^2+(1-y){m_\pi^2}-y(1-y)M^2.
\end{align}
The latter one of Eq.~(\ref{LCWFs}) have the following expressions:
\begin{align}\label{later}		\notag\psi{_+}{_+}(x,y,\boldsymbol{k}_T,\boldsymbol{r}_T)&=\frac{my}{\sqrt{x(y-x)}}\phi_2,\\ \notag\psi{_+}{_-}(x,y,\boldsymbol{k}_T,\boldsymbol{r}_T)&=\frac{y(k_1-ik_2)-x(r_1-ir_2)}{\sqrt{x(y-x)}}\phi_2, \\ \notag\psi{_-}{_+}(x,y,\boldsymbol{k}_T,\boldsymbol{r}_T)&=\frac{y(k_1+ik_2)-x(r_1+ir_2)}{\sqrt{x(y-x)}}\phi_2, \\		\psi{_-}{_-}(x,y,\boldsymbol{k}_T,\boldsymbol{r}_T)&=\frac{-my}{\sqrt{x(y-x)}}\phi_2,
\end{align} 
Here, $m$ is the mass of quark and the sea quark. 
Again, $\phi_2(x,y,\boldsymbol{k}_T,\boldsymbol{r}_T)$ is the momentum space wave function of the $|q\bar{q}\rangle$ Fock state 
\begin{align}	     	
\phi_2(x,y,\boldsymbol{k}_T,\boldsymbol{r}_T)=
    -\frac{g(k^2)\sqrt{\frac{x}{y}(1-\frac{x}{y})}}{(\boldsymbol{k}_T
    -\frac{x}{y}\boldsymbol{r}_T)^2+L_2^2(m^2)},
\end{align}
$g(k^2)$ is the form factor for the coupling of the pion meson-quark-sea quark vertex, and
\begin{align}		
    	L_2^2(m^2)=\frac{x}{y}m^2+\left(1-\frac{x}{y}\right)m^2
    	-\frac{x}{y}\left(1-\frac{x}{y}\right){m_\pi}^2.
\end{align}

For the form factors $g(r^2)$ and $g(k^2)$, we adopt the dipolar form
\begin{align}
    	g(r^2)&=-g_1(1-y)\frac{\boldsymbol{r}_T^2+L_1^2(m_\pi^2)}
    	{[\boldsymbol{r}_T^2+L_1^2(\Lambda^2_\pi)]^2},\label{eq15}	\\		g(k^2)&=-g_2(1-\frac{x}{y})\frac{(\boldsymbol{k}_T-\frac{x}{y}\boldsymbol{r}_T)^2+L_2^2(m^2)}{[(\boldsymbol{k}_T-\frac{x}{y}\boldsymbol{r}_T)^2+L_2^2(\Lambda^2_{\bar{q}})]^2}.\label{eq16}	
\end{align}
    
If the quark is treated as the active parton and the antiquark as the spectator, the transverse momentum and longitudinal momentum fraction of the active quark are $-\boldsymbol{k}_T$ and $-x$, respectively. Accordingly, the transverse momentum and longitudinal momentum fraction of the spectator antiquark are $\boldsymbol{r}_T+\boldsymbol{k}_T$ and $y+x$, respectively. We then find that the LCWFs corresponding to these two cases are related by
    \begin{align}\label{relation1}	\notag\psi{_+}{_+}(-x,y,-\boldsymbol{k}_T,\boldsymbol{r}_T)&=
    	\psi{_+}{_+}(x,y,\boldsymbol{k}_T,\boldsymbol{r}_T),\\	\notag\psi{_+}{_-}(-x,y,-\boldsymbol{k}_T,\boldsymbol{r}_T)&=-
    	\psi{_+}{_-}(x,y,\boldsymbol{k}_T,\boldsymbol{r}_T), \\      	\notag\psi{_-}{_+}(-x,y,-\boldsymbol{k}_T,\boldsymbol{r}_T)&=-
    	\psi{_-}{_+}(x,y,\boldsymbol{k}_T,\boldsymbol{r}_T), \\   	\psi{_-}{_-}(-x,y,-\boldsymbol{k}_T,\boldsymbol{r}_T)&=\psi{_-}{_-}(x,y,\boldsymbol{k}_T,\boldsymbol{r}_T).
    \end{align} 

The antiquark distributions are defined through the conjugate correlation function, which is related to the quark-quark correlation function by~\cite{Kumano:2021fem}
    \begin{align}\label{relation2}
    	\notag&W^{C[\Gamma]}_{\Lambda^{\prime}\Lambda}(x, \boldsymbol{k}_{T}, t; m_{q},m_{\bar{q}})
    	\\&=\left\{\begin{array}{ll}
    		-W^{[\Gamma]}_{\Lambda^{\prime}\Lambda}(-x, -\boldsymbol{k}_{T}, t; m_{\bar{q}},m_{q})  \text { for } \Gamma =\gamma^{j}, \gamma^{-}, i \sigma^{i j} \gamma_{5}, i \sigma^{j-} \gamma_{5} \\
    		+W^{[\Gamma]}_{\Lambda^{\prime}\Lambda}(-x, -\boldsymbol{k}_{T}, t; m_{\bar{q}},m_{q})  \text { for } \Gamma = \mathbf{1}, \gamma^{j} \gamma_{5}
    	\end{array}\right..
    \end{align}
Using Eq.~(\ref{correlation}) as well as the relations in Eq.~(\ref{relation1}) and Eq.~(\ref{relation2}),  we can obtain the overlap representation of the generalized correlator in terms of LCWFs for antiquark
    \begin{widetext}
    \begin{align}
    	\notag W^{[\gamma^{-}]}_{\Lambda^{\prime}\Lambda} &=-\int_{x}^{1}\frac{dy}{y}\int\frac{d^2\boldsymbol{r}_T}{16\pi^3}\sum_{{\lambda_B}{\lambda_q}} \left[\frac{m^{2}+\boldsymbol{k}_{T}^{2}-\frac{1}{4}\boldsymbol{\Delta}_{T}^{2}+i\boldsymbol{k}_{T} \times\boldsymbol{\Delta}_{T}}{2(xP^{+})^{2}}\psi^{\Lambda^{\prime}*}_{{\lambda_B}{\lambda_q}{-}}(x,y,\boldsymbol{k}^{\prime\prime}_T,\boldsymbol{r}^{\prime\prime}_T) \psi^{\Lambda}_{{\lambda_B}{\lambda_q}{-}}(x,y,\boldsymbol{k}^{\prime}_T,\boldsymbol{r}^{\prime}_T)
    	\right.\displaybreak[0] \\\notag&\left.
    	+\frac{m^{2}+\boldsymbol{k}_{T}^{2}-\frac{1}{4}\boldsymbol{\Delta}_{T}^{2}-i\boldsymbol{k}_{T} \times\boldsymbol{\Delta}_{T}}{2(xP^{+})^{2}}\psi^{\Lambda^{\prime}*}_{{\lambda_B}{\lambda_q}{+}}(x,y,\boldsymbol{k}^{\prime\prime}_T,\boldsymbol{r}^{\prime\prime}_T) \psi^{\Lambda}_{{\lambda_B}{\lambda_q}{+}}(x,y,\boldsymbol{k}^{\prime}_T,\boldsymbol{r}^{\prime}_T)
    	\right.\displaybreak[0] \\\notag&\left.
    	+\frac{-m(\Delta_1-i\Delta_2)}{2(xP^{+})^{2}}  \psi^{\Lambda^{\prime}*}_{{\lambda_B}{\lambda_q}{-}}(x,y,\boldsymbol{k}^{\prime\prime}_T,\boldsymbol{r}^{\prime\prime}_T) \psi^{\Lambda}_{{\lambda_B}{\lambda_q}{+}}(x,y,\boldsymbol{k}^{\prime}_T,\boldsymbol{r}^{\prime}_T)
    	\right.\displaybreak[0] \\&\left.
    	+ \frac{m(\Delta_1+i\Delta_2)}{2(xP^{+})^{2}}  \psi^{\Lambda^{\prime}*}_{{\lambda_B}{\lambda_q}{+}}(x,y,\boldsymbol{k}^{\prime\prime}_T,\boldsymbol{r}^{\prime\prime}_T) \psi^{\Lambda}_{{\lambda_B}{\lambda_q}{-}}(x,y,\boldsymbol{k}^{\prime}_T,\boldsymbol{r}^{\prime}_T) \right],\label{W4}
    	\\
    	\notag W^{[\gamma^{-}\gamma_{5}]}_{\Lambda^{\prime}\Lambda}&=\int_{x}^{1}\frac{dy}{y}\int\frac{d^2\boldsymbol{r}_T}{16\pi^3}\sum_{{\lambda_B}{\lambda_q}} \left[-\frac{m^{2}-\boldsymbol{k}_{T}^{2}+\frac{1}{4}\boldsymbol{\Delta}_{T}^{2}-i\boldsymbol{k}_{T} \times\boldsymbol{\Delta}_{T}}{2(xP^{+})^{2}}\psi^{\Lambda^{\prime}*}_{{\lambda_B}{\lambda_q}{-}}(x,y,\boldsymbol{k}^{\prime\prime}_T,\boldsymbol{r}^{\prime\prime}_T) \psi^{\Lambda}_{{\lambda_B}{\lambda_q}{-}}(x,y,\boldsymbol{k}^{\prime}_T,\boldsymbol{r}^{\prime}_T)
    	\right.\displaybreak[0] \\\notag&\left.
    	+\frac{m^{2}-\boldsymbol{k}_{T}^{2}+\frac{1}{4}\boldsymbol{\Delta}_{T}^{2}+i\boldsymbol{k}_{T} \times\boldsymbol{\Delta}_{T}}{2(xP^{+})^{2}}\psi^{\Lambda^{\prime}*}_{{\lambda_B}{\lambda_q}{+}}(x,y,\boldsymbol{k}^{\prime\prime}_T,\boldsymbol{r}^{\prime\prime}_T) \psi^{\Lambda}_{{\lambda_B}{\lambda_q}{+}}(x,y,\boldsymbol{k}^{\prime}_T,\boldsymbol{r}^{\prime}_T)
    	\right.\displaybreak[0] \\\notag&\left.
    	+\frac{m(k_1-ik_2)}{(xP^{+})^{2}}  \psi^{\Lambda^{\prime}*}_{{\lambda_B}{\lambda_q}{-}}(x,y,\boldsymbol{k}^{\prime\prime}_T,\boldsymbol{r}^{\prime\prime}_T) \psi^{\Lambda}_{{\lambda_B}{\lambda_q}{+}}(x,y,\boldsymbol{k}^{\prime}_T,\boldsymbol{r}^{\prime}_T)
    	\right.\displaybreak[0] \\&\left.
    	+ \frac{m(k_1+ik_2)}{(xP^{+})^{2}}  \psi^{\Lambda^{\prime}*}_{{\lambda_B}{\lambda_q}{+}}(x,y,\boldsymbol{k}^{\prime\prime}_T,\boldsymbol{r}^{\prime\prime}_T) \psi^{\Lambda}_{{\lambda_B}{\lambda_q}{-}}(x,y,\boldsymbol{k}^{\prime}_T,\boldsymbol{r}^{\prime}_T) \right],\label{W5}
    	\\
    	\notag W^{[i \sigma^{1 -} \gamma_{5}]}_{\Lambda^{\prime}\Lambda}  &=-\int_{x}^{1}\frac{dy}{y}\int\frac{d^2\boldsymbol{r}_T}{16\pi^3}\sum_{{\lambda_B}{\lambda_q}} \left[\frac{m^{2}-({k}_{1}-i{k}_{2})^{2}+\frac{1}{4}({\Delta}_{1}-i{\Delta}_{2})^{2}}{(xP^{+})^{2}}\psi^{\Lambda^{\prime}*}_{{\lambda_B}{\lambda_q}{-}}(x,y,\boldsymbol{k}^{\prime\prime}_T,\boldsymbol{r}^{\prime\prime}_T) \psi^{\Lambda}_{{\lambda_B}{\lambda_q}{+}}(x,y,\boldsymbol{k}^{\prime}_T,\boldsymbol{r}^{\prime}_T)
    	\right.\displaybreak[0] \\\notag&\left.
    	+\frac{m^{2}-({k}_{1}+i{k}_{2})^{2}+\frac{1}{4}({\Delta}_{1}+i{\Delta}_{2})^{2}}{(xP^{+})^{2}}\psi^{\Lambda^{\prime}*}_{{\lambda_B}{\lambda_q}{+}}(x,y,\boldsymbol{k}^{\prime\prime}_T,\boldsymbol{r}^{\prime\prime}_T) \psi^{\Lambda}_{{\lambda_B}{\lambda_q}{-}}(x,y,\boldsymbol{k}^{\prime}_T,\boldsymbol{r}^{\prime}_T) 
    	\right.\displaybreak[0] \\\notag&\left.
    	+\frac{m(2k_1+i\Delta_2)}{(xP^{+})^{2}}  \psi^{\Lambda^{\prime}*}_{{\lambda_B}{\lambda_q}{-}}(x,y,\boldsymbol{k}^{\prime\prime}_T,\boldsymbol{r}^{\prime\prime}_T) \psi^{\Lambda}_{{\lambda_B}{\lambda_q}{-}}(x,y,\boldsymbol{k}^{\prime}_T,\boldsymbol{r}^{\prime}_T)
    	\right.\displaybreak[0] \\&\left.
    	+ \frac{-m(2k_1-i\Delta_2)}{(xP^{+})^{2}}  \psi^{\Lambda^{\prime}*}_{{\lambda_B}{\lambda_q}{+}}(x,y,\boldsymbol{k}^{\prime\prime}_T,\boldsymbol{r}^{\prime\prime}_T) \psi^{\Lambda}_{{\lambda_B}{\lambda_q}{+}}(x,y,\boldsymbol{k}^{\prime}_T,\boldsymbol{r}^{\prime}_T) \right],\label{W6}
    	\\
    	\notag W^{[i \sigma^{2 -} \gamma_{5}]}_{\Lambda^{\prime}\Lambda}  &=-\int_{x}^{1}\frac{dy}{y} \int\frac{d^2\boldsymbol{r}_T}{16\pi^3}\sum_{{\lambda_B}{\lambda_q}} \left[\frac{-i\left(m^{2}+({k}_{1}-i{k}_{2})^{2}-\frac{1}{4}({\Delta}_{1}-i{\Delta}_{2})^{2}\right)}{(xP^{+})^{2}}\psi^{\Lambda^{\prime}*}_{{\lambda_B}{\lambda_q}{-}}(x,y,\boldsymbol{k}^{\prime\prime}_T,\boldsymbol{r}^{\prime\prime}_T) \psi^{\Lambda}_{{\lambda_B}{\lambda_q}{+}}(x,y,\boldsymbol{k}^{\prime}_T,\boldsymbol{r}^{\prime}_T)
    	\right.\displaybreak[0] \\\notag&\left.
    	+\frac{i\left(m^{2}+({k}_{1}+i{k}_{2})^{2}-\frac{1}{4}({\Delta}_{1}+i{\Delta}_{2})^{2}\right)}{(xP^{+})^{2}}\psi^{\Lambda^{\prime}*}_{{\lambda_B}{\lambda_q}{+}}(x,y,\boldsymbol{k}^{\prime\prime}_T,\boldsymbol{r}^{\prime\prime}_T) \psi^{\Lambda}_{{\lambda_B}{\lambda_q}{-}}(x,y,\boldsymbol{k}^{\prime}_T,\boldsymbol{r}^{\prime}_T) 
    	\right.\displaybreak[0] \\\notag&\left.
    	+\frac{m(2k_2-i\Delta_1)}{(xP^{+})^{2}}  \psi^{\Lambda^{\prime}*}_{{\lambda_B}{\lambda_q}{-}}(x,y,\boldsymbol{k}^{\prime\prime}_T,\boldsymbol{r}^{\prime\prime}_T) \psi^{\Lambda}_{{\lambda_B}{\lambda_q}{-}}(x,y,\boldsymbol{k}^{\prime}_T,\boldsymbol{r}^{\prime}_T)
    	\right.\displaybreak[0] \\&\left.
    	+ \frac{-m(2k_2+i\Delta_1)}{(xP^{+})^{2}}  \psi^{\Lambda^{\prime}*}_{{\lambda_B}{\lambda_q}{+}}(x,y,\boldsymbol{k}^{\prime\prime}_T,\boldsymbol{r}^{\prime\prime}_T) \psi^{\Lambda}_{{\lambda_B}{\lambda_q}{+}}(x,y,\boldsymbol{k}^{\prime}_T,\boldsymbol{r}^{\prime}_T) \right].\label{W7}
    \end{align} 

    Based on the above LCWFs in Eqs.~(\ref{former}) (\ref{later}) and the overlap representation for the generalized correlator in Eqs.~(\ref{W4})-(\ref{W7}), we obtain the analytic results of the twist-4 F-type GTMDs of the sea quark in proton at $\xi=0$
    \begin{align}
    	\notag &F_{3,1}^{\overline{q}/P}(x,\boldsymbol{k}_{T},\boldsymbol{\Delta}_{T})=-\frac{1}{x^2M^2}\int_{x}^{1}\frac{dy}{y} \int d^2\boldsymbol{r}_T \frac{g_1^2g_2^2y(1-y)^2(1-\frac{x}{y})^2 \left[(M_B-(1-y)M)^2+\boldsymbol{r}_T^2-\frac{1}{4}(1-y)^2\boldsymbol{\Delta}_T^2\right]}{4(2\pi)^6 D_1(y,\boldsymbol{r}_T,\boldsymbol{\Delta}_T) D_2(\frac{x}{y},\boldsymbol{k}_T-\frac{x}{y}\boldsymbol{r}_T,\boldsymbol{\Delta}_T)} 
    	\\&\times \left[ \left(m^2+(\boldsymbol{k}_T-\frac{x}{y}  \boldsymbol{r}_T)^2-\frac{1}{4}(1-\frac{x}{y})^2\boldsymbol{\Delta}_T^2\right)(m^2+\boldsymbol{k}_T^2-\frac{1}{4}\boldsymbol{\Delta}_T^2)+(1-\frac{x}{y}) (\boldsymbol{k}_T\times\boldsymbol{\Delta}_{T})(\boldsymbol{k}_T-\frac{x}{y}  \boldsymbol{r}_T)\times\boldsymbol{\Delta}_{T}+(1-\frac{x}{y})m^2\boldsymbol{\Delta}_T^2\right],
    	\\
    	&F_{3,2}^{\overline{q}/P}(x,\boldsymbol{k}_{T},\boldsymbol{\Delta}_{T})=0,
    	\\\notag &F_{3,3}^{\overline{q}/P}(x,\boldsymbol{k}_{T},\boldsymbol{\Delta}_{T})=-\frac{1}{x^2M}\int_{x}^{1}\frac{dy}{y} \int d^2\boldsymbol{r}_T \frac{g_1^2g_2^2y(1-y)^2(1-\frac{x}{y})^2}{16(2\pi)^6 D_1(y,\boldsymbol{r}_T,\boldsymbol{\Delta}_T) D_2(\frac{x}{y},\boldsymbol{k}_T-\frac{x}{y}\boldsymbol{r}_T,\boldsymbol{\Delta}_T)}   \\\notag&\times\left[\frac{1}{2M }\left((M_B-(1-y)M)^2+\boldsymbol{r}_T^2-\frac{1}{4}(1-y)^2\boldsymbol{\Delta}_T^2\right)-(1-y)(M_B-(1-y)M)\right]
    	\\&\times \left[ \left(m^2+(\boldsymbol{k}_T-\frac{x}{y}  \boldsymbol{r}_T)^2-\frac{1}{4}(1-\frac{x}{y})^2\boldsymbol{\Delta}_T^2\right)(m^2+\boldsymbol{k}_T^2-\frac{1}{4}\boldsymbol{\Delta}_T^2)+(1-\frac{x}{y}) (\boldsymbol{k}_T\times\boldsymbol{\Delta}_{T})(\boldsymbol{k}_T-\frac{x}{y}  \boldsymbol{r}_T)\times\boldsymbol{\Delta}_{T}+(1-\frac{x}{y})m^2\boldsymbol{\Delta}_T^2\right],
    	\\\notag &F_{3,4}^{\overline{q}/P}(x,\boldsymbol{k}_{T},\boldsymbol{\Delta}_{T})=-\frac{1}{x^2}\int_{x}^{1}\frac{dy}{y} \int{d^2\boldsymbol{r}_T}  \frac{g_1^2g_2^2y(1-y)^3(1-\frac{x}{y})^2 }{4(2\pi)^6 D_1(y,\boldsymbol{r}_T,\boldsymbol{\Delta}_T) D_2(\frac{x}{y},\boldsymbol{k}_T-\frac{x}{y}\boldsymbol{r}_T,\boldsymbol{\Delta}_T)}\cdot \frac{\boldsymbol{r}_T\times\boldsymbol{\Delta}_{T}}{\boldsymbol{k}_T\times\boldsymbol{\Delta}_{T}}
    	\\&\times \left[ \left(m^2+(\boldsymbol{k}_T-\frac{x}{y}  \boldsymbol{r}_T)^2-\frac{1}{4}(1-\frac{x}{y})^2\boldsymbol{\Delta}_T^2\right)(m^2+\boldsymbol{k}_T^2-\frac{1}{4}\boldsymbol{\Delta}_T^2)+(1-\frac{x}{y}) (\boldsymbol{k}_T\times\boldsymbol{\Delta}_{T})(\boldsymbol{k}_T-\frac{x}{y}  \boldsymbol{r}_T)\times\boldsymbol{\Delta}_{T}+(1-\frac{x}{y})m^2\boldsymbol{\Delta}_T^2\right],
    \end{align}
    the analytic results of the twist-4 G-type GTMDs
    \begin{align}
    	\notag
    	&G_{3,1}^{\overline{q}/P}(x,\boldsymbol{k}_{T},\boldsymbol{\Delta}_{T})=\frac{1}{x^2}\int_{x}^{1}\frac{dy}{y} \int{d^2\boldsymbol{r}_T} \frac{g_1^2 g_2^2 y(1-y)^2 (1-\frac{x}{y})^2 \left[(M_B-(1-y)M)^2+\boldsymbol{r}_T^2-\frac{1}{4}(1-y)^2\boldsymbol{\Delta}_T^2\right] }{4(2\pi)^6 D_1(y,\boldsymbol{r}_T,\boldsymbol{\Delta}_T) D_2(\frac{x}{y},\boldsymbol{k}_T-\frac{x}{y}\boldsymbol{r}_T,\boldsymbol{\Delta}_T)} 
    	\\&\times \left[ m^2+(\boldsymbol{k}_T-\frac{x}{y}  \boldsymbol{r}_T)^2-\frac{1}{4}(1-\frac{x}{y})^2\boldsymbol{\Delta}_T^2-2(1-\frac{x}{y})m^2+(1-\frac{x}{y})(m^2-\boldsymbol{k}_T^2+\frac{1}{4}\boldsymbol{\Delta}_T^2)(\boldsymbol{k}_T-\frac{x}{y}  \boldsymbol{r}_T)\times\boldsymbol{\Delta}_{T}/(\boldsymbol{k}_{T}\times\boldsymbol{\Delta}_{T})\right],
    	\\\notag
    	&G_{3,2}^{\overline{q}/P}(x,\boldsymbol{k}_{T},\boldsymbol{\Delta}_{T})
    	=\frac{1}{x^2M}\int_{x}^{1}\frac{dy}{y} \int{d^2\boldsymbol{r}_T} \frac{ g_1^2 g_2^2 y(1-y)^2 (1-\frac{x}{y})^2\boldsymbol{\Delta}_T^2}{2(2\pi)^6D_1(y,\boldsymbol{r}_T,\boldsymbol{\Delta}_T) D_2(\frac{x}{y},\boldsymbol{k}_T-\frac{x}{y}\boldsymbol{r}_T,\boldsymbol{\Delta}_T)} 
    	\\\notag&\times\left[\frac{1}{2M}\left((M_B-(1-y)M)^2+\boldsymbol{r}_T^2-\frac{1}{4}(1-y)^2\boldsymbol{\Delta}_T^2\right)-(1-y)(M_B-(1-y)M)\right]
    	\\&\times \left[ m^2+(\boldsymbol{k}_T-\frac{x}{y}  \boldsymbol{r}_T)^2-\frac{1}{4}(1-\frac{x}{y})^2\boldsymbol{\Delta}_T^2-2(1-\frac{x}{y})m^2+(1-\frac{x}{y})(m^2-\boldsymbol{k}_T^2+\frac{1}{4}\boldsymbol{\Delta}_T^2)(\boldsymbol{k}_T-\frac{x}{y}  \boldsymbol{r}_T)\times\boldsymbol{\Delta}_{T}/(\boldsymbol{k}_{T}\times\boldsymbol{\Delta}_{T})\right],
    	\\\notag
    	&G_{3,3}^{\overline{q}/P}(x,\boldsymbol{k}_{T},\boldsymbol{\Delta}_{T})
    	=-\frac{1}{x^2M}\int_{x}^{1}\frac{dy}{y} \int{d^2\boldsymbol{r}_T}\frac{ g_1^2 g_2^2 y(1-y)^2 (1-\frac{x}{y})^2\boldsymbol{k}_T\cdot\boldsymbol{\Delta}_T }{2(2\pi)^6D_1(y,\boldsymbol{r}_T,\boldsymbol{\Delta}_T) D_2(\frac{x}{y},\boldsymbol{k}_T-\frac{x}{y}\boldsymbol{r}_T,\boldsymbol{\Delta}_T)}
    	\\\notag&\times\left[\frac{1}{2M }\left((M_B-(1-y)M)^2+\boldsymbol{r}_T^2-\frac{1}{4}(1-y)^2\boldsymbol{\Delta}_T^2\right)-(1-y)(M_B-(1-y)M)\right]
    	\\&\times \left[ m^2+(\boldsymbol{k}_T-\frac{x}{y}  \boldsymbol{r}_T)^2-\frac{1}{4}(1-\frac{x}{y})^2\boldsymbol{\Delta}_T^2-2(1-\frac{x}{y})m^2+(1-\frac{x}{y})(m^2-\boldsymbol{k}_T^2+\frac{1}{4}\boldsymbol{\Delta}_T^2)(\boldsymbol{k}_T-\frac{x}{y}  \boldsymbol{r}_T)\times\boldsymbol{\Delta}_{T}/(\boldsymbol{k}_{T}\times\boldsymbol{\Delta}_{T})\right],
    	\\\notag
    	&G_{3,4}^{\overline{q}/P}(x,\boldsymbol{k}_{T},\boldsymbol{\Delta}_{T})=\frac{1}{x^2 M^2}\int_{x}^{1}\frac{dy}{y} \int{d^2\boldsymbol{r}_T}  \frac{g_1^2 g_2^2 y(1-y)^3 (1-\frac{x}{y})^2 \boldsymbol{\Delta}_T\times\boldsymbol{r}_T}{4(2\pi)^6D_1(y,\boldsymbol{r}_T,\boldsymbol{\Delta}_T) D_2(\frac{x}{y},\boldsymbol{k}_T-\frac{x}{y}\boldsymbol{r}_T,\boldsymbol{\Delta}_T)}
    	\\&\times \left[ \left(m^2+(\boldsymbol{k}_T-\frac{x}{y}  \boldsymbol{r}_T)^2-\frac{1}{4}(1-\frac{x}{y})^2\boldsymbol{\Delta}_T^2-2(1-\frac{x}{y})m^2\right)(\boldsymbol{k}_T\times\boldsymbol{\Delta}_T)+(1-\frac{x}{y})(m^2-\boldsymbol{k}_T^2+\frac{1}{4}\boldsymbol{\Delta}_T^2)(\boldsymbol{k}_T-\frac{x}{y}  \boldsymbol{r}_T)\times\boldsymbol{\Delta}_{T}\right],
    \end{align}
    and the analytic results of the twist-4 H-type GTMDs
    \begin{align}
    	\notag
    	&H_{3,1}^{\overline{q}/P}(x,\boldsymbol{k}_{T},\boldsymbol{\Delta}_{T})=\frac{1}{x M}\int_{x}^{1}\frac{dy}{y} \int{d^2\boldsymbol{r}_T} \frac{g_1^2 g_2^2 (1-y)^2 (1-\frac{x}{y})^3 m (\boldsymbol{k}_T\cdot\boldsymbol{r}_T)}{(2\pi)^6D_1(y,\boldsymbol{r}_T,\boldsymbol{\Delta}_T) D_2(\frac{x}{y},\boldsymbol{k}_T-\frac{x}{y}\boldsymbol{r}_T,\boldsymbol{\Delta}_T)}\cdot \frac{\boldsymbol{r}_T\times\boldsymbol{\Delta}_T}{\boldsymbol{k}_T\times\boldsymbol{\Delta}_T}
    	\\&\times \left[(M_B-(1-y)M)^2+\boldsymbol{r}_T^2-\frac{1}{4}(1-y)^2\boldsymbol{\Delta}_T^2\right],
    	\\\notag
    	&H_{3,2}^{\overline{q}/P}(x,\boldsymbol{k}_{T},\boldsymbol{\Delta}_{T})
    	=-\frac{1}{x^2M}\int_{x}^{1}\frac{dy}{y} \int{d^2\boldsymbol{r}_T} \frac{ g_1^2 g_2^2 y(1-y)^2 (1-\frac{x}{y})^2m \left[(M_B-(1-y)M)^2+\boldsymbol{r}_T^2-\frac{1}{4}(1-y)^2\boldsymbol{\Delta}_T^2\right] }{2(2\pi)^6D_1(y,\boldsymbol{r}_T,\boldsymbol{\Delta}_T) D_2(\frac{x}{y},\boldsymbol{k}_T-\frac{x}{y}\boldsymbol{r}_T,\boldsymbol{\Delta}_T)}
    	\\&\times \left[ \left(m^2+(\boldsymbol{k}_T-\frac{x}{y}  \boldsymbol{r}_T)^2-\frac{1}{4}(1-\frac{x}{y})^2\boldsymbol{\Delta}_T^2-(1-\frac{x}{y})(m^2+\boldsymbol{k}_T^2-\frac{1}{4}\boldsymbol{\Delta}_T^2)\right)+2(1-\frac{x}{y})\boldsymbol{k}_T^2\cdot\frac{\boldsymbol{r}_T\times\boldsymbol{\Delta}_T}{\boldsymbol{k}_T\times\boldsymbol{\Delta}_T}\right],
    	\\\notag
    	&H_{3,3}^{\overline{q}/P}(x,\boldsymbol{k}_{T},\boldsymbol{\Delta}_{T})
    	=\frac{1}{x^2M^2}\int_{x}^{1}\frac{dy}{y} \int{d^2\boldsymbol{r}_T}\frac{ g_1^2 g_2^2 y(1-y)^2 (1-\frac{x}{y})^2m}{2(2\pi)^6 D_1(y,\boldsymbol{r}_T,\boldsymbol{\Delta}_T) D_2(\frac{x}{y},\boldsymbol{k}_T-\frac{x}{y}\boldsymbol{r}_T,\boldsymbol{\Delta}_T)} \\\notag&\times\left[\frac{1}{2M	}\left((M_B-(1-y)M)^2+\boldsymbol{r}_T^2-\frac{1}{4}(1-y)^2\boldsymbol{\Delta}_T^2\right)-(1-y)(M_B-(1-y)M)\right]
    	\\\notag&\times \left[ \left(m^2+(\boldsymbol{k}_T-\frac{x}{y}  \boldsymbol{r}_T)^2-\frac{1}{4}(1-\frac{x}{y})^2\boldsymbol{\Delta}_T^2-(1-\frac{x}{y})(m^2+\boldsymbol{k}_T^2
    -\frac{1}{4}\boldsymbol{\Delta}_T^2)\right)\boldsymbol{\Delta}_{T}^{2}+2\frac{x}{y}(1-\frac{x}{y})( \boldsymbol{r}_T\times\boldsymbol{\Delta}_{T})( \boldsymbol{k}_T\times\boldsymbol{\Delta}_{T})\right],
    	\\\notag
    	&H_{3,4}^{\overline{q}/P}(x,\boldsymbol{k}_{T},\boldsymbol{\Delta}_{T})=0
    \end{align}
    \begin{align}
    	\notag
    	&H_{3,5}^{\overline{q}/P}(x,\boldsymbol{k}_{T},\boldsymbol{\Delta}_{T})
    	=-\frac{1}{x^2M^2}\int_{x}^{1}\frac{dy}{y} \int{d^2\boldsymbol{r}_T}\frac{ g_1^2 g_2^2 y(1-y)^2 (1-\frac{x}{y})^2m}{2(2\pi)^6 D_1(y,\boldsymbol{r}_T,\boldsymbol{\Delta}_T) D_2(\frac{x}{y},\boldsymbol{k}_T-\frac{x}{y}\boldsymbol{r}_T,\boldsymbol{\Delta}_T)} \\\notag&\times\left[\frac{1}{2M    	}\left((M_B-(1-y)M)^2+\boldsymbol{r}_T^2-\frac{1}{4}(1-y)^2\boldsymbol{\Delta}_T^2\right)-(1-y)(M_B-(1-y)M)\right]
    	\\\notag&\times \left[2\frac{x}{y}(1-\frac{x}{y})( \boldsymbol{r}_T\times\boldsymbol{\Delta}_{T})( \boldsymbol{k}_T\cdot\boldsymbol{\Delta}_{T})\right],
    	\\\notag
    	&H_{3,6}^{\overline{q}/P}(x,\boldsymbol{k}_{T},\boldsymbol{\Delta}_{T})
    	=-\frac{1}{x^2M^2}\int_{x}^{1}\frac{dy}{y} \int{d^2\boldsymbol{r}_T}\frac{ g_1^2 g_2^2 y(1-y)^2 (1-\frac{x}{y})^2m}{2(2\pi)^6 D_1(y,\boldsymbol{r}_T,\boldsymbol{\Delta}_T) D_2(\frac{x}{y},\boldsymbol{k}_T-\frac{x}{y}\boldsymbol{r}_T,\boldsymbol{\Delta}_T)} \\\notag&\times\left[\frac{1}{2M}\left((M_B-(1-y)M)^2+\boldsymbol{r}_T^2-\frac{1}{4}(1-y)^2
    \boldsymbol{\Delta}_T^2\right)-(1-y)(M_B-(1-y)M)\right]
    	\\\notag&\times \left[ \left(m^2+(\boldsymbol{k}_T-\frac{x}{y}  \boldsymbol{r}_T)^2-\frac{1}{4}(1-\frac{x}{y})^2\boldsymbol{\Delta}_T^2-(1-\frac{x}{y})(m^2+\boldsymbol{k}_T^2-\frac{1}{4}\boldsymbol{\Delta}_T^2)\right)M^{2}
    	\right.\\&\left.
    	+2\frac{x}{y}(1-\frac{x}{y})( \boldsymbol{r}_T\times\boldsymbol{\Delta}_{T})(  \frac{M^{2} \boldsymbol{k}_T\times\boldsymbol{\Delta}_{T} }{\boldsymbol{\Delta}_{T}^{2}}-\frac{\boldsymbol{k}_T\cdot\boldsymbol{\Delta}_{T}}{\boldsymbol{\Delta}_{T}^{2}})\right],
    	\\\notag
    	&H_{3,7}^{\overline{q}/P}(x,\boldsymbol{k}_{T},\boldsymbol{\Delta}_{T})
    	=\frac{1}{x^2M}\int_{x}^{1}\frac{dy}{y} \int{d^2\boldsymbol{r}_T}\frac{ g_1^2 g_2^2 y(1-y)^3 (1-\frac{x}{y})^2m(\boldsymbol{r}_T\times\boldsymbol{\Delta}_T)}{4(2\pi)^6 D_1(y,\boldsymbol{r}_T,\boldsymbol{\Delta}_T) D_2(\frac{x}{y},\boldsymbol{k}_T-\frac{x}{y}\boldsymbol{r}_T,\boldsymbol{\Delta}_T)} 
    	\\&\times \left[ \left(m^2+(\boldsymbol{k}_T-\frac{x}{y}  \boldsymbol{r}_T)^2-\frac{1}{4}(1-\frac{x}{y})^2\boldsymbol{\Delta}_T^2-(1-\frac{x}{y})(m^2+\boldsymbol{k}_T^2-\frac{1}{4}\boldsymbol{\Delta}_T^2)\right)\frac{\boldsymbol{\Delta}_{T}^{2}}{\boldsymbol{k}_T\times\boldsymbol{\Delta}_{T}}+2\frac{x}{y}(1-\frac{x}{y})( \boldsymbol{r}_T\times\boldsymbol{\Delta}_{T})\right],
    	\\\notag
    	&H_{3,8}^{\overline{q}/P}(x,\boldsymbol{k}_{T},\boldsymbol{\Delta}_{T})=-\frac{1}{x^2 M}\int_{x}^{1}\frac{dy}{y} \int{d^2\boldsymbol{r}_T}  \frac{g_1^2 g_2^2 y(1-y)^3 (1-\frac{x}{y})^2 m(\boldsymbol{r}_T\times\boldsymbol{\Delta}_T)}{2(2\pi)^6D_1(y,\boldsymbol{r}_T,\boldsymbol{\Delta}_T) D_2(\frac{x}{y},\boldsymbol{k}_T-\frac{x}{y}\boldsymbol{r}_T,\boldsymbol{\Delta}_T)}
    	\\&\times\left[m^2+(\boldsymbol{k}_T-\frac{x}{y}  \boldsymbol{r}_T)^2-\frac{1}{4}(1-\frac{x}{y})^2\boldsymbol{\Delta}_T^2-(1-\frac{x}{y})(m^2+\boldsymbol{k}_T^2-\frac{1}{4}\boldsymbol{\Delta}_T^2)\right]\frac{\boldsymbol{k}_T\cdot\boldsymbol{\Delta}_T}{\boldsymbol{k}_T\times\boldsymbol{\Delta}_T}.
    \end{align}
    where
    \begin{align}      	 
    D_1(y,\boldsymbol{r}_T,\boldsymbol{\Delta}_T)&=\left[(\boldsymbol{r}_T-\frac{1}{2}(1-y)
    	\boldsymbol{\Delta}_T)^2+L_1^2\right]^2 \left[(\boldsymbol{r}_T+\frac{1}{2}(1-y)\boldsymbol{\Delta}_T)^2+L_1^2\right]^2,\\
    	D_2(\frac{x}{y},\boldsymbol{k}_T-\frac{x}{y}\boldsymbol{r}_T,\boldsymbol{\Delta}_T)
    	&=\left[[(\boldsymbol{k}_T-\frac{x}{y}\boldsymbol{r}_T)-\frac{1}{2}(1-\frac{x}{y})
    	\boldsymbol{\Delta}_T]^2+L_2^2\right]^2
    	\left[[(\boldsymbol{k}_T-\frac{x}{y}\boldsymbol{r}_T)
    	+\frac{1}{2}(1-\frac{x}{y})\boldsymbol{\Delta}_T]^2+L_2^2\right]^2.
    \end{align}
  
    \begin{figure*}[htbp]
    	\centering
    	\subfigure{\begin{minipage}[b]{0.245\linewidth}
    			\centering
    			\includegraphics[width=\linewidth]{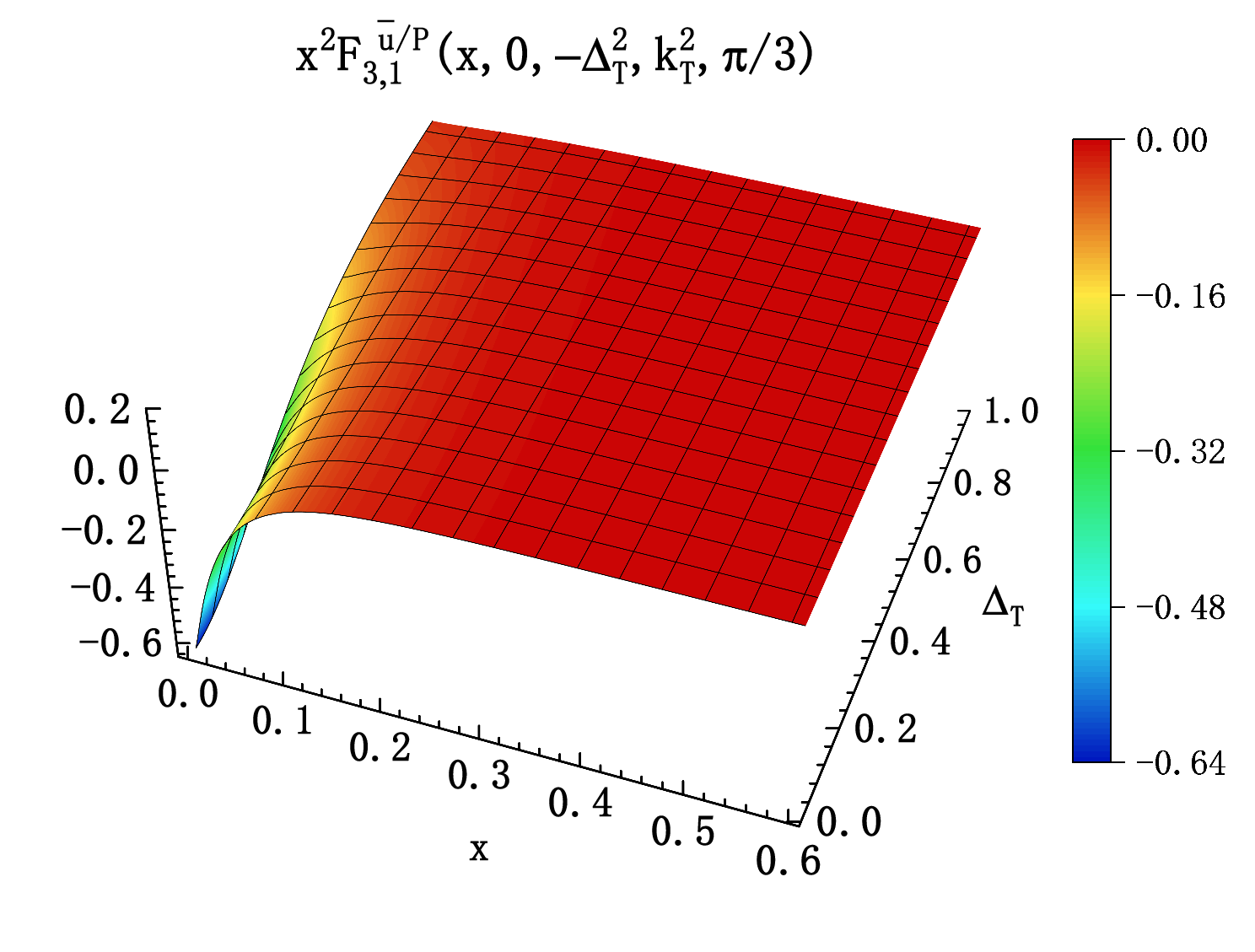}
    	\end{minipage}}
    	\subfigure{\begin{minipage}[b]{0.245\linewidth}
    			\centering
    			\includegraphics[width=\linewidth]{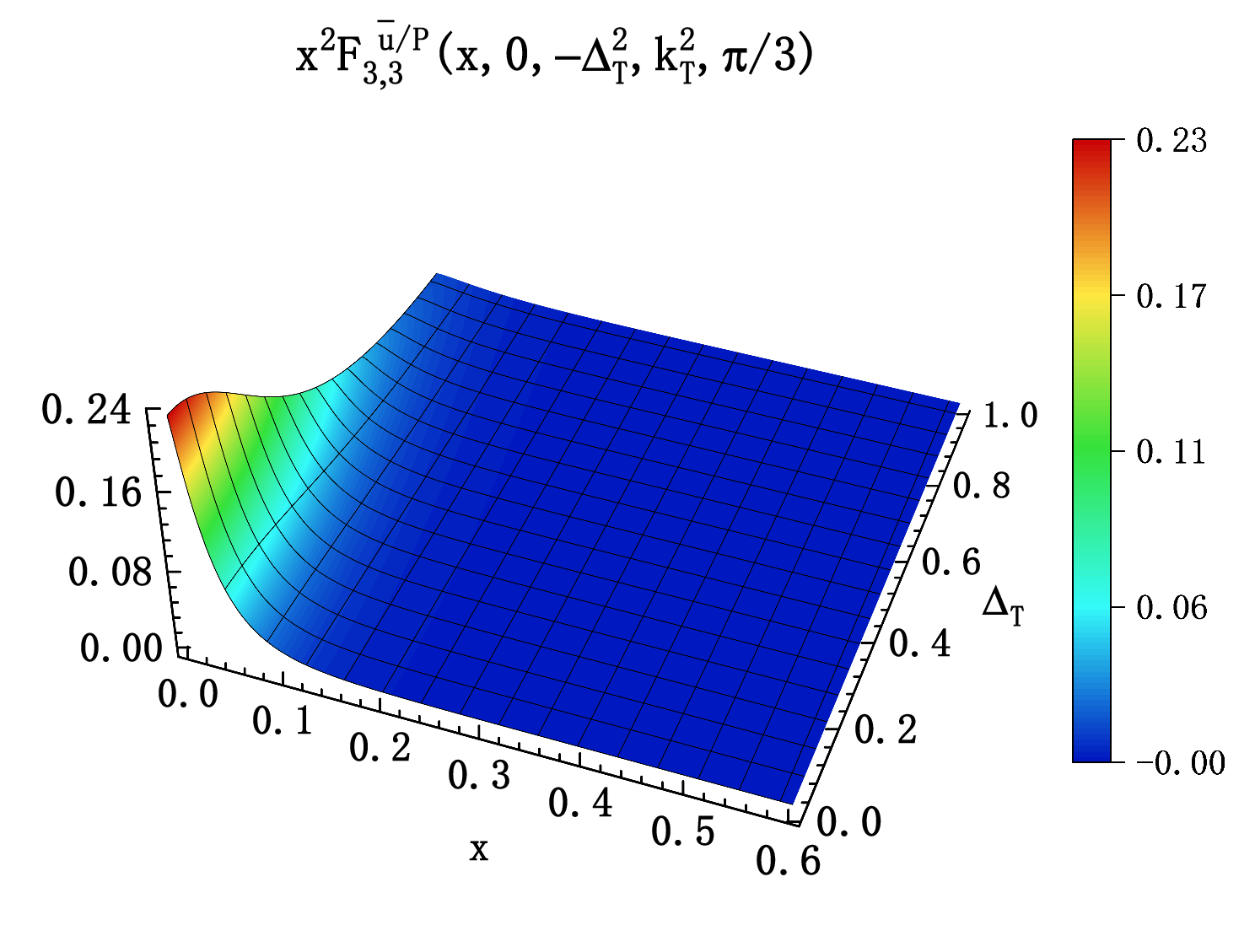}    	
    	\end{minipage}}
    	\subfigure{\begin{minipage}[b]{0.245\linewidth}
    			\centering
    			\includegraphics[width=\linewidth]{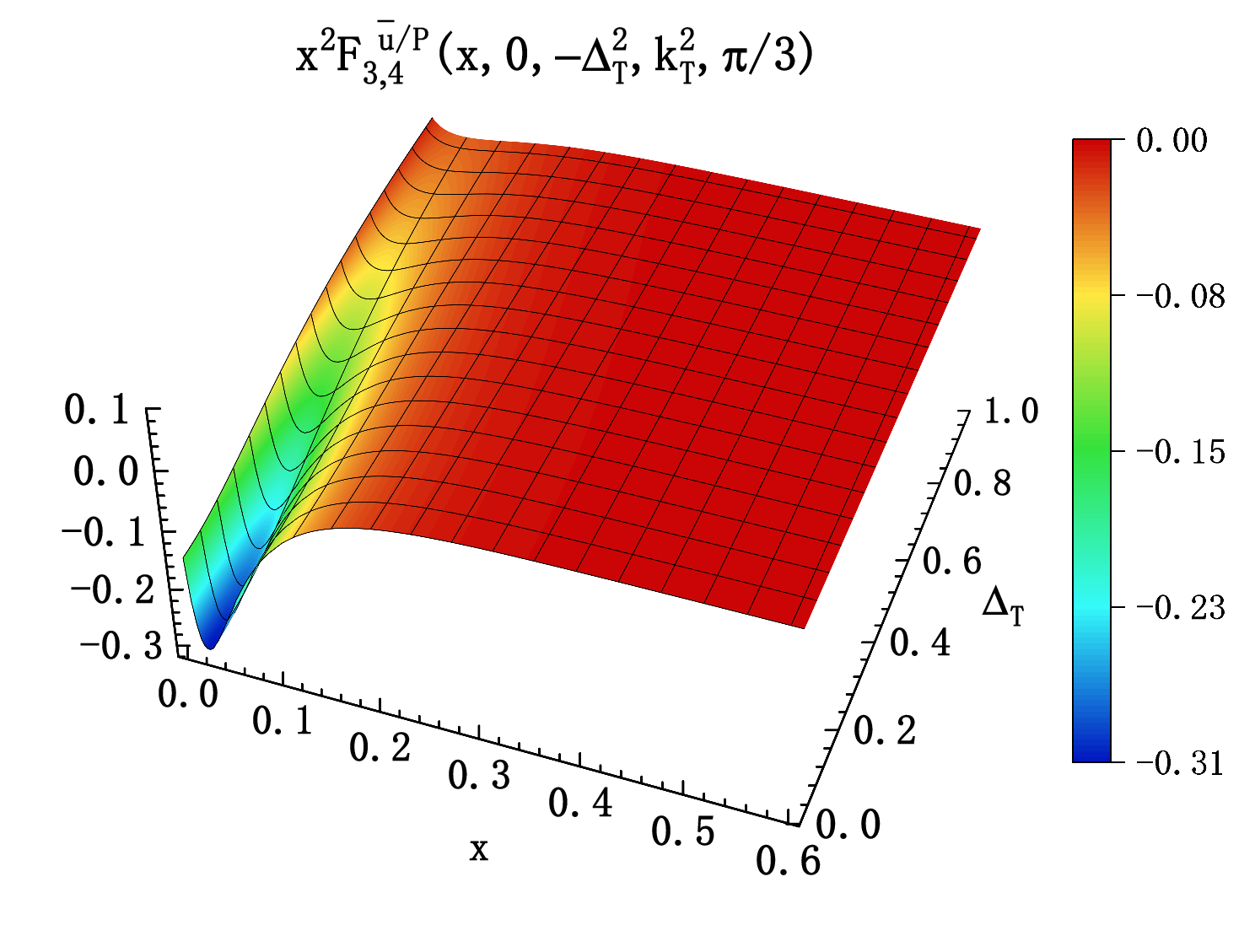}
    	\end{minipage}}

    	\subfigure{\begin{minipage}[b]{0.245\linewidth}
    			\centering
    			\includegraphics[width=\linewidth]{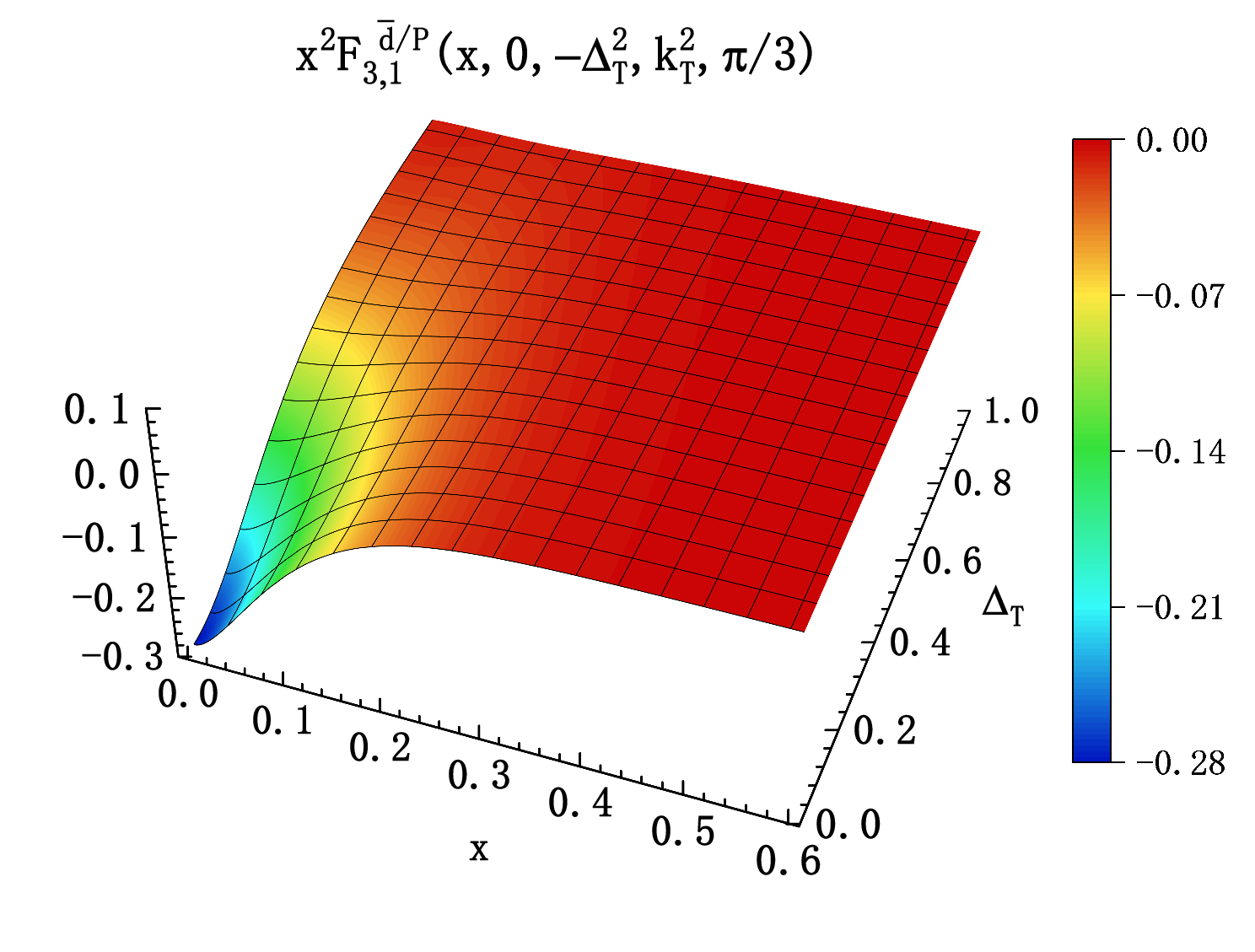}
    	\end{minipage}}	
    	\subfigure{\begin{minipage}[b]{0.245\linewidth}
    			\centering
    			\includegraphics[width=\linewidth]{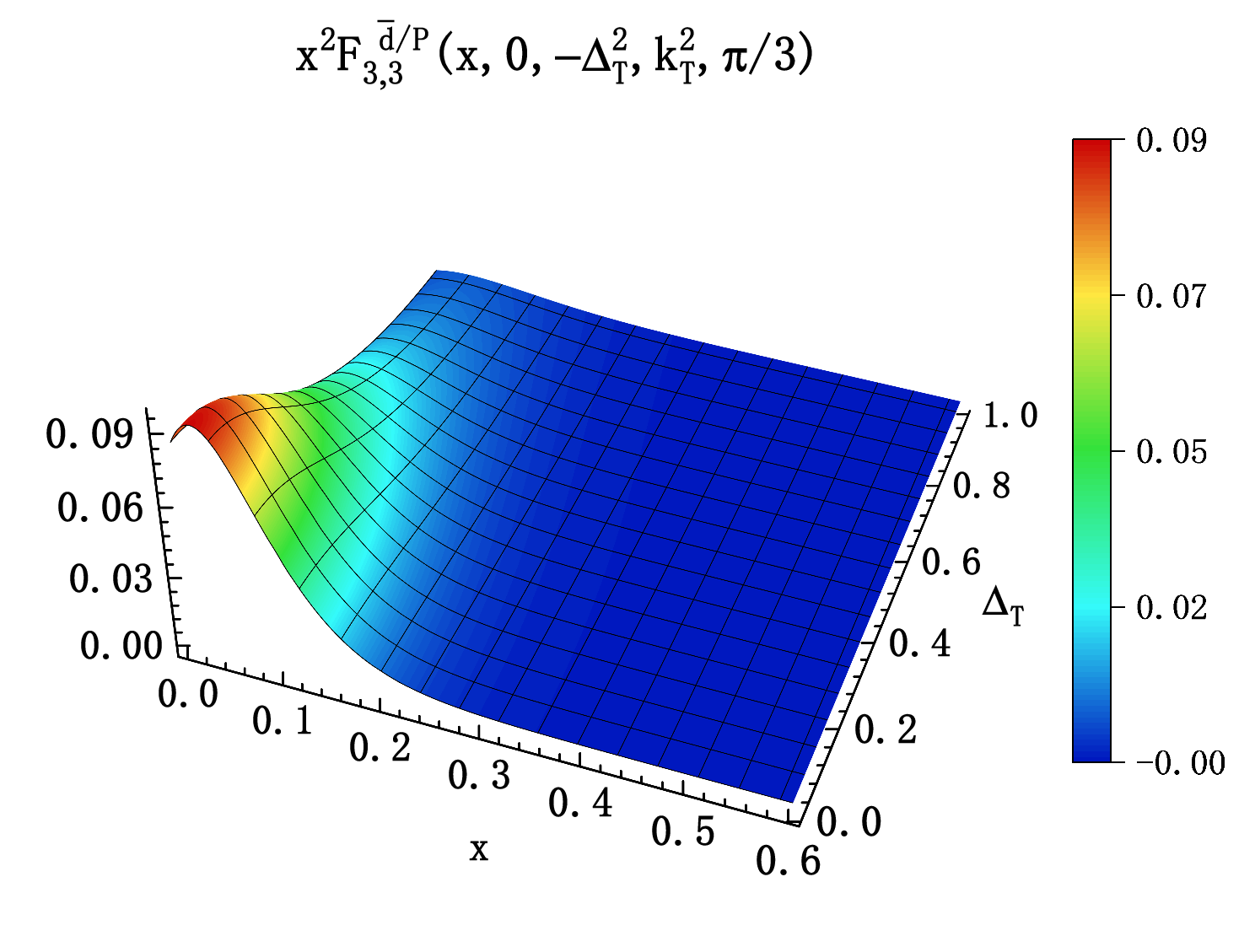}
    	\end{minipage}}
    	\subfigure{\begin{minipage}[b]{0.245\linewidth}
    			\centering
    			\includegraphics[width=\linewidth]{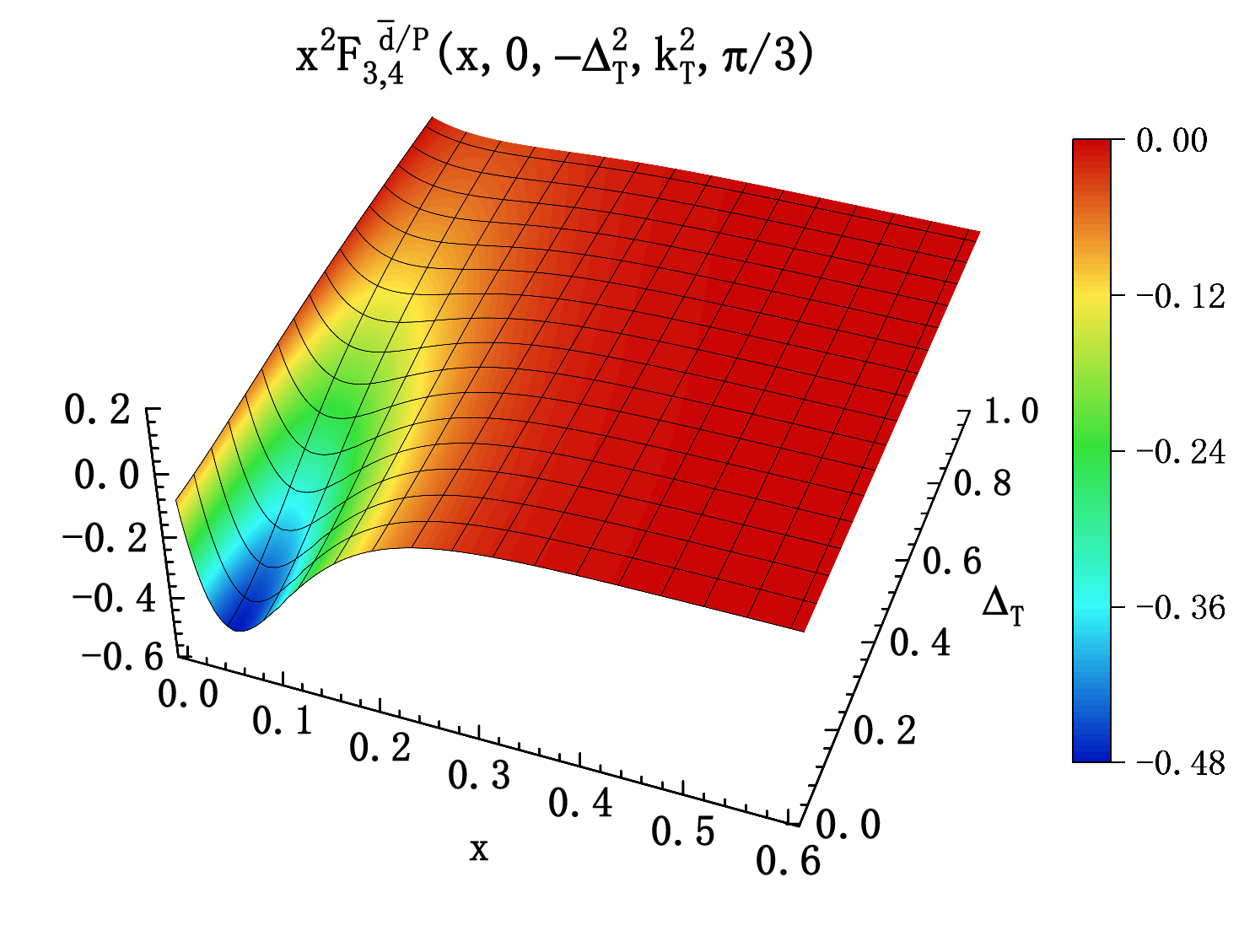}    	
    	\end{minipage}}
    	\caption{The twist-4 GTMDs (multiplied by $x^2$) for unpolarized sea quarks are plotted as functions of $x$ and ${\Delta}_T$ at fixed ${k}_T=0.1$ GeV and $\theta=\pi/3$.
} \label{udbarF}      
    \end{figure*}

    \begin{figure*}[htbp]
    	\centering
    	\subfigure{\begin{minipage}[b]{0.245\linewidth}
    			\centering
    			\includegraphics[width=\linewidth]{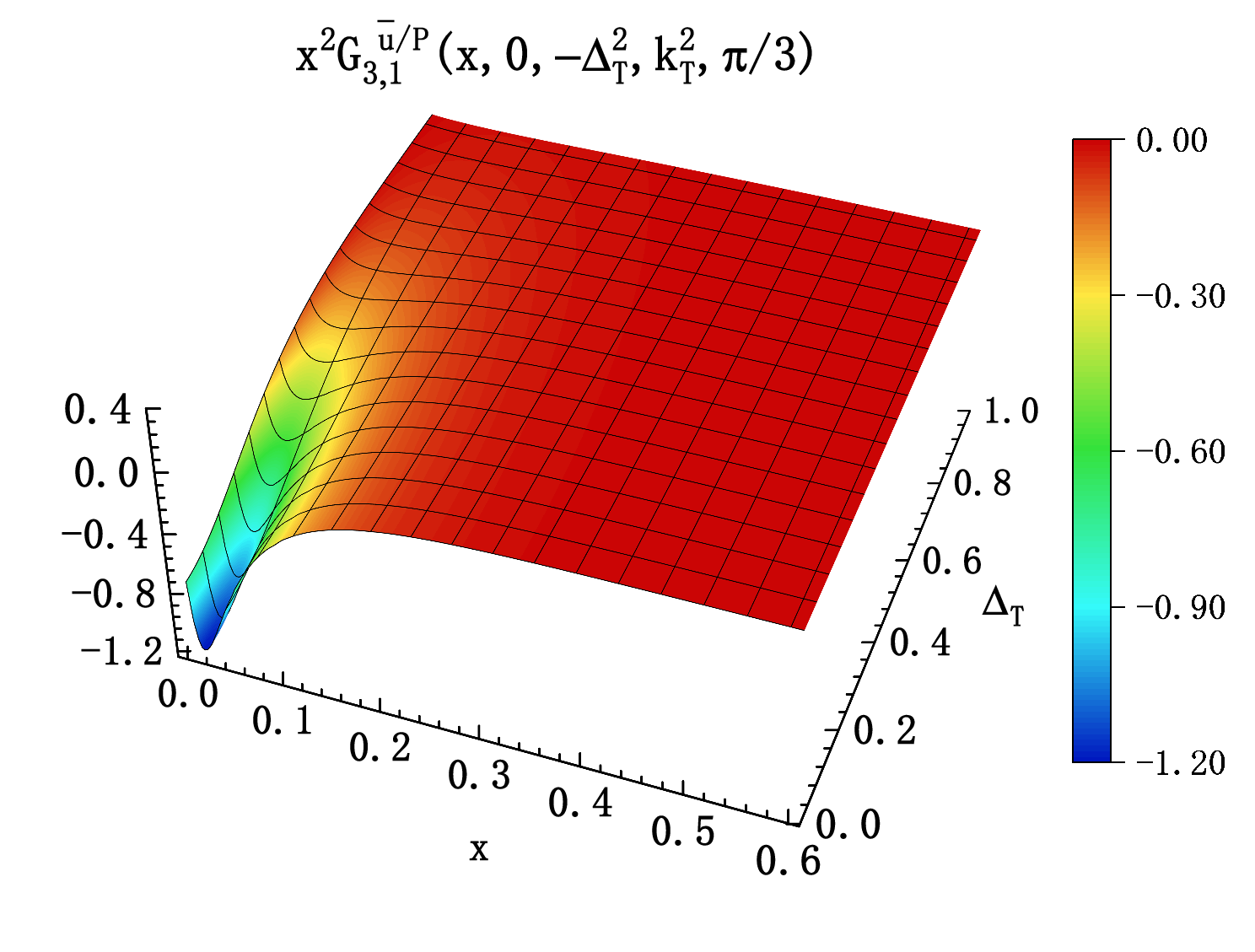}
    	\end{minipage}}
    	\subfigure{\begin{minipage}[b]{0.245\linewidth}
    			\centering
    			\includegraphics[width=\linewidth]{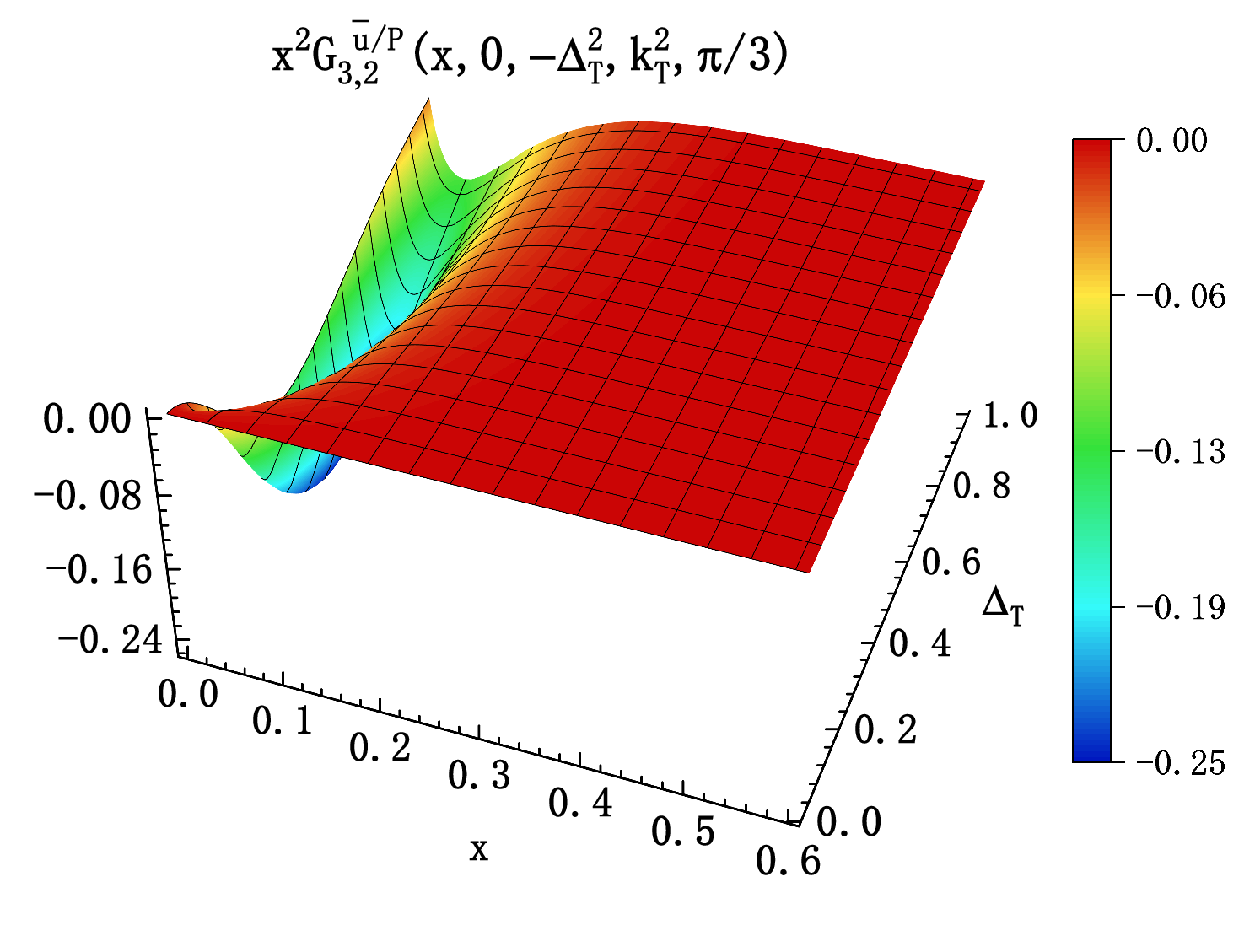}    	
    	\end{minipage}}
    	\subfigure{\begin{minipage}[b]{0.245\linewidth}
    			\centering
    			\includegraphics[width=\linewidth]{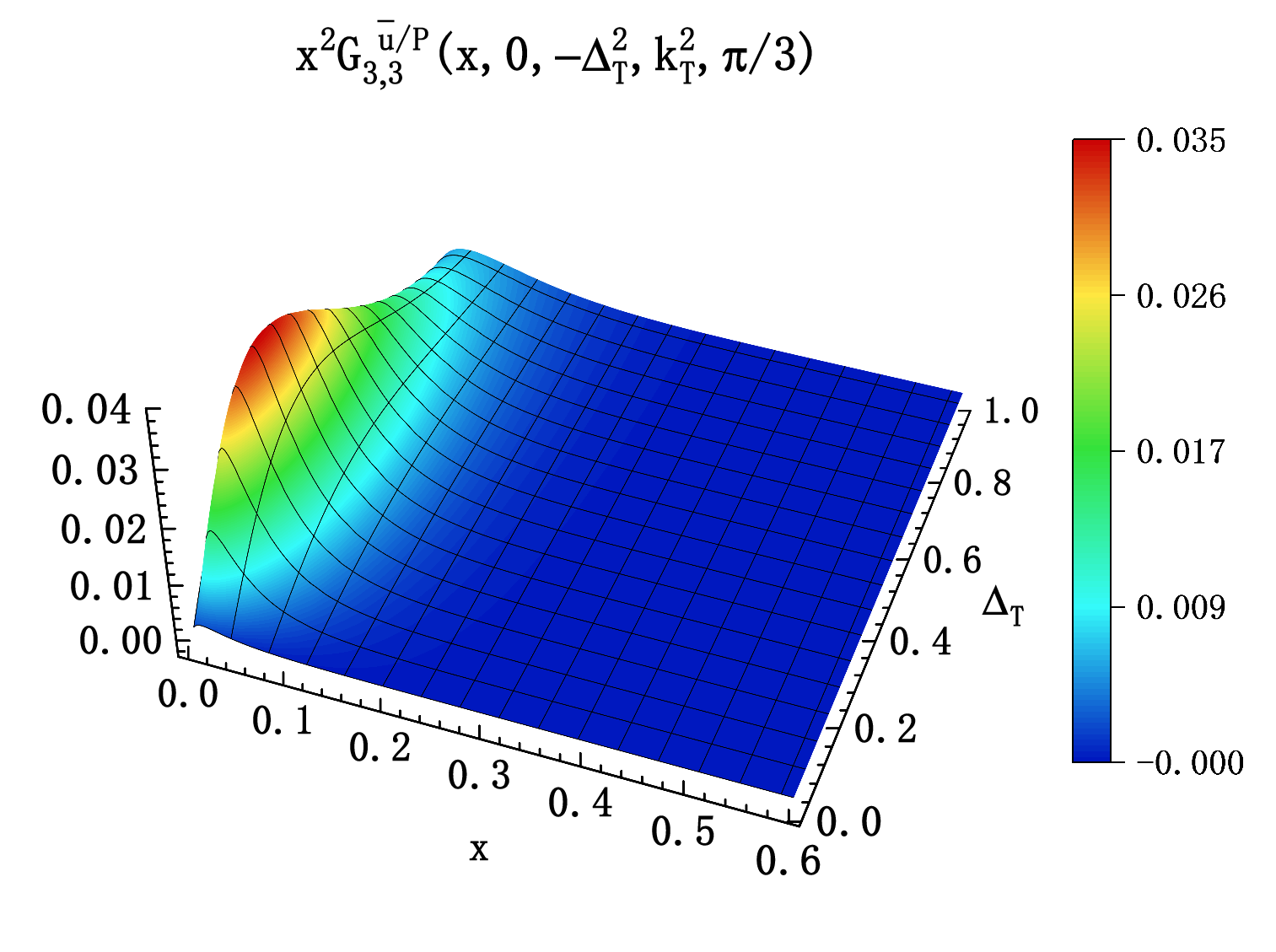}
    	\end{minipage}}
        \subfigure{\begin{minipage}[b]{0.245\linewidth}
        		\centering
        		\includegraphics[width=\linewidth]{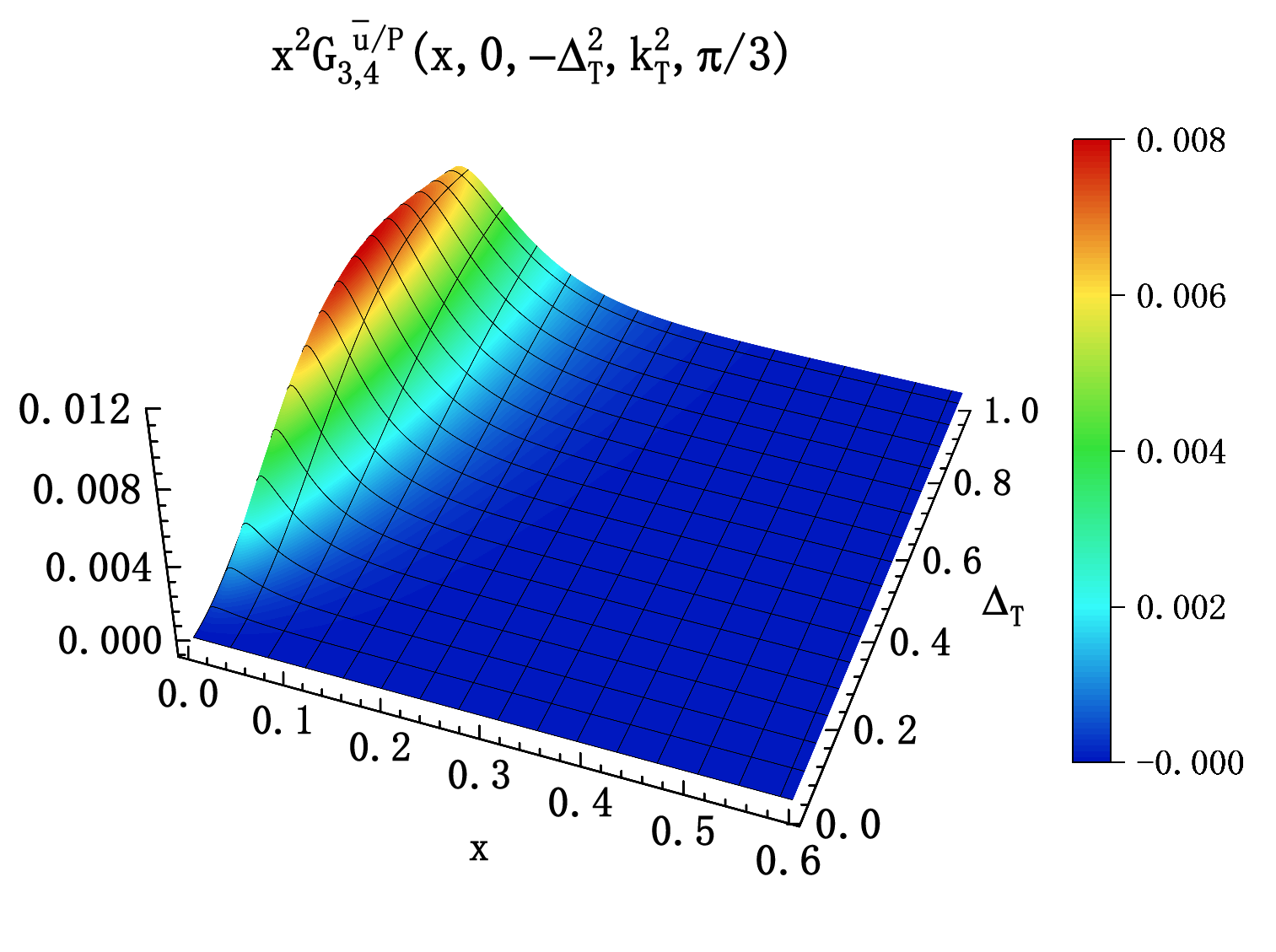}
        \end{minipage}}
    	\subfigure{\begin{minipage}[b]{0.245\linewidth}
    			\centering
    			\includegraphics[width=\linewidth]{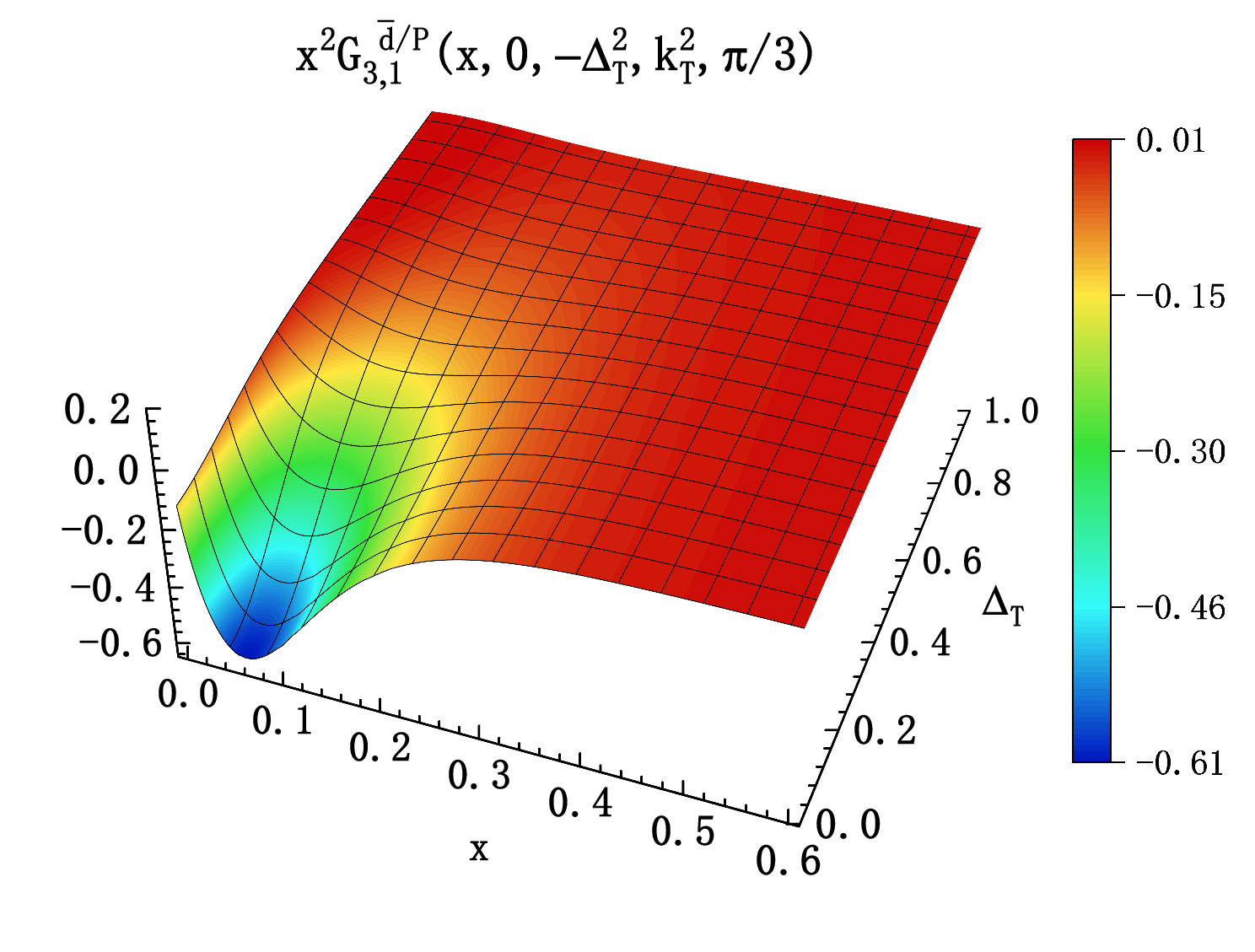}    	
    	\end{minipage}}	
    	\subfigure{\begin{minipage}[b]{0.245\linewidth}
    			\centering
    			\includegraphics[width=\linewidth]{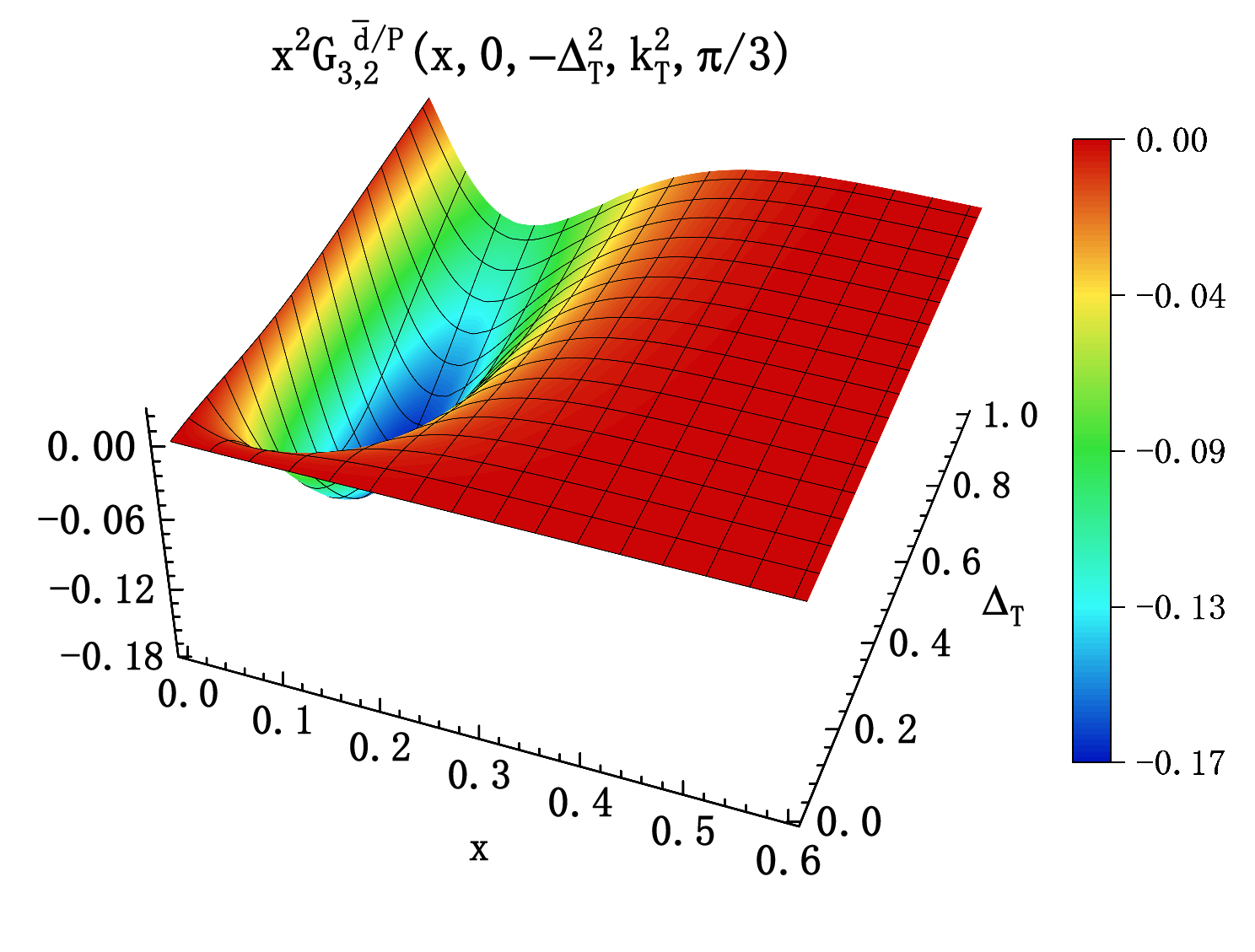}    
    	\end{minipage}}	
    	\centering
    	\subfigure{\begin{minipage}[b]{0.245\linewidth}
    			\centering
    			\includegraphics[width=\linewidth]{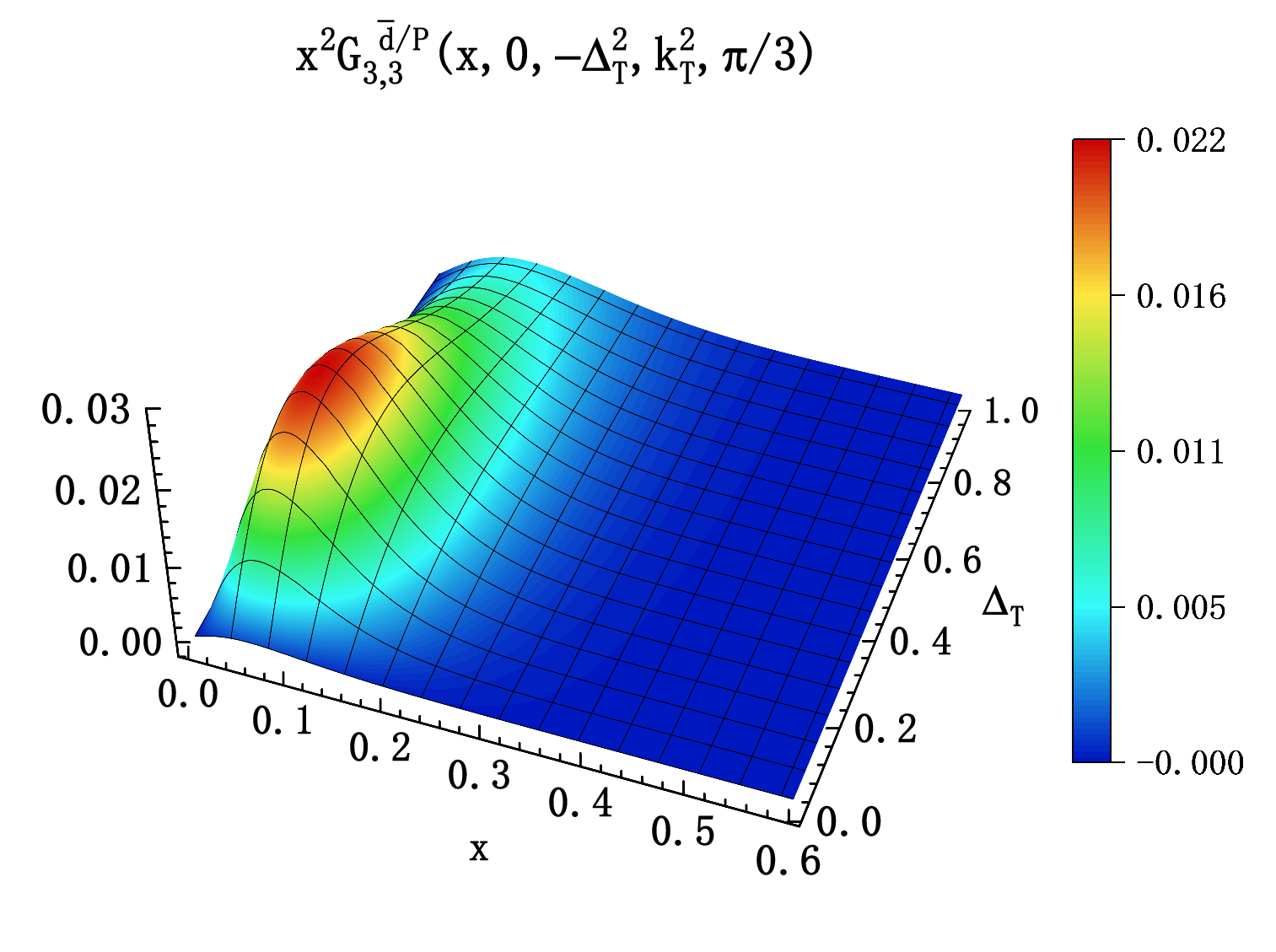}
    	\end{minipage}}
    	\subfigure{\begin{minipage}[b]{0.245\linewidth}
    			\centering
    			\includegraphics[width=\linewidth]{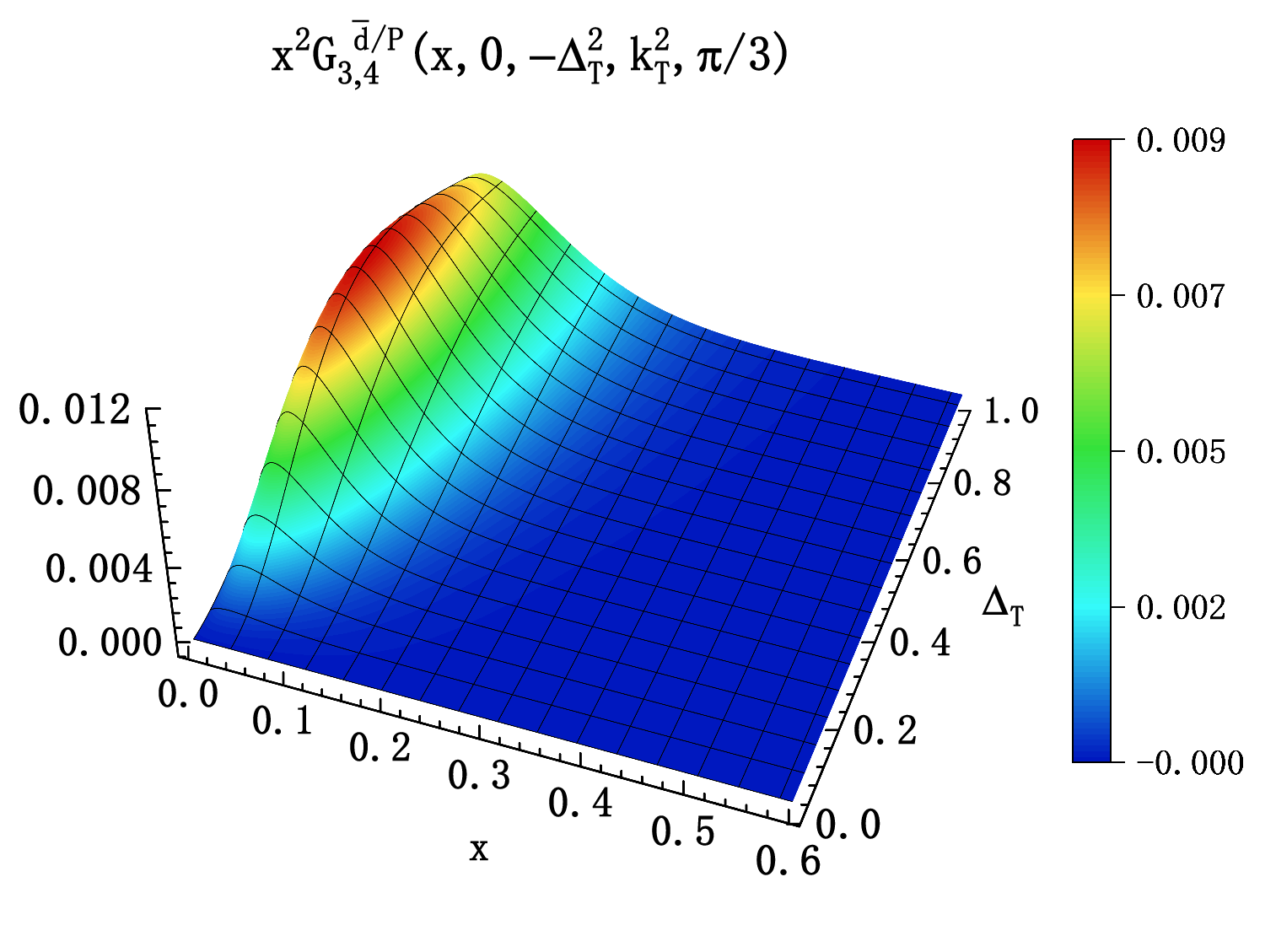}    
    	\end{minipage}}
    	\caption{Similar to Fig~\ref{udbarF}, but for longitudinally polarized sea quarks.} \label{udbarG}      
    \end{figure*}
   
   \begin{figure*}[htbp]
    	\centering
    	\subfigure{\begin{minipage}[b]{0.245\linewidth}
    			\centering
    			\includegraphics[width=\linewidth]{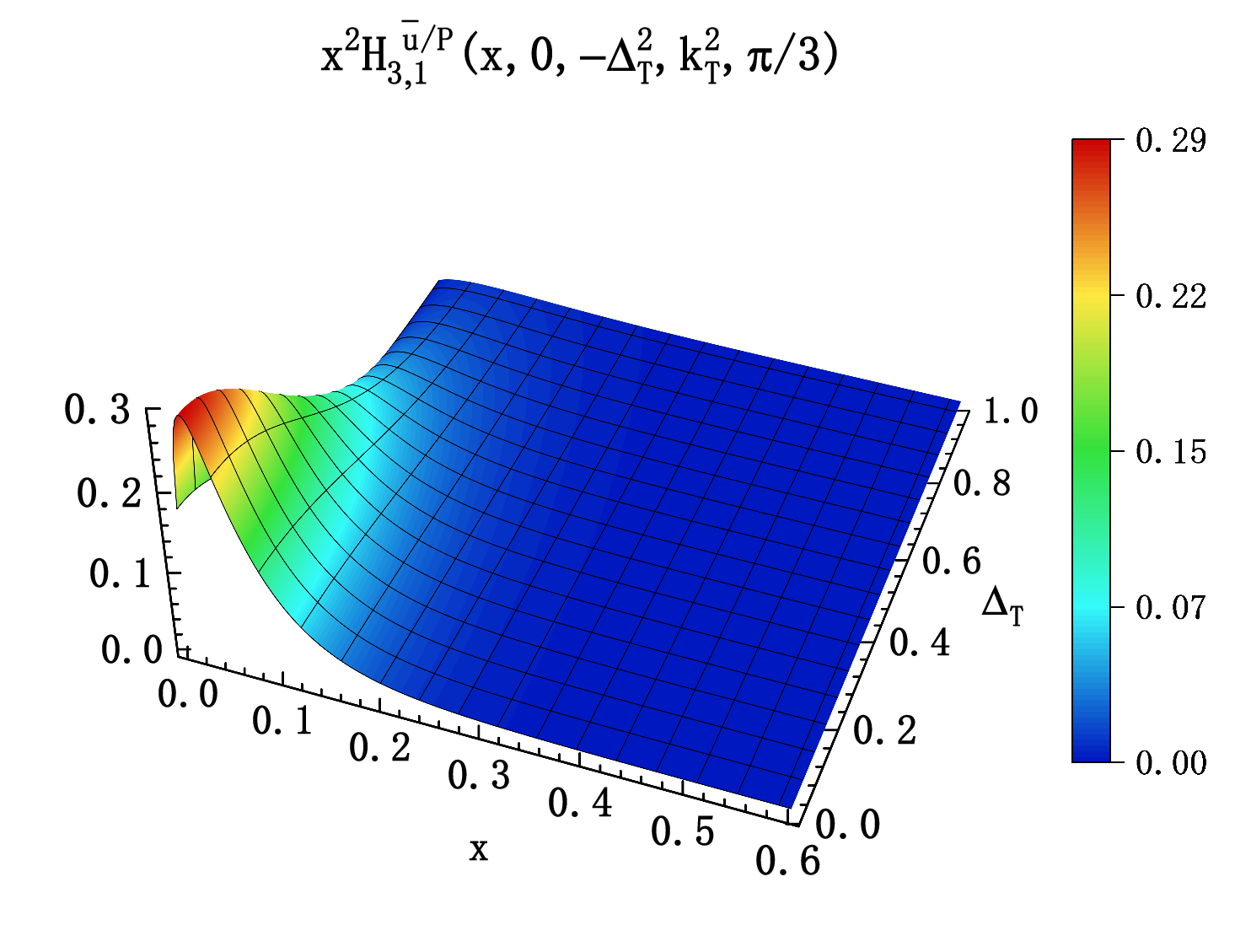}
    	\end{minipage}}
    	\subfigure{\begin{minipage}[b]{0.245\linewidth}
    			\centering
    			\includegraphics[width=\linewidth]{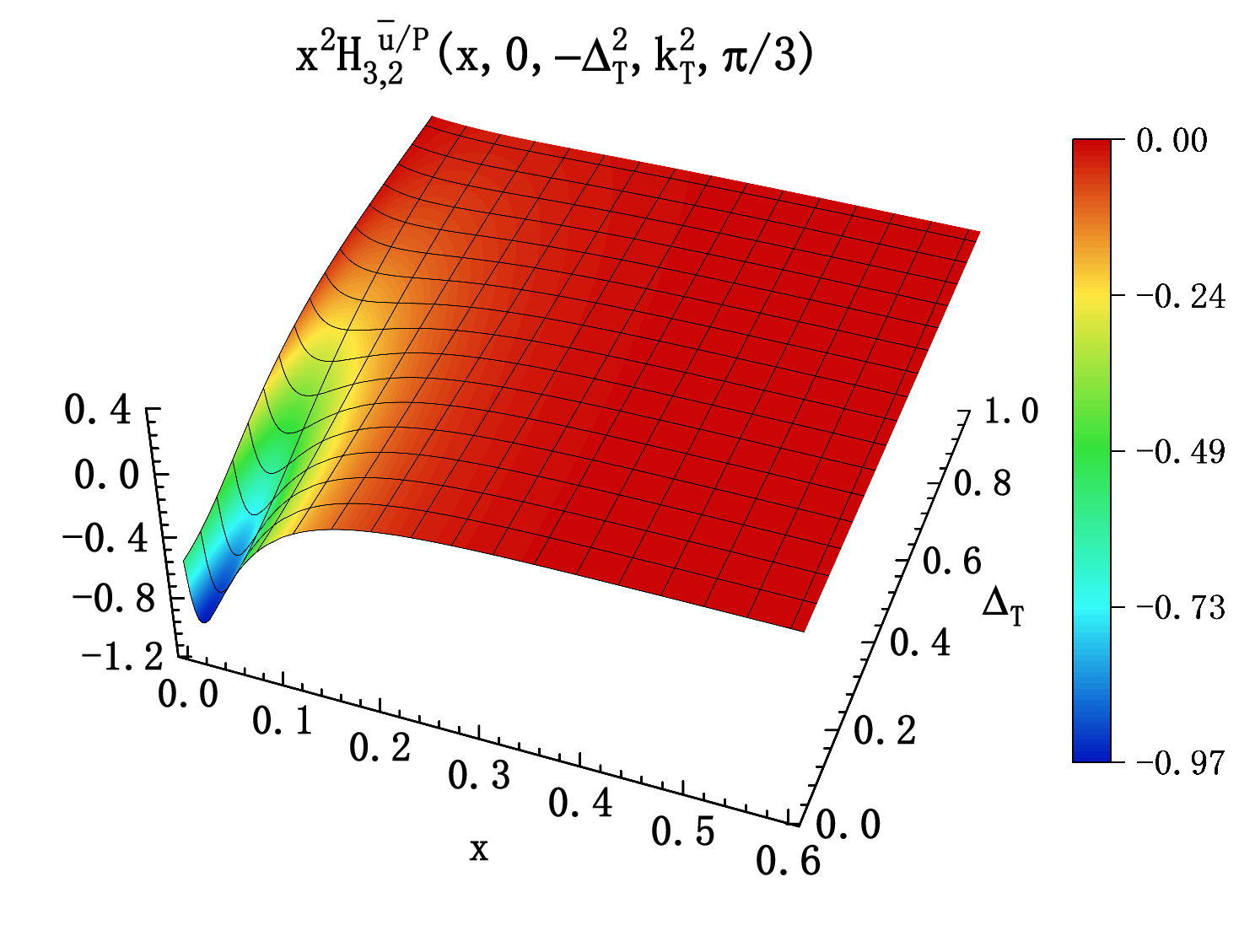}    	
    	\end{minipage}}
    	\subfigure{\begin{minipage}[b]{0.245\linewidth}
    			\centering
    			\includegraphics[width=\linewidth]{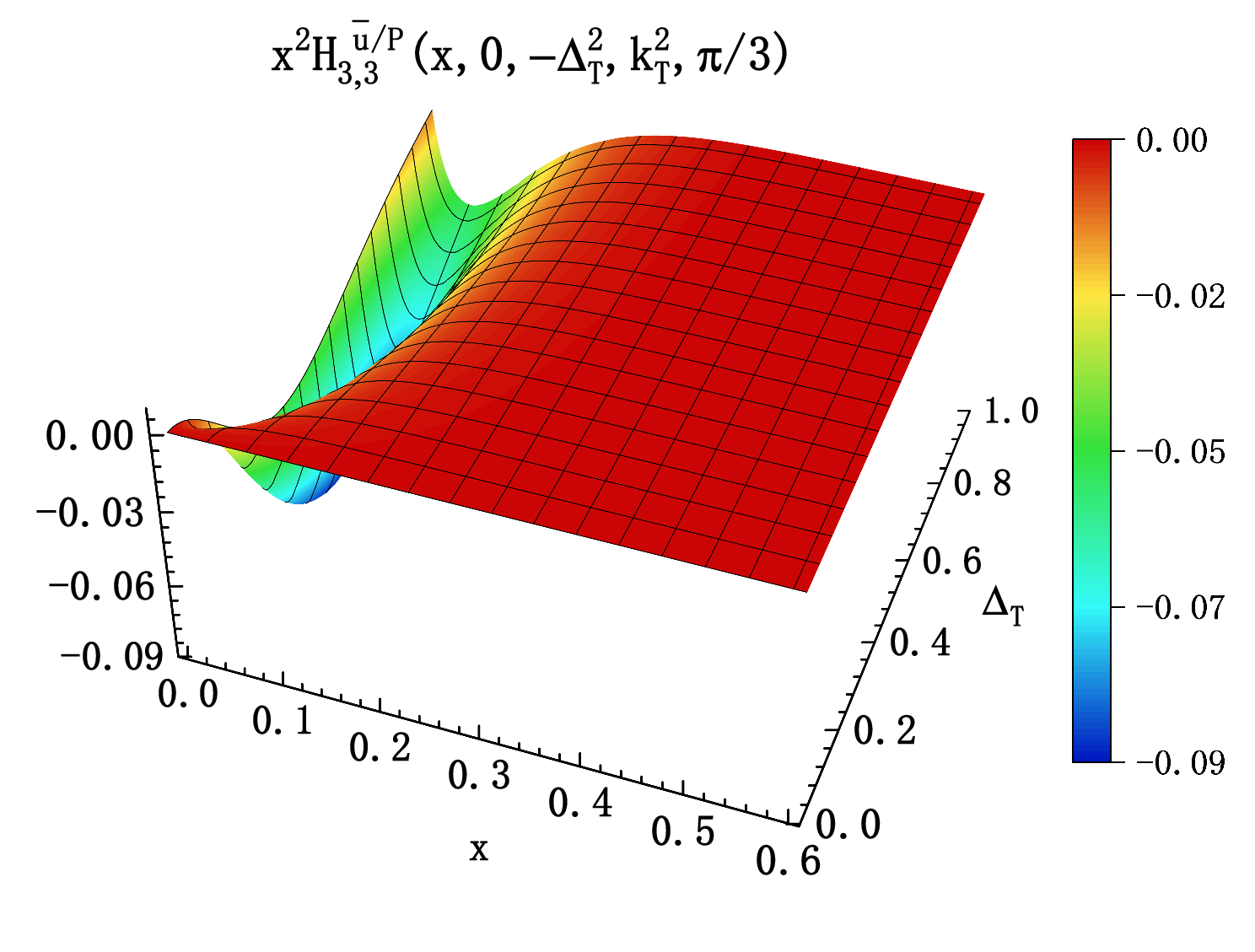}
    	\end{minipage}}
    	\subfigure{\begin{minipage}[b]{0.245\linewidth}
    			\centering
    			\includegraphics[width=\linewidth]{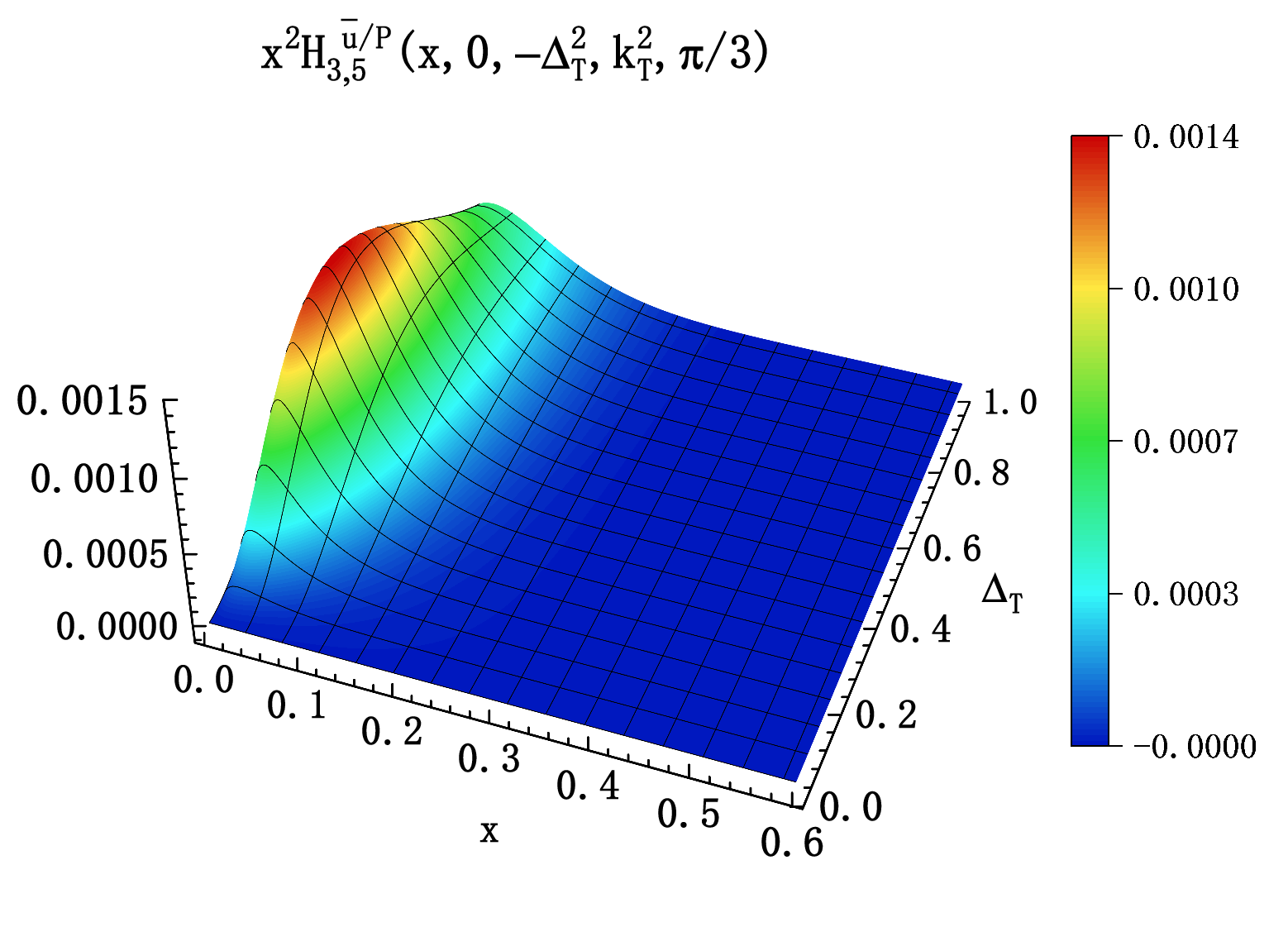}
    	\end{minipage}}
        \subfigure{\begin{minipage}[b]{0.245\linewidth}
        		\centering
        		\includegraphics[width=\linewidth]{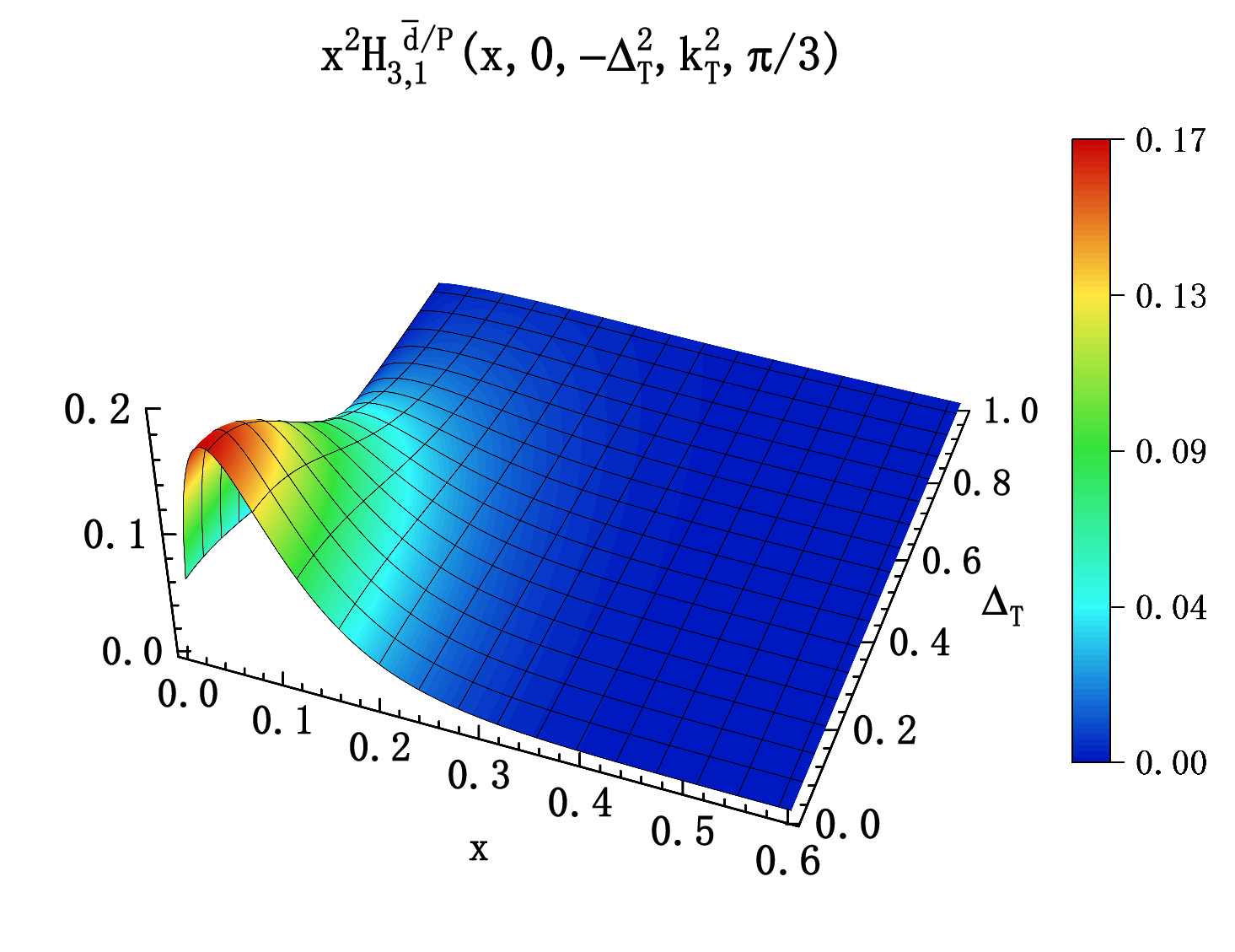}
        \end{minipage}}
        \subfigure{\begin{minipage}[b]{0.245\linewidth}
        		\centering
        		\includegraphics[width=\linewidth]{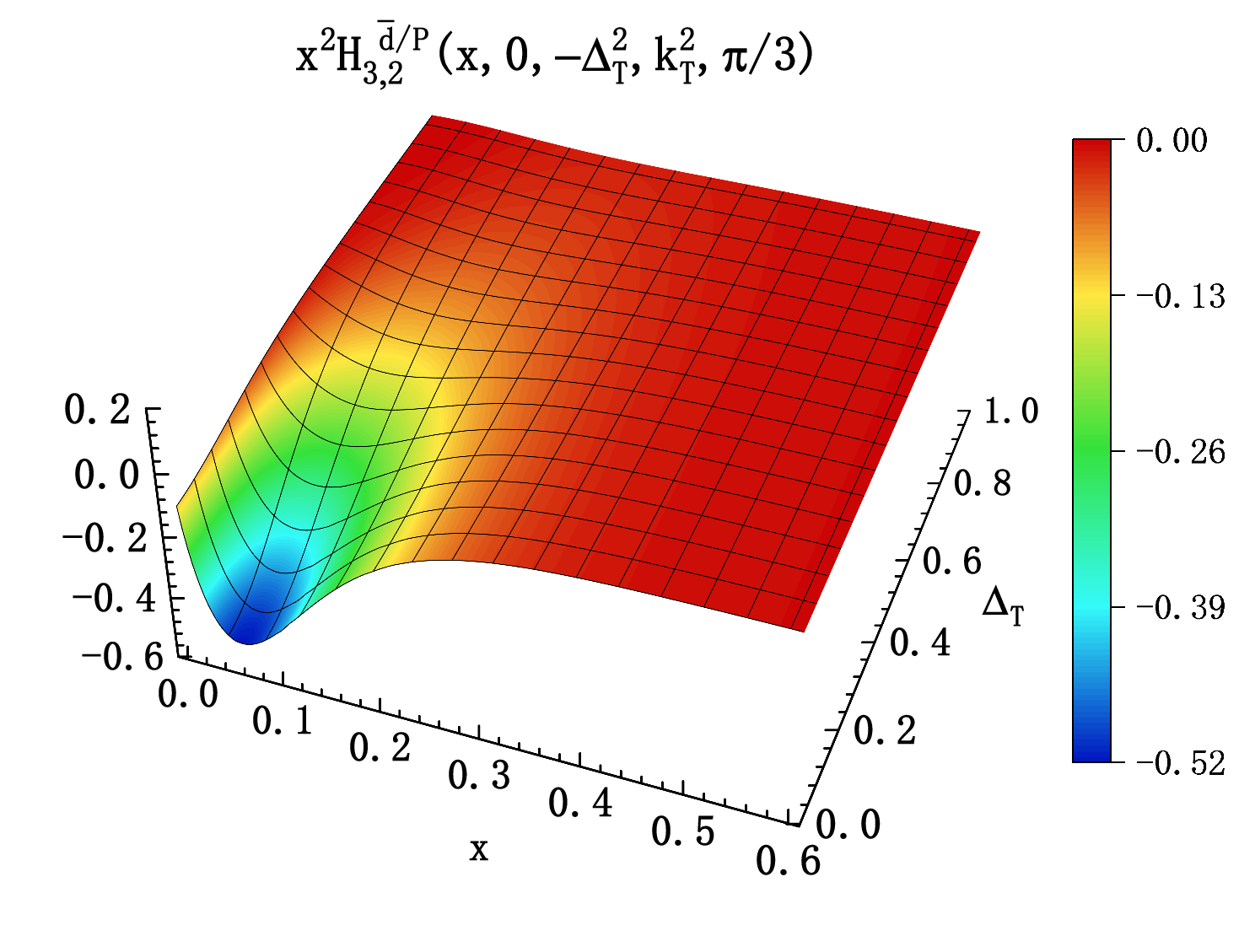}    	
        \end{minipage}}
        \subfigure{\begin{minipage}[b]{0.245\linewidth}
        		\centering
        		\includegraphics[width=\linewidth]{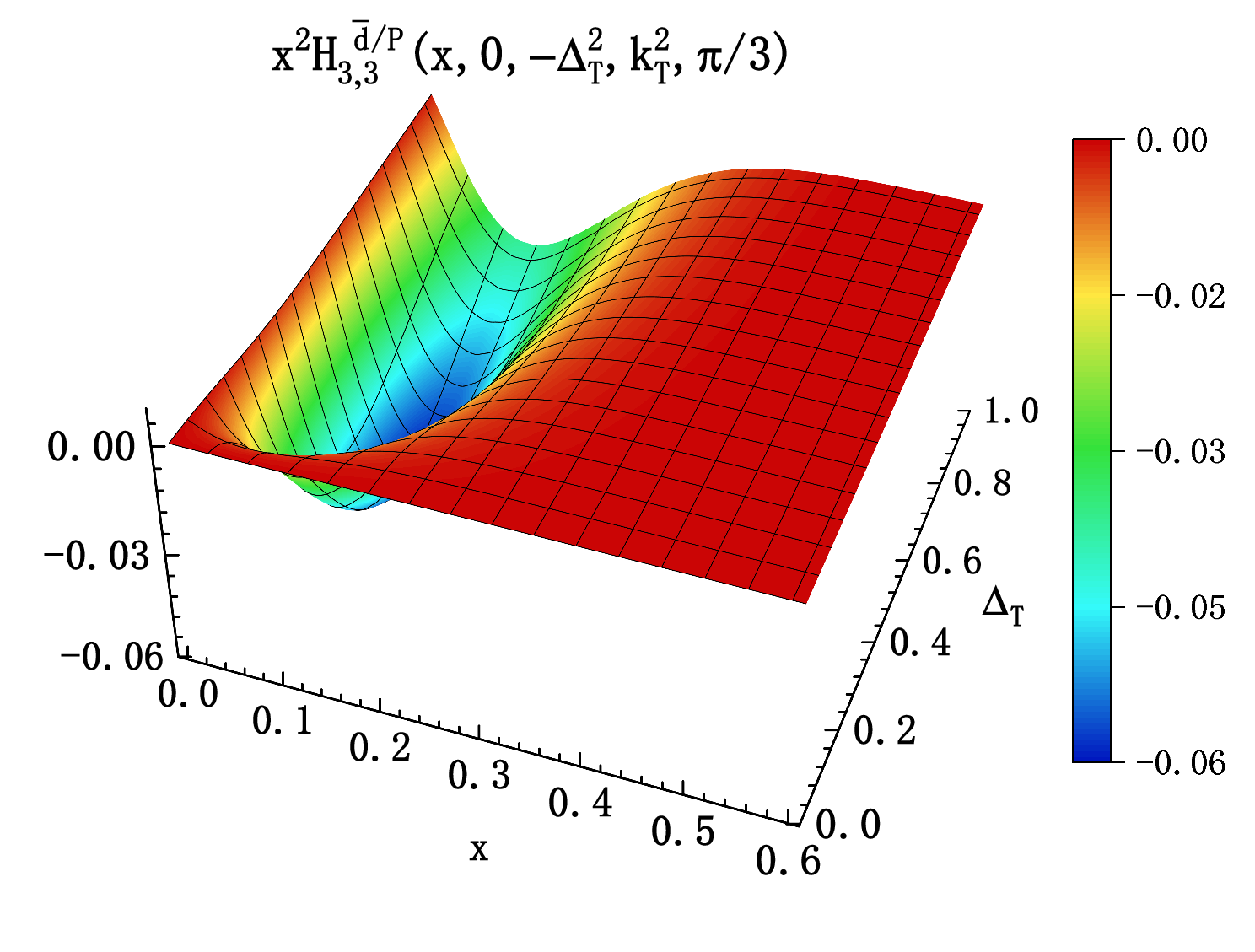}
        \end{minipage}}
        \subfigure{\begin{minipage}[b]{0.245\linewidth}
        		\centering
        		\includegraphics[width=\linewidth]{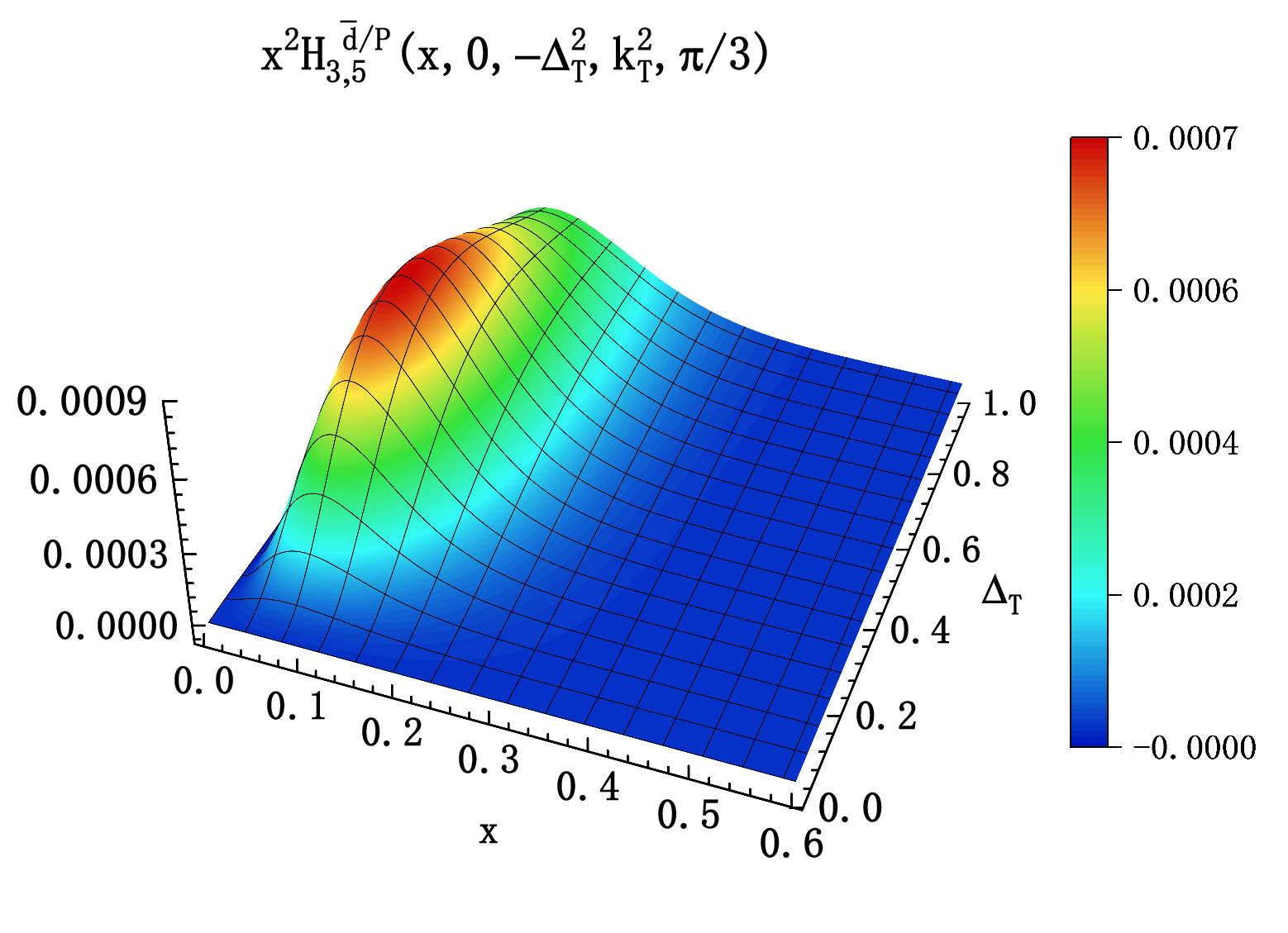}
        \end{minipage}}
    	\subfigure{\begin{minipage}[b]{0.245\linewidth}
    			\centering
    			\includegraphics[width=\linewidth]{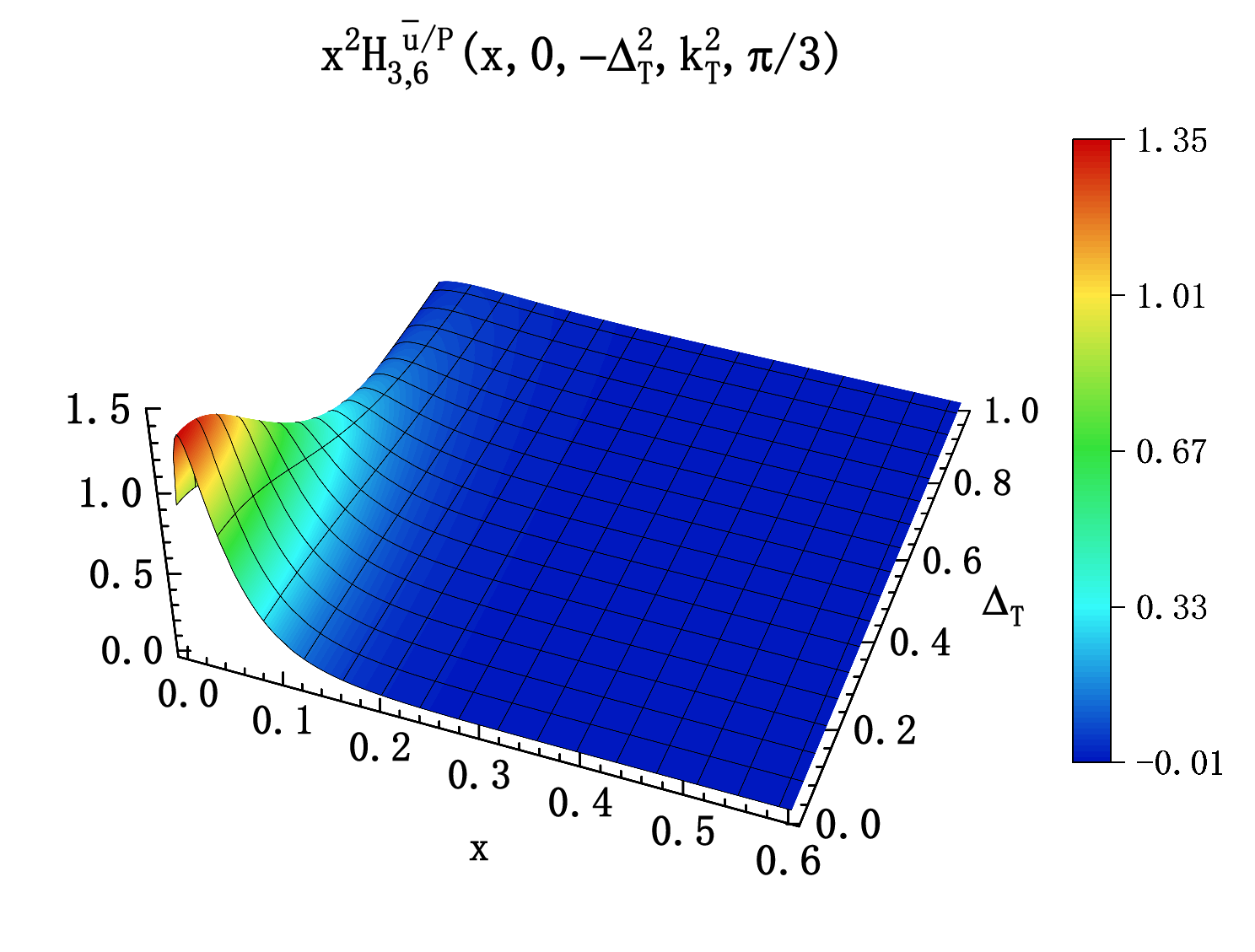}    	
    	\end{minipage}}	
    	\subfigure{\begin{minipage}[b]{0.245\linewidth}
    			\centering
    			\includegraphics[width=\linewidth]{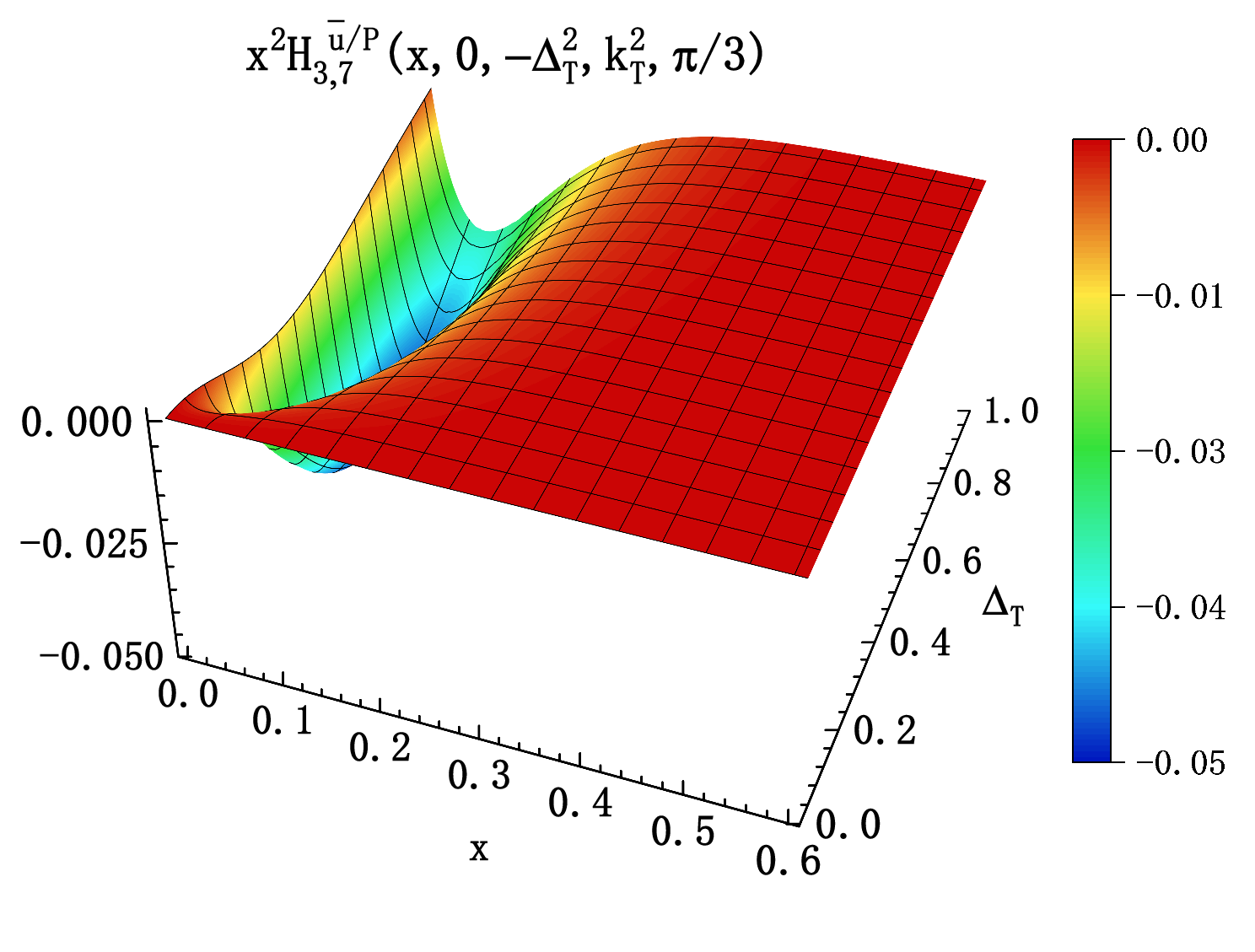}    
    	\end{minipage}}	
    	\subfigure{\begin{minipage}[b]{0.245\linewidth}
    			\centering
    			\includegraphics[width=\linewidth]{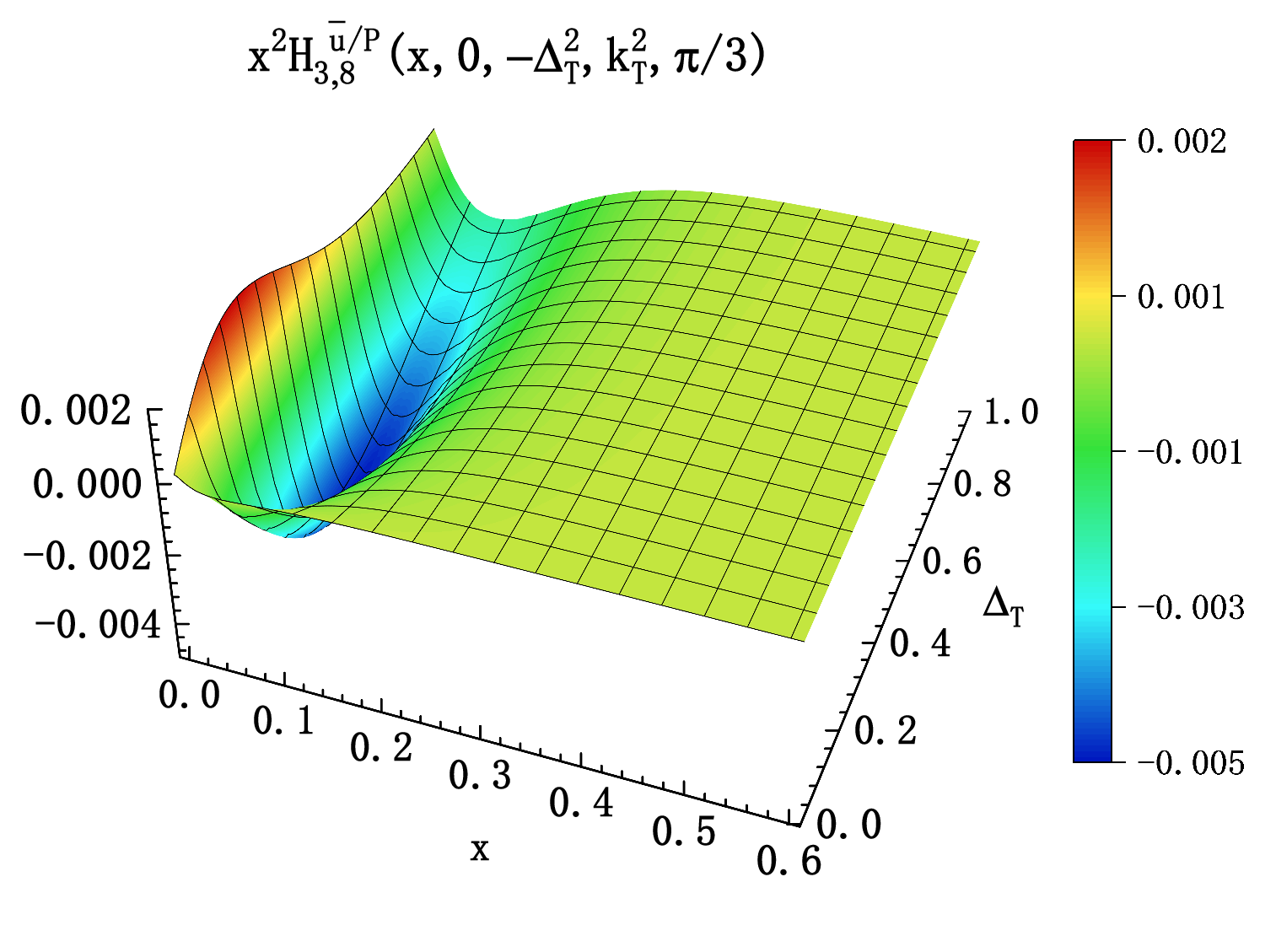}
    	\end{minipage}}
    
    	\subfigure{\begin{minipage}[b]{0.245\linewidth}
    			\centering
    			\includegraphics[width=\linewidth]{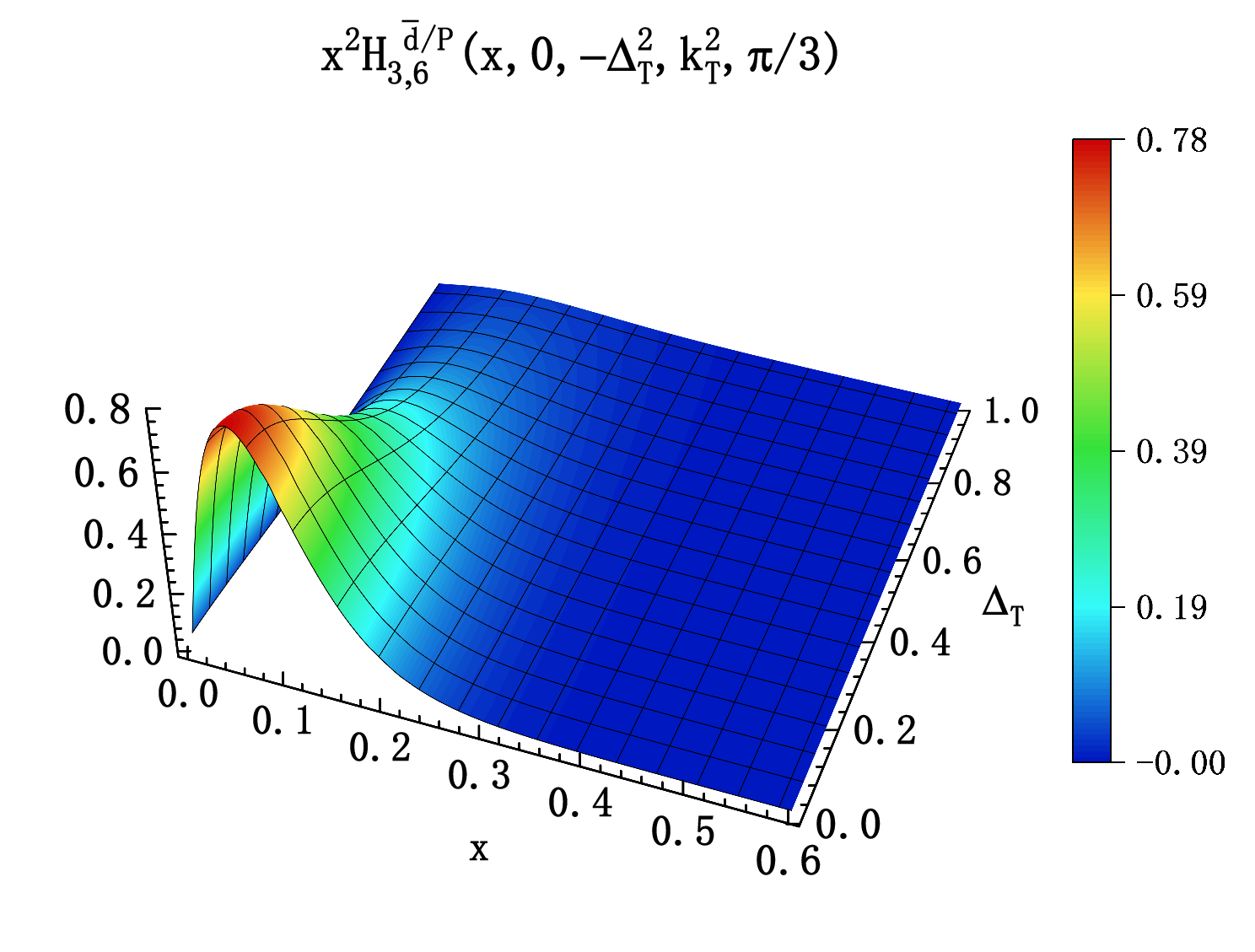}    	
    	\end{minipage}}	
    	\subfigure{\begin{minipage}[b]{0.245\linewidth}
    			\centering
    			\includegraphics[width=\linewidth]{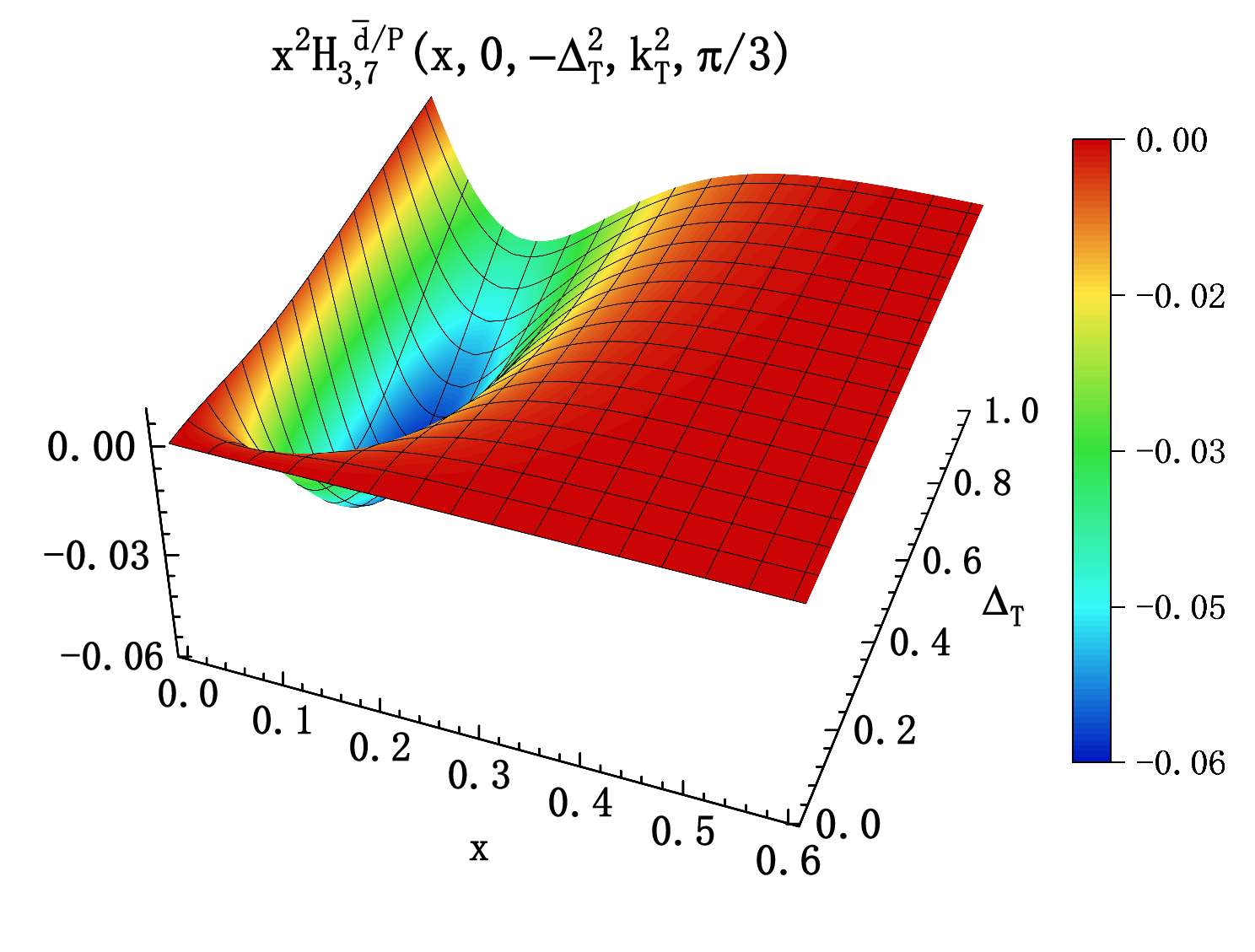}    
    	\end{minipage}}	
    	\centering
    	\subfigure{\begin{minipage}[b]{0.245\linewidth}
    			\centering
    			\includegraphics[width=\linewidth]{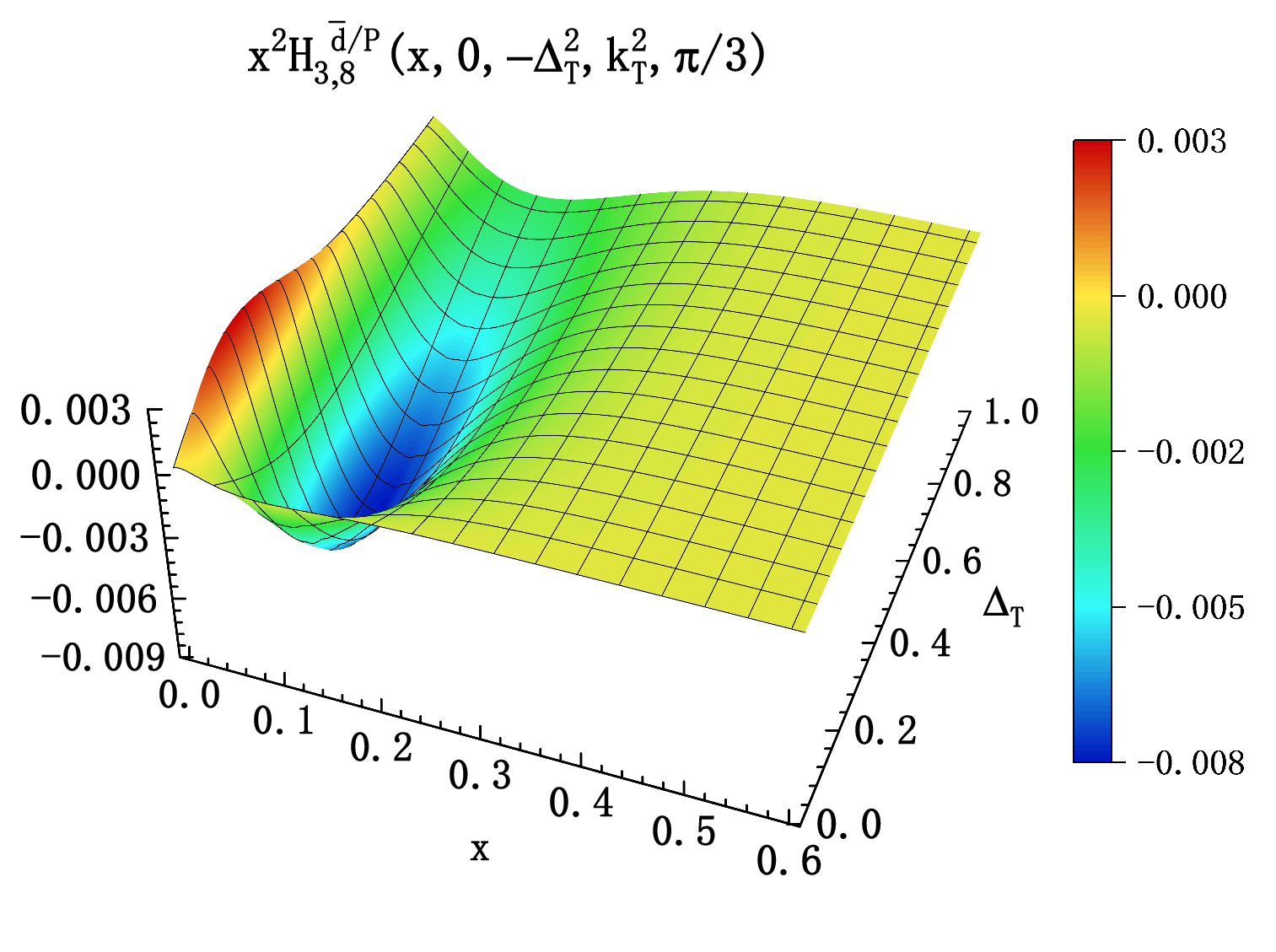}
    	\end{minipage}}
    	\caption{Similar to Fig~\ref{udbarF}, but for  transversely polarized sea quarks.} \label{udbarH}      
    \end{figure*}

   \subsection{Relation with twist-4 GPDS}
In what follows, we focus on the GPD limit obtained upon integrating over the transverse parton momentum. For zero skewness, the GPD correlator can be expressed in terms of the fully unintegrated quark–quark GTMD correlator:
\begin{align}
   	F_{\Lambda^{\prime} \Lambda}^{[\Gamma]}\left(x, \boldsymbol{\Delta}_{T}^{2}\right)=\int d^{2} \boldsymbol{k}_{T} W_{\Lambda^{\prime} \Lambda}^{[\Gamma]}\left(x, \boldsymbol{k}_{T}^{2}, \boldsymbol{\Delta}_{T}^{2}, \boldsymbol{k}_{T} \cdot \boldsymbol{\Delta}_{T}\right) .
   \end{align}
Following Ref.~\cite{Meissner:2009ww}, the twist-4 GPDs can be straightforwardly written as $\boldsymbol{k}_{T}$ integrals of twist-4 GTMDs:   
   	\begin{align}
   		H_{3}(x, 0, t) & = \int d^{2} \boldsymbol{k}_T F_{3,1},\label{H3}
   		\\
   		E_{3}(x, 0, t) & = \int d^{2} \boldsymbol{k}_T\left[-F_{3,1}+2\left(\frac{\boldsymbol{k}_T \cdot \boldsymbol{\Delta}_T}{\boldsymbol{\Delta}_T^{2}} F_{3,2}+F_{3,3}\right)\right],
   		\\
   		\tilde{H}_{3}(x, 0, t) & = \int d^{2} \boldsymbol{k}_T G_{3,4}, 
   		\\
   		H_{3 T}(x, 0, t) & = \int d^{2} \boldsymbol{k}_T\left[H_{3,3}+\frac{\boldsymbol{\Delta}_T^{2}}{M^{2}}\left(\frac{\left(\boldsymbol{k}_T \cdot \boldsymbol{\Delta}_T\right)^{2}}{\left(\boldsymbol{\Delta}_T^{2}\right)^{2}} H_{3,4}+\frac{\boldsymbol{k}_T \cdot \boldsymbol{\Delta}_T}{\boldsymbol{\Delta}_T^{2}} H_{3,5}+H_{3,6}\right)\right],
   	\end{align}
   	\begin{align}
   		E_{3 T}(x, 0, t) + 2\tilde{H}_{3 T}(x, 0, t)& = \int d^{2} \boldsymbol{k}_T\left[2\left(\frac{\boldsymbol{k}_T \cdot \boldsymbol{\Delta}_T}{\boldsymbol{\Delta}_T^{2}} H_{3,1}+H_{3,2}\right)\right], 
   		\\\notag
   		{H}_{3 T}(x, 0, t)+\frac{\boldsymbol{\Delta}_T^{2}}{4M^{2}}\tilde{H}_{3 T}(x, 0, t) & = \int d^{2} \boldsymbol{k}_T\left[\frac{\boldsymbol{\Delta}_T^{2}}{4M^{2}}\left(\frac{\boldsymbol{k}_T \cdot \boldsymbol{\Delta}_T}{\boldsymbol{\Delta}_T^{2}} H_{3,1}+H_{3,2}\right)+H_{3,3}\right. 
   		\\&\left.+\frac{\boldsymbol{\Delta}_T^{2}}{2M^{2}}\left(\frac{\left(\boldsymbol{k}_T \cdot \boldsymbol{\Delta}_T\right)^{2}}{\left(\boldsymbol{\Delta}_T^{2}\right)^{2}} H_{3, 4}+\frac{\boldsymbol{k}_T \cdot \boldsymbol{\Delta}_T}{\boldsymbol{\Delta}_T^{2}} H_{3,5}+H_{3,6}\right) \right].\label{H3T}
   	\end{align}
    From Eqs.~(\ref{W1})–(\ref{W3}), we readily derive the relations
   	\begin{align}
   		&-F_{3,1}+2\left(\frac{\boldsymbol{k}_T \cdot \boldsymbol{\Delta}_T}{\boldsymbol{\Delta}_T^{2}} F_{3,2}+F_{3,3}\right)=\frac{(P^{+})^{2}}{M\boldsymbol{\Delta}_T^{2}}\left[\Delta_{1}\left(W_{-+}^{\left[\gamma^{-}\right]}-W_{+-}^{\left[\gamma^{-}\right]}\right)-i\Delta_{2}\left(W_{+-}^{\left[\gamma^{-}\right]}+W_{-+}^{\left[\gamma^{-}\right]}\right)\right],\label{F}
   		\\
   		&\frac{\boldsymbol{k}_T \cdot \boldsymbol{\Delta}_T}{\boldsymbol{\Delta}_T^{2}} H_{3,1}+H_{3,2}= \frac{-i(P^{+})^{2}}{2M\boldsymbol{\Delta}_T^{2}}\left[\Delta_{2}\left(W_{++}^{\left[i \sigma^{1 -} \gamma_{5}\right]}+W_{--}^{\left[i \sigma^{1 -} \gamma_{5}\right]}\right)-\Delta_{1}\left(W_{++}^{\left[i \sigma^{2 -} \gamma_{5}\right]}+W_{--}^{\left[i \sigma^{2 -} \gamma_{5}\right]}\right)\right],  
   		\\\notag
   		&H_{3,3}+\frac{\boldsymbol{\Delta}_T^{2}}{M^{2}}\left(\frac{\left(\boldsymbol{k}_T \cdot \boldsymbol{\Delta}_T\right)^{2}}{\left(\boldsymbol{\Delta}_T^{2}\right)^{2}} H_{3,4}+\frac{\boldsymbol{k}_T \cdot \boldsymbol{\Delta}_T}{\boldsymbol{\Delta}_T^{2}} H_{3,5}+H_{3,6}\right)\\&=\frac{(P^{+})^{2}}{2M^{2}\boldsymbol{\Delta}_T^{2}}\left[(\Delta_{1}+i\Delta_{2})\left(\Delta_{1}W_{+-}^{\left[i \sigma^{1 -} \gamma_{5}\right]}+\Delta_{2}W_{+-}^{\left[i \sigma^{2 -} \gamma_{5}\right]}\right)
   		+(\Delta_{1}-i\Delta_{2})\left(\Delta_{1}W_{-+}^{\left[i \sigma^{1 -} \gamma_{5}\right]}+\Delta_{2}W_{-+}^{\left[i \sigma^{2 -} \gamma_{5}\right]}\right)\right],
   		\\\notag
   		&\frac{\boldsymbol{\Delta}_T^{2}}{4M^{2}}\left(\frac{\boldsymbol{k}_T \cdot \boldsymbol{\Delta}_T}{\boldsymbol{\Delta}_T^{2}} H_{3,1}+H_{3,2}\right)+
   		H_{3,3}+\frac{\boldsymbol{\Delta}_T^{2}}{2M^{2}}\left(\frac{\left(\boldsymbol{k}_T \cdot \boldsymbol{\Delta}_T\right)^{2}}{\left(\boldsymbol{\Delta}_T^{2}\right)^{2}} H_{3,4}+\frac{\boldsymbol{k}_T \cdot \boldsymbol{\Delta}_T}{\boldsymbol{\Delta}_T^{2}} H_{3,5}+H_{3,6}\right)
   		\\&= \frac{(P^{+})^{2}}{4M^{2}}\left[\left(W_{+-}^{\left[i \sigma^{1 -} \gamma_{5}\right]}+W_{-+}^{\left[i \sigma^{1 -} \gamma_{5}\right]}\right)+i\left(W_{+-}^{\left[i \sigma^{2 -} \gamma_{5}\right]}-W_{-+}^{\left[i \sigma^{2 -} \gamma_{5}\right]}\right)\right]. \label{H}
   	\end{align}
   
Substituting these relations into Eqs.~(\ref{H3})–(\ref{H3T}) and Eqs.~(\ref{F})–(\ref{H}), together with the LCWFs from Eqs.~(\ref{former}), (\ref{later}) and the overlap representation in Eqs.~(\ref{W4})–(\ref{W7}), we obtain the expressions for the twist-4 sea-quark GPDs within our model:
\begin{align}
   		\notag &H_3^{\overline{q}/P}(x,0,t)=-\frac{1}{x^2M^2}\int_{x}^{1}\frac{dy}{y}\int d^2\boldsymbol{k}_T\int{d^2\boldsymbol{r}_T} \frac{g_1^2g_2^2y(1-y)^2(1-\frac{x}{y})^2 \left[(M_B-(1-y)M)^2+\boldsymbol{r}_T^2-\frac{1}{4}(1-y)\boldsymbol{\Delta}_T^2\right]}{4(2\pi)^6D_1(y,\boldsymbol{r}_T,\boldsymbol{\Delta}_T) D_2(\frac{x}{y},\boldsymbol{k}_T-\frac{x}{y}\boldsymbol{r}_T,\boldsymbol{\Delta}_T)}
   		\\&\times \left[ \left(m^2+(\boldsymbol{k}_T-\frac{x}{y}  \boldsymbol{r}_T)^2-\frac{1}{4}(1-\frac{x}{y})^2\boldsymbol{\Delta}_T^2\right)(m^2+\boldsymbol{k}_T^2-\frac{1}{4}\boldsymbol{\Delta}_T^2)+(1-\frac{x}{y})(\boldsymbol{k}_T\times\boldsymbol{\Delta}_{T}) (\boldsymbol{k}_T-\frac{x}{y}  \boldsymbol{r}_T)\times\boldsymbol{\Delta}_{T} +(1-\frac{x}{y})m^2\boldsymbol{\Delta}_T^2\right],
   		\\\notag
   		&E_3^{\overline{q}/P}(x,0,t)=-\frac{1}{x^2 M}\int_{x}^{1}\frac{dy}{y}\int d^2\boldsymbol{k}_T\int{d^2\boldsymbol{r}_T} \frac{g_1^2 g_2^2 y(1-y)^3 (1-\frac{x}{y})^2 \left[M_B-(1-y)M\right]}{2(2\pi)^6D_1(y,\boldsymbol{r}_T,\boldsymbol{\Delta}_T) D_2(\frac{x}{y},\boldsymbol{k}_T-\frac{x}{y}\boldsymbol{r}_T,\boldsymbol{\Delta}_T)}
   		\\&\times \left[ \left(m^2+(\boldsymbol{k}_T-\frac{x}{y}  \boldsymbol{r}_T)^2-\frac{1}{4}(1-\frac{x}{y})^2\boldsymbol{\Delta}_T^2\right)(m^2+\boldsymbol{k}_T^2-\frac{1}{4}\boldsymbol{\Delta}_T^2)+(1-\frac{x}{y})(\boldsymbol{k}_T\times\boldsymbol{\Delta}_{T})(\boldsymbol{k}_T-\frac{x}{y}  \boldsymbol{r}_T)\times\boldsymbol{\Delta}_{T} +(1-\frac{x}{y})m^2\boldsymbol{\Delta}_T^2\right],
   		\\\notag
   		&\widetilde{H}_3^{\overline{q}/P}(x,0,t)=\frac{1}{x^2 M^2}\int_{x}^{1}\frac{dy}{y}\int d^2\boldsymbol{k}_T\int{d^2\boldsymbol{r}_T} \frac{g_1^2 g_2^2 y(1-y)^3 (1-\frac{x}{y})^2 \boldsymbol{\Delta}_T\times\boldsymbol{r}_T}{4(2\pi)^6D_1(y,\boldsymbol{r}_T,\boldsymbol{\Delta}_T) D_2(\frac{x}{y},\boldsymbol{k}_T-\frac{x}{y}\boldsymbol{r}_T,\boldsymbol{\Delta}_T)}
   		\\&\times \left[ \left(m^2+(\boldsymbol{k}_T-\frac{x}{y}  \boldsymbol{r}_T)^2-\frac{1}{4}(1-\frac{x}{y})^2\boldsymbol{\Delta}_T^2-2(1-\frac{x}{y})m^2\right)(\boldsymbol{k}_T\times\boldsymbol{\Delta}_T)+(1-\frac{x}{y})(m^2-\boldsymbol{k}_T^2+\frac{1}{4}\boldsymbol{\Delta}_T^2)(\boldsymbol{k}_T-\frac{x}{y}  \boldsymbol{r}_T)\times\boldsymbol{\Delta}_{T}\right],
   		\\\notag
   		&\widetilde{H}_{3T}^{\overline{q}/P}(x,0,t)=-\frac{1}{x^2 }\int_{x}^{1}\frac{dy}{y}\int d^2\boldsymbol{k}_T\int{d^2\boldsymbol{r}_T} \frac{g_1^2 g_2^2 y(1-y)^3 (1-\frac{x}{y})^2 m\left[M_B-(1-y)M\right]}{(2\pi)^6D_1(y,\boldsymbol{r}_T,\boldsymbol{\Delta}_T) D_2(\frac{x}{y},\boldsymbol{k}_T-\frac{x}{y}\boldsymbol{r}_T,\boldsymbol{\Delta}_T)}
   		\\&\times \left[ m^2+(\boldsymbol{k}_T-\frac{x}{y}  \boldsymbol{r}_T)^2-\frac{1}{4}(1-\frac{x}{y})^2\boldsymbol{\Delta}_T^2-(1-\frac{x}{y})(m^2+\boldsymbol{k}_T^2-\frac{1}{4}\boldsymbol{\Delta}_T^2)+2\frac{x}{y}(1-\frac{x}{y})(\boldsymbol{r}_T\times\boldsymbol{\Delta}_T)\frac{\boldsymbol{k}_T\times\boldsymbol{\Delta}_T}{\boldsymbol{\Delta}_T^{2}}\right],
   	\end{align}
   \begin{align}
   	    \notag
   		&E_{3T}^{\overline{q}/P}(x,0,t)=-\frac{1}{x^2M }\int_{x}^{1}\frac{dy}{y}\int d^2\boldsymbol{k}_T\int{d^2\boldsymbol{r}_T}\frac{g_1^2 g_2^2 y(1-y)^2 (1-\frac{x}{y})^2 m}{(2\pi)^6D_1(y,\boldsymbol{r}_T,\boldsymbol{\Delta}_T) D_2(\frac{x}{y},\boldsymbol{k}_T-\frac{x}{y}\boldsymbol{r}_T,\boldsymbol{\Delta}_T)}   		\\&\notag\times\left[(M_B-(1-y)M)^2+\boldsymbol{r}_T^2-\frac{1}{4}(1-y)^2\boldsymbol{\Delta}_T^2-2(1-y)M\left(M_B-(1-y)M\right)\right]
   		\\&\times\left[ m^2+(\boldsymbol{k}_T-\frac{x}{y}  \boldsymbol{r}_T)^2-\frac{1}{4}(1-\frac{x}{y})^2\boldsymbol{\Delta}_T^2-(1-\frac{x}{y})(m^2+\boldsymbol{k}_T^2-\frac{1}{4}\boldsymbol{\Delta}_T^2)+2\frac{x}{y}(1-\frac{x}{y})(\boldsymbol{r}_T\times\boldsymbol{\Delta}_T)\frac{\boldsymbol{k}_T\times\boldsymbol{\Delta}_T}{\boldsymbol{\Delta}_T^{2}}\right],
   		\\
   		&H_{3T}^{\overline{q}/P}(x,0,t)=0.
   	\end{align}
   \end{widetext}

\section{Numerical results and discussion}\label{Sec4}
     
    \begin{figure*}[htbp]
    	\centering
    	\subfigure{\begin{minipage}[b]{0.245\linewidth}
    			\centering
    			\includegraphics[width=\linewidth]{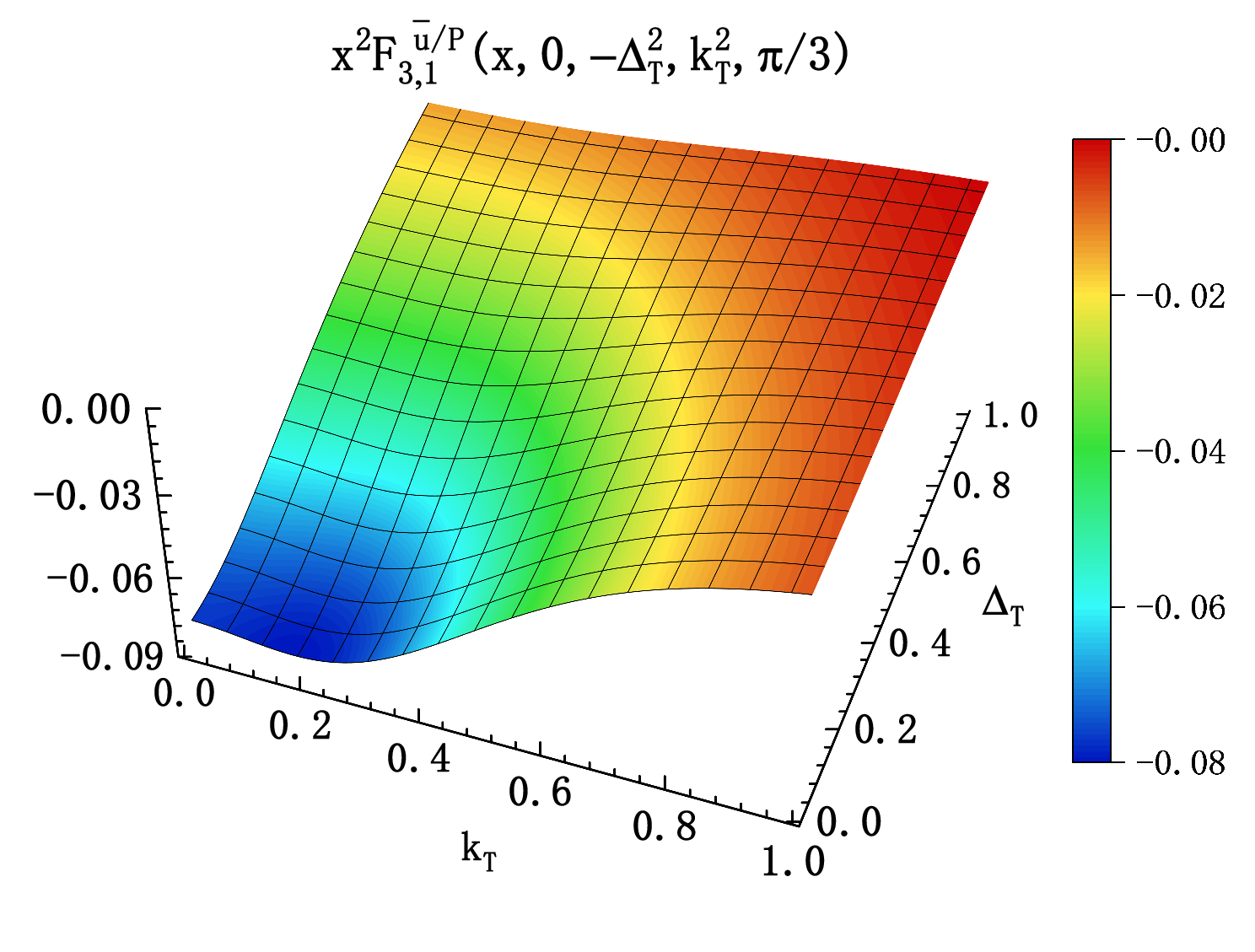}
    	\end{minipage}}
    	\subfigure{\begin{minipage}[b]{0.245\linewidth}
    			\centering
    			\includegraphics[width=\linewidth]{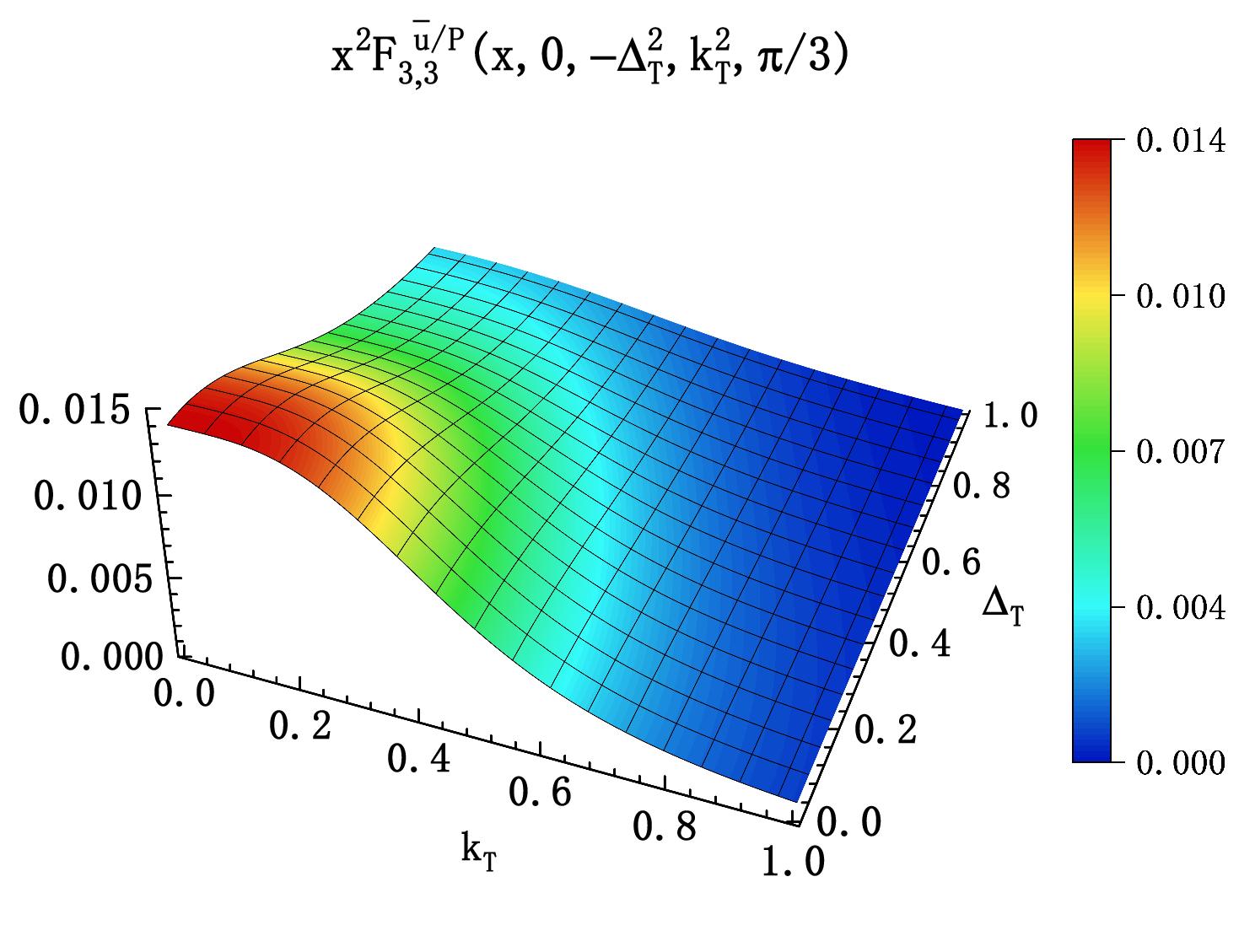}    	
    	\end{minipage}}
    	\subfigure{\begin{minipage}[b]{0.245\linewidth}
    			\centering
    			\includegraphics[width=\linewidth]{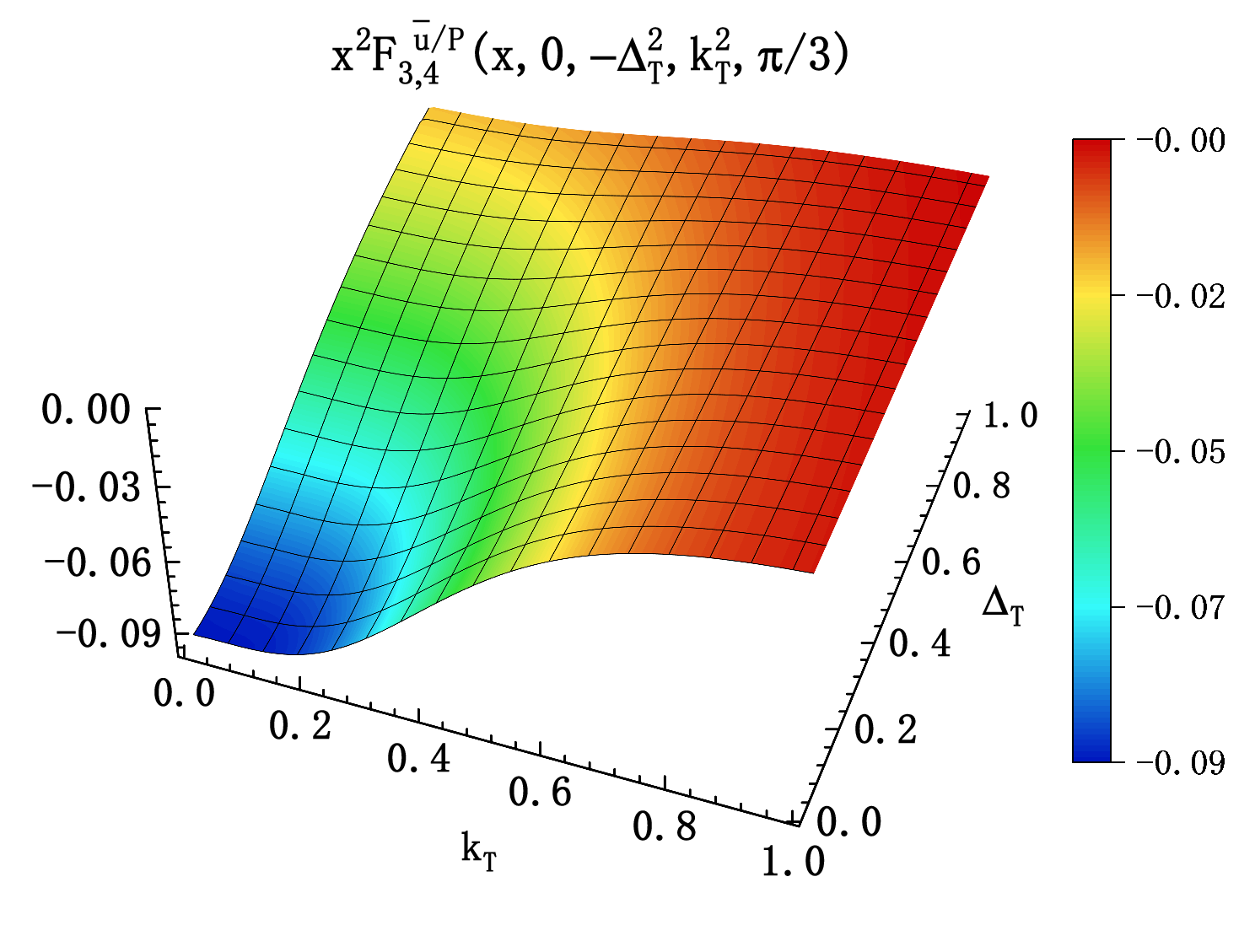}
    	\end{minipage}}
    
    	\centering
    	\subfigure{\begin{minipage}[b]{0.245\linewidth}
    			\centering
    			\includegraphics[width=\linewidth]{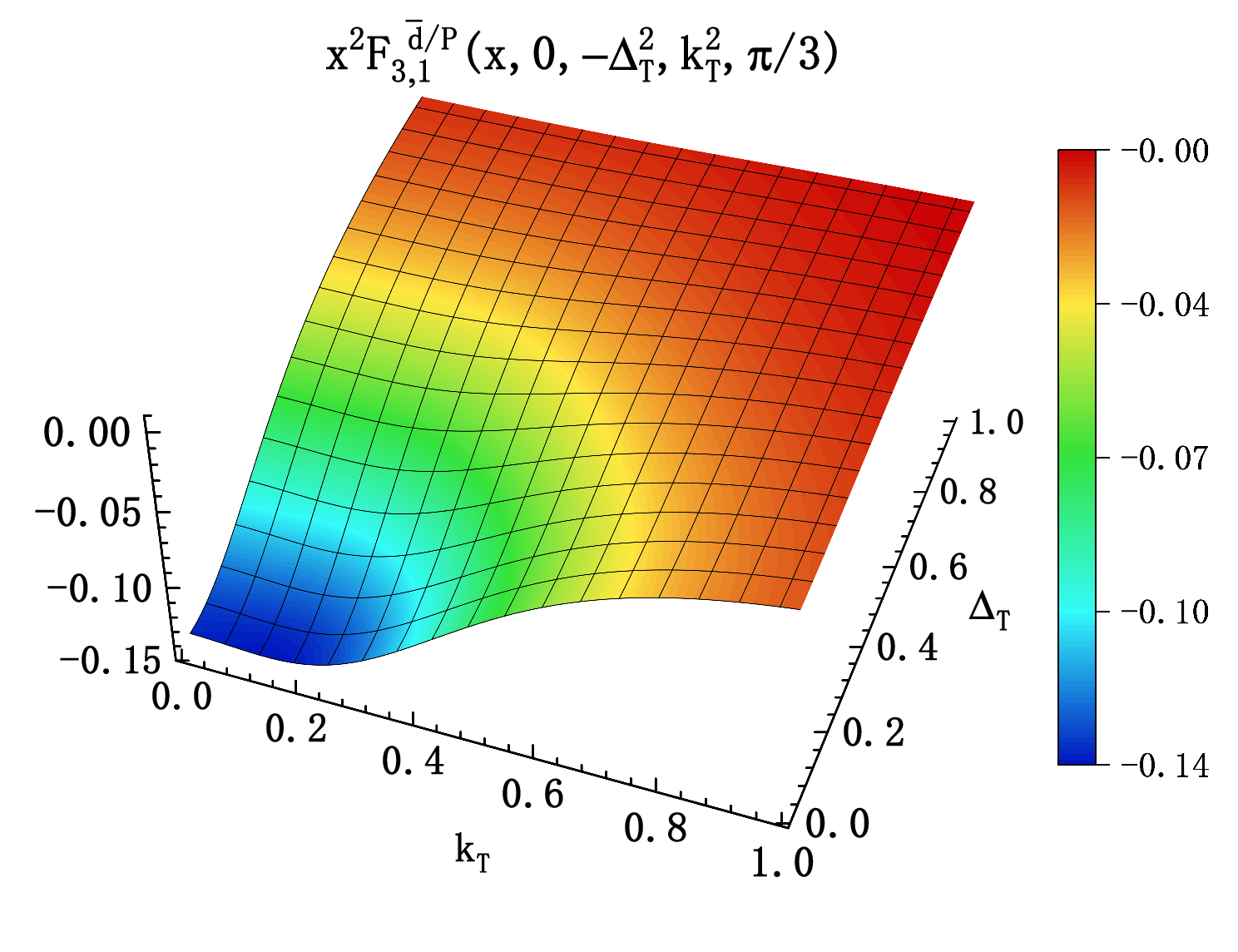}
    	\end{minipage}}	
    	\subfigure{\begin{minipage}[b]{0.245\linewidth}
    			\centering
    			\includegraphics[width=\linewidth]{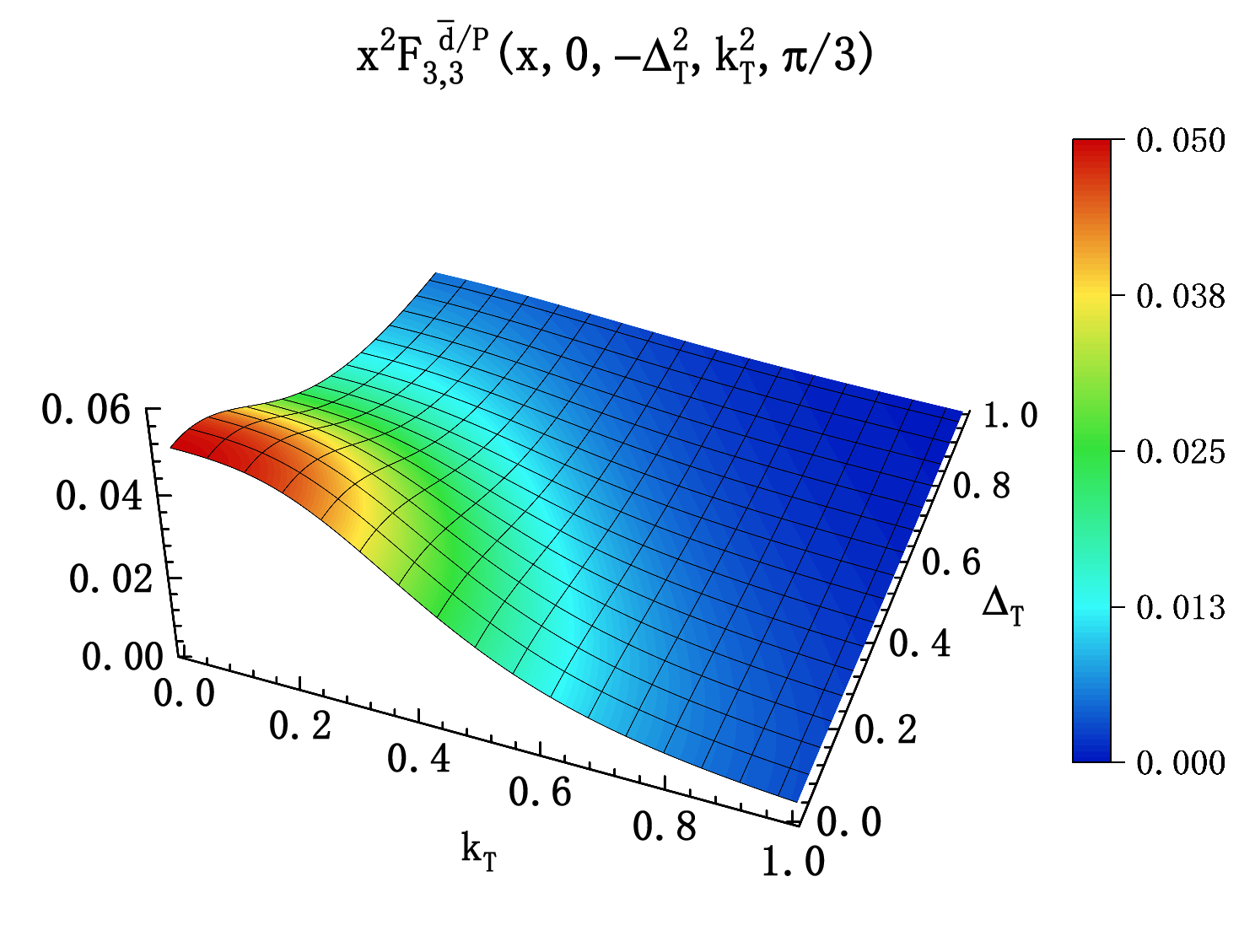}
    	\end{minipage}}
    	\subfigure{\begin{minipage}[b]{0.245\linewidth}
    			\centering
    			\includegraphics[width=\linewidth]{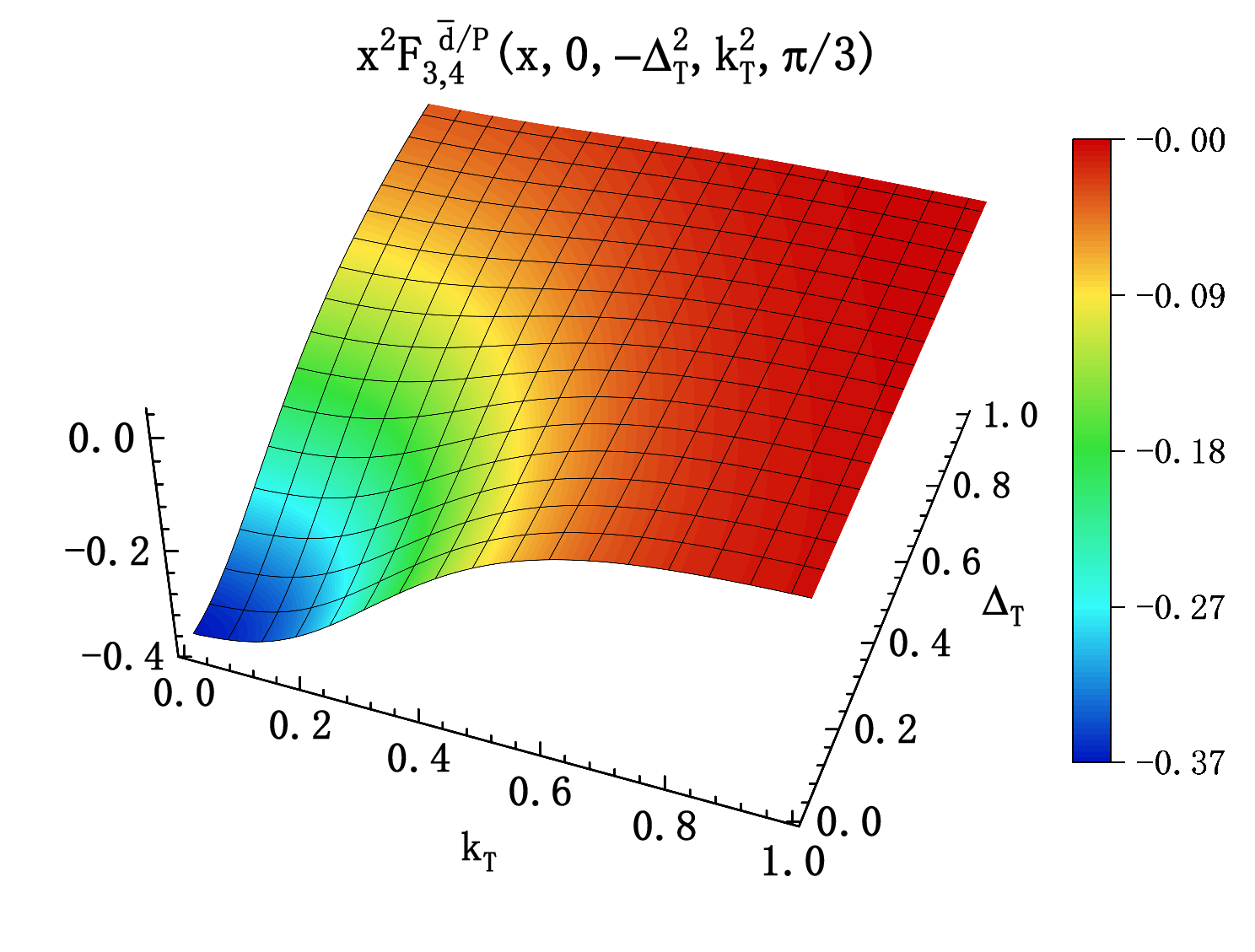}    	
    	\end{minipage}}
    	\caption{The twist-4 GTMDs (multiplied by $x^2$) for unpolarized sea quarks as functions of $k_T$ and $\Delta_T$ at fixed $x=0.1$ and $\theta=\pi/3$.} \label{udbarF-2}      
    \end{figure*}
    
    \begin{figure*}[htbp]
    	\centering
    	\subfigure{\begin{minipage}[b]{0.245\linewidth}
    			\centering
    			\includegraphics[width=\linewidth]{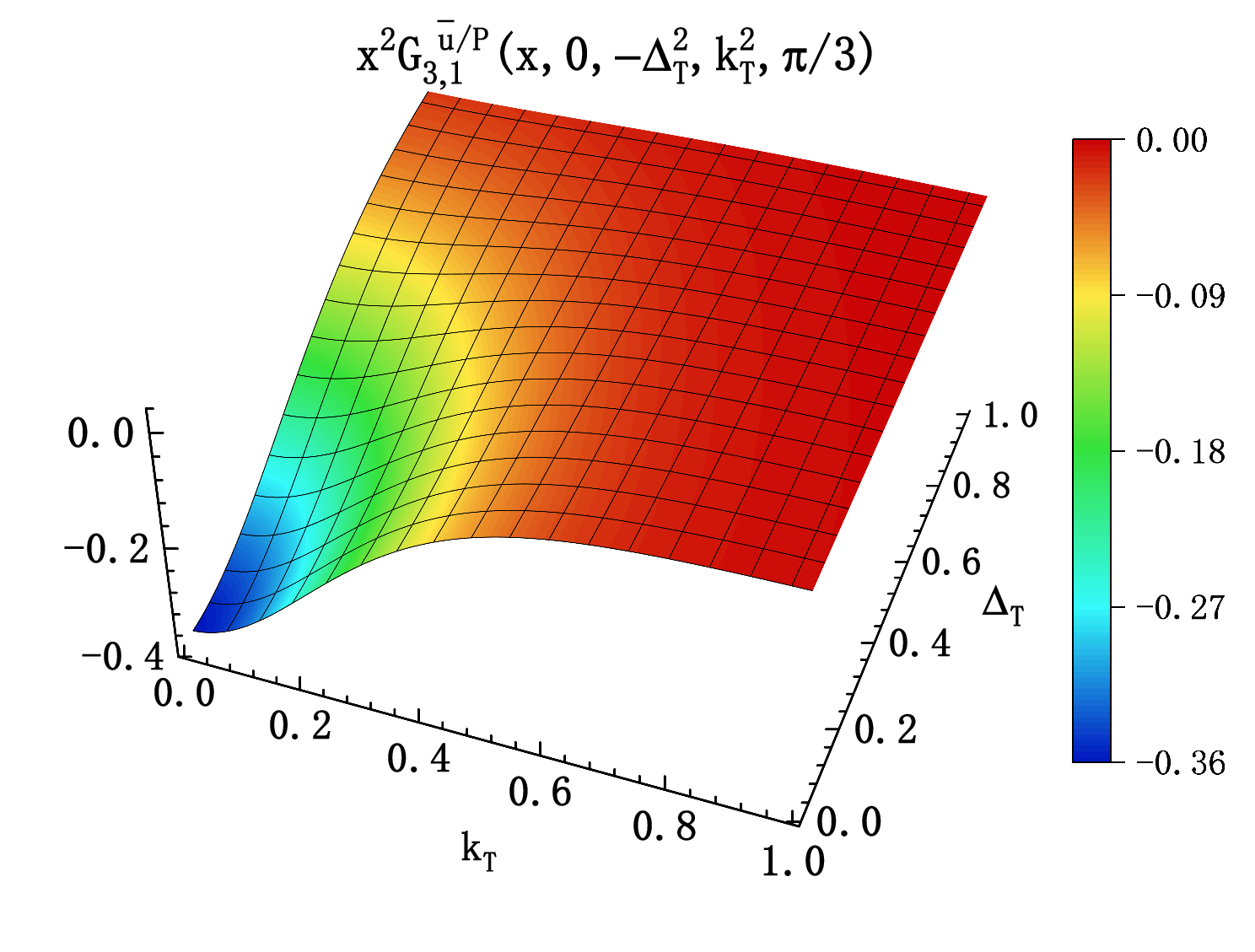}
    	\end{minipage}}
    	\subfigure{\begin{minipage}[b]{0.245\linewidth}
    			\centering
    			\includegraphics[width=\linewidth]{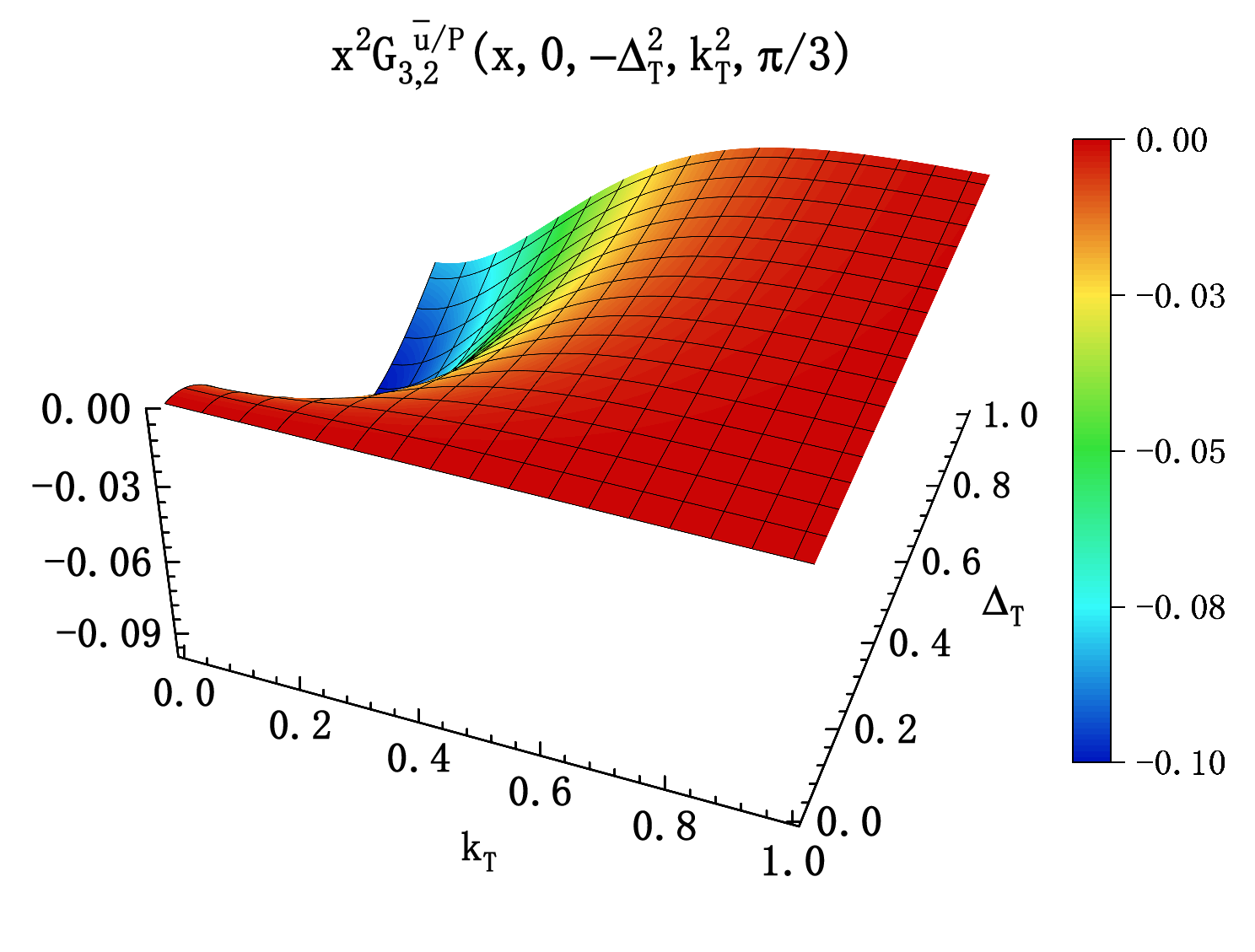}  	
    	\end{minipage}}
    	\subfigure{\begin{minipage}[b]{0.245\linewidth}
    			\centering
    			\includegraphics[width=\linewidth]{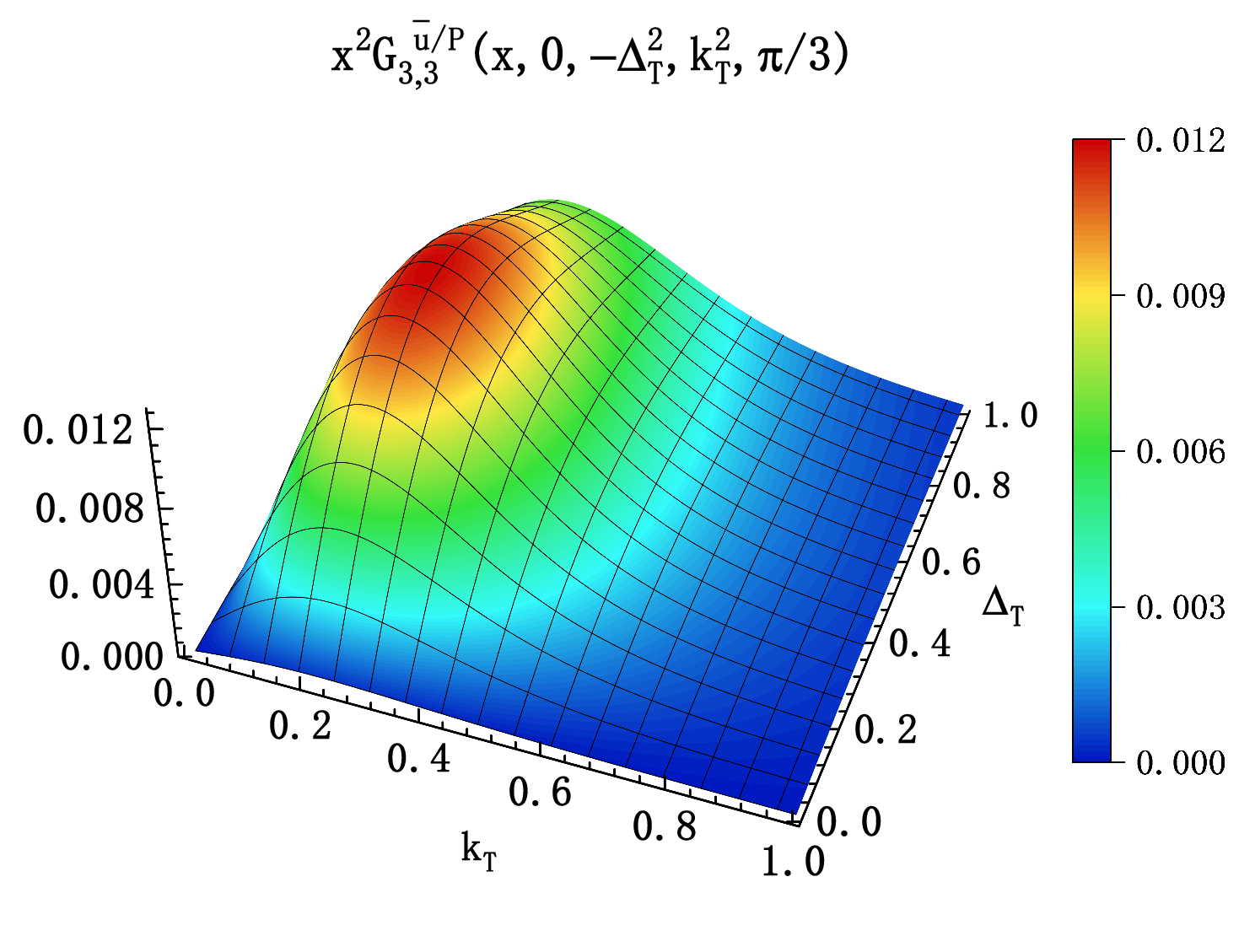}
    	\end{minipage}}
    	\subfigure{\begin{minipage}[b]{0.245\linewidth}
    			\centering
    			\includegraphics[width=\linewidth]{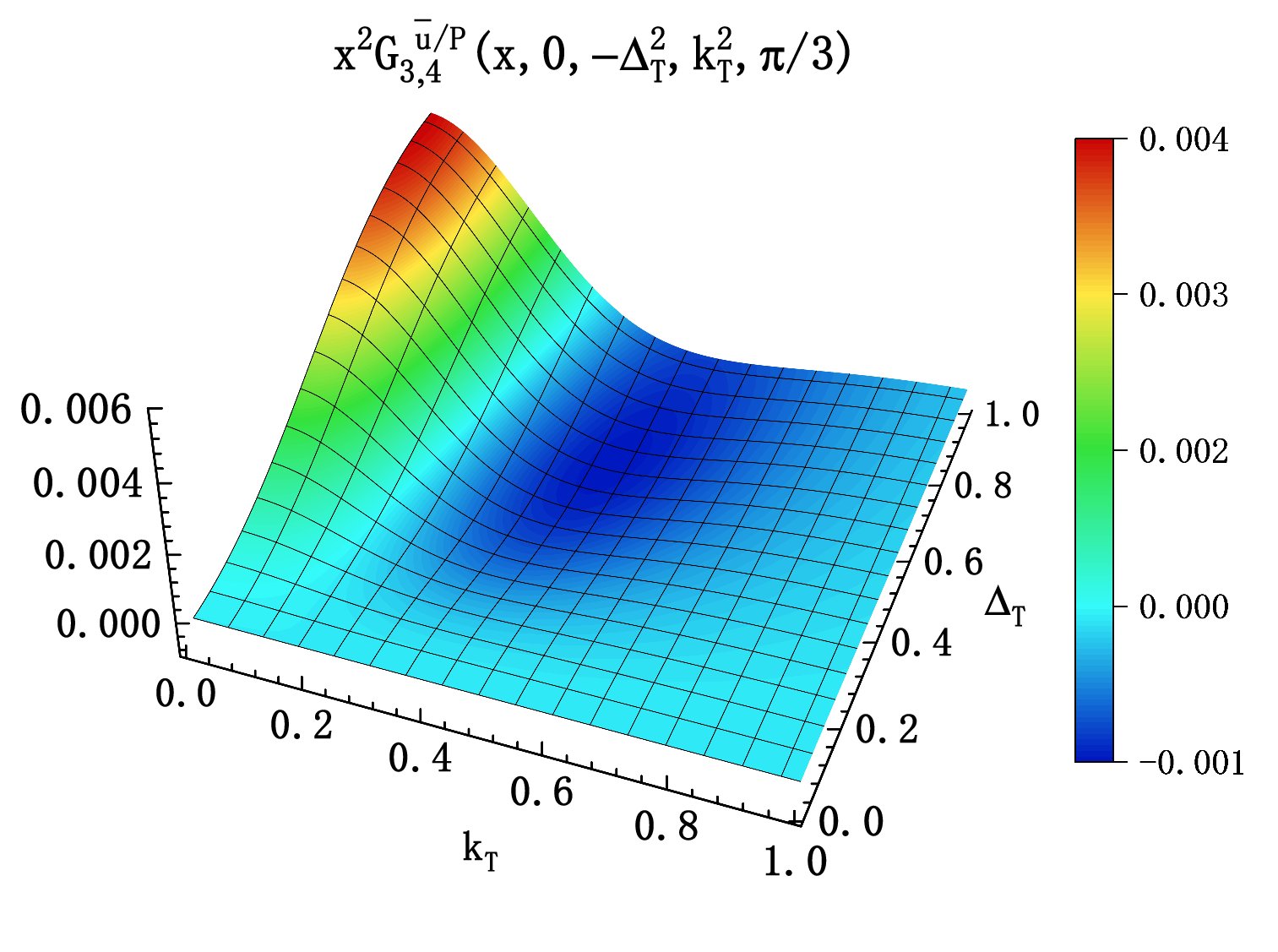}
    	\end{minipage}}
    	\subfigure{\begin{minipage}[b]{0.245\linewidth}
    			\centering
    			\includegraphics[width=\linewidth]{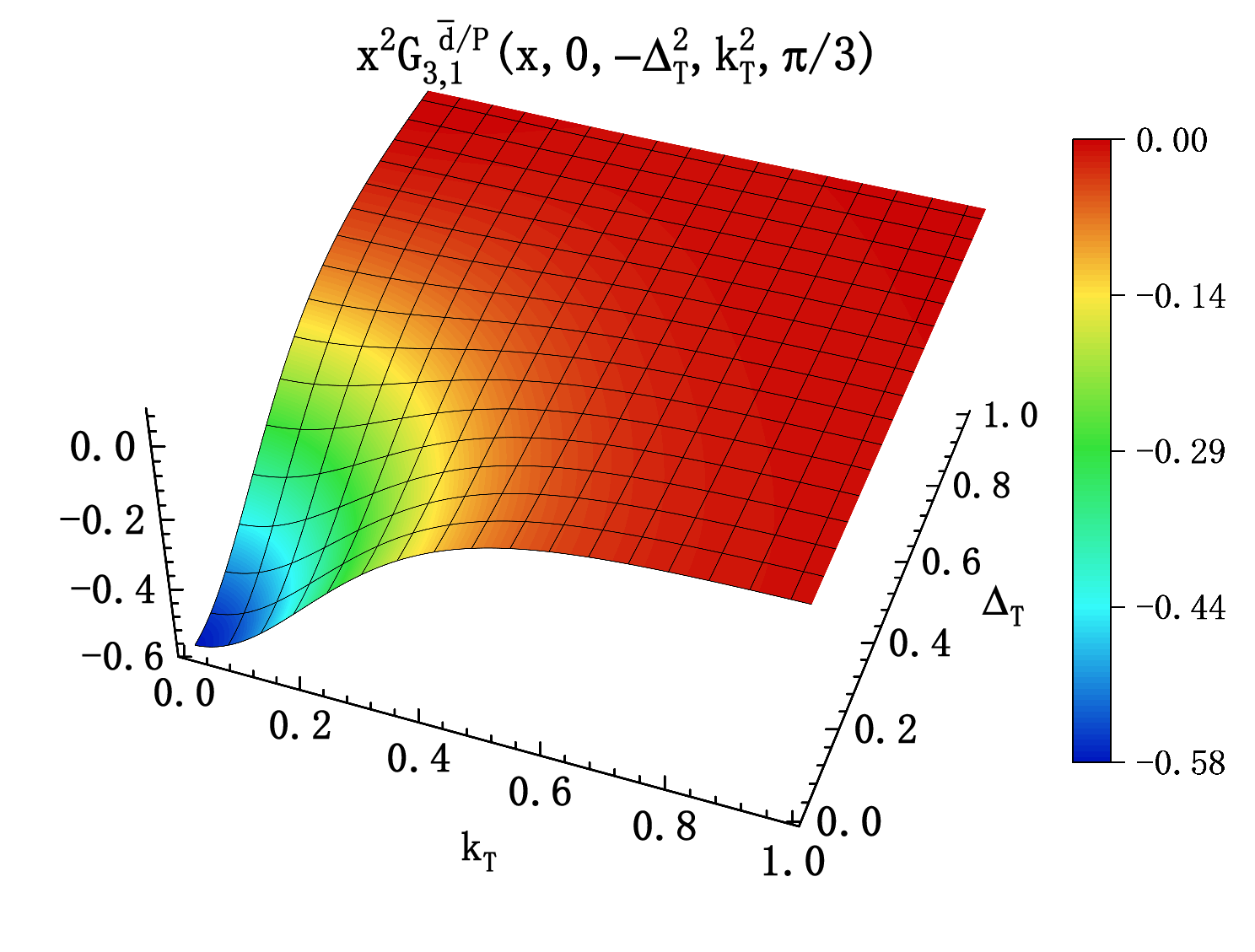}    	
    	\end{minipage}}	
    	\subfigure{\begin{minipage}[b]{0.245\linewidth}
    			\centering
    			\includegraphics[width=\linewidth]{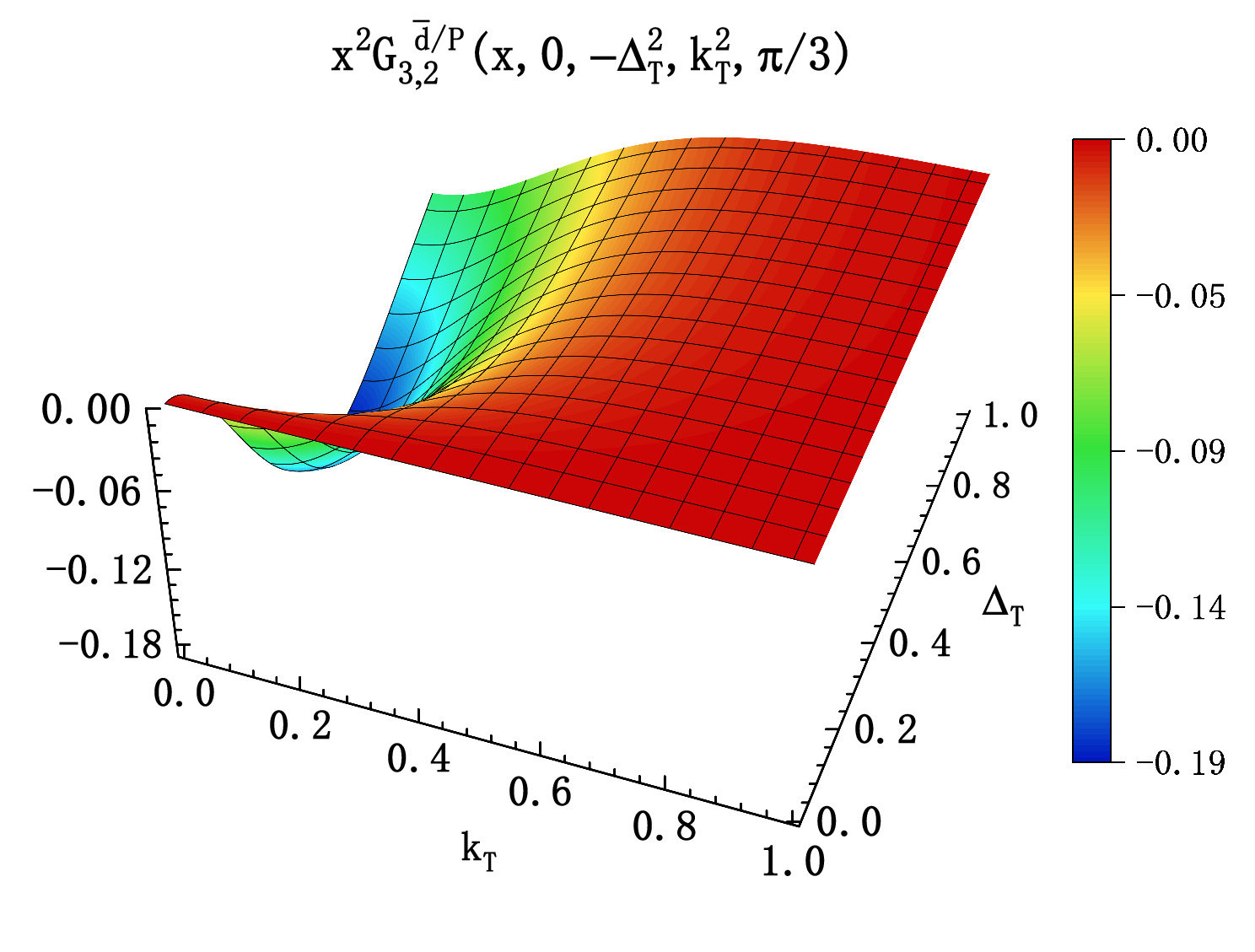}    
    	\end{minipage}}	
    	\centering
    	\subfigure{\begin{minipage}[b]{0.245\linewidth}
    			\centering
    			\includegraphics[width=\linewidth]{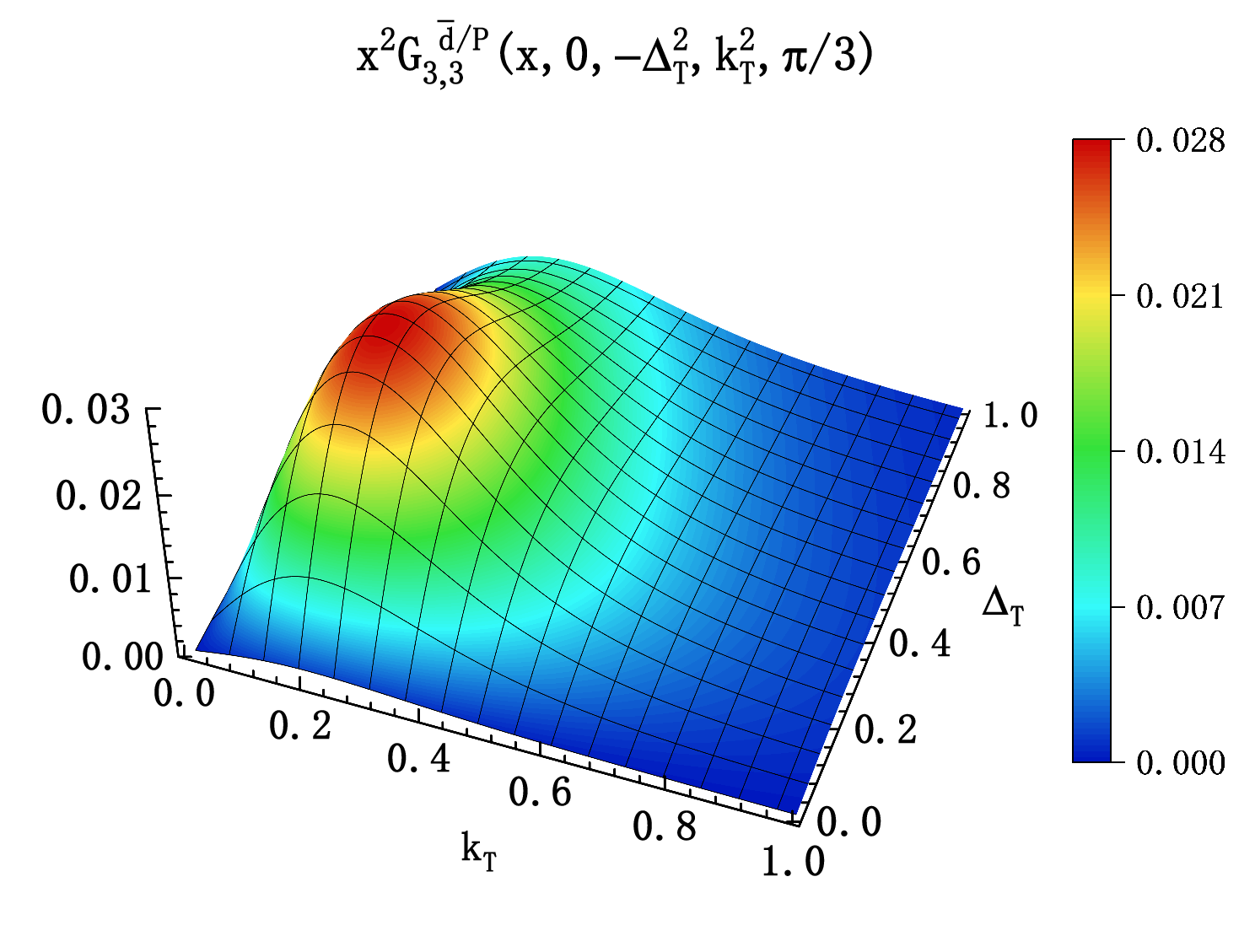}
    	\end{minipage}}
    	\subfigure{\begin{minipage}[b]{0.245\linewidth}
    			\centering
    			\includegraphics[width=\linewidth]{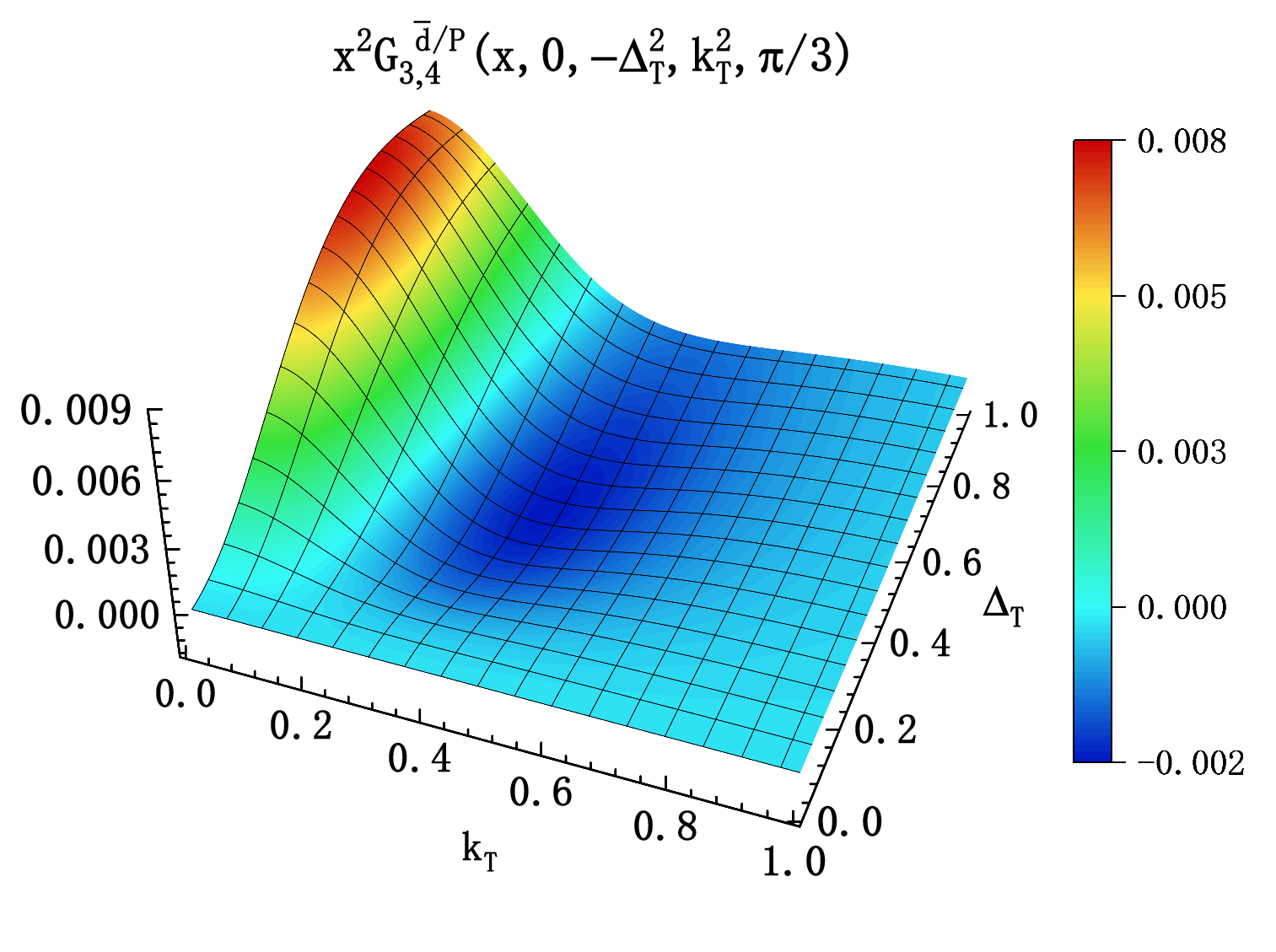}    
    	\end{minipage}}
    	\caption{Similar to Fig~\ref{udbarF-2}, but for longitudinally polarized sea quarks.} \label{udbarG-2}      
    \end{figure*}
    
    \begin{figure*}[htbp]
    	\centering
    	\subfigure{\begin{minipage}[b]{0.245\linewidth}
    			\centering
    			\includegraphics[width=\linewidth]{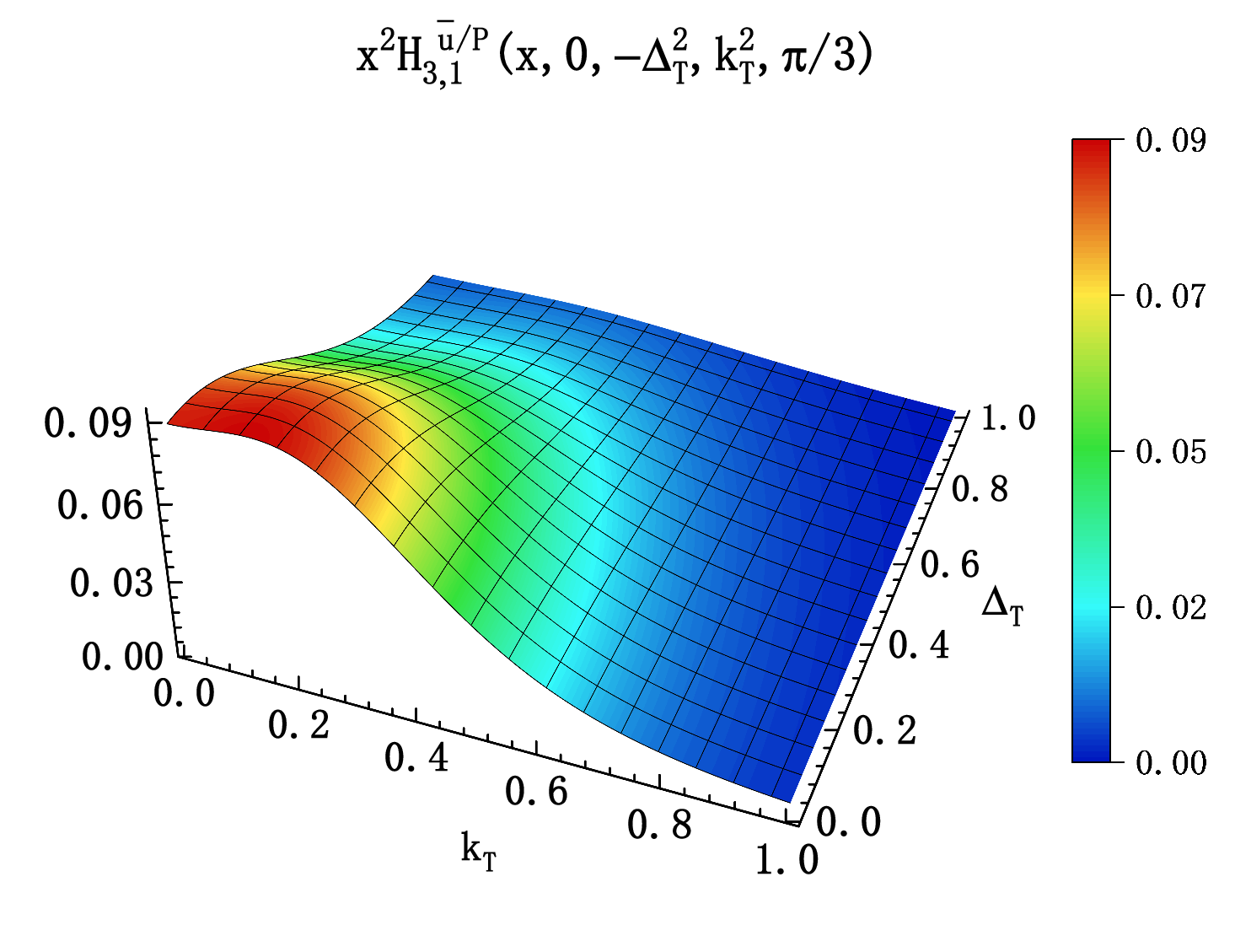}
    	\end{minipage}}
    	\subfigure{\begin{minipage}[b]{0.245\linewidth}
    			\centering
    			\includegraphics[width=\linewidth]{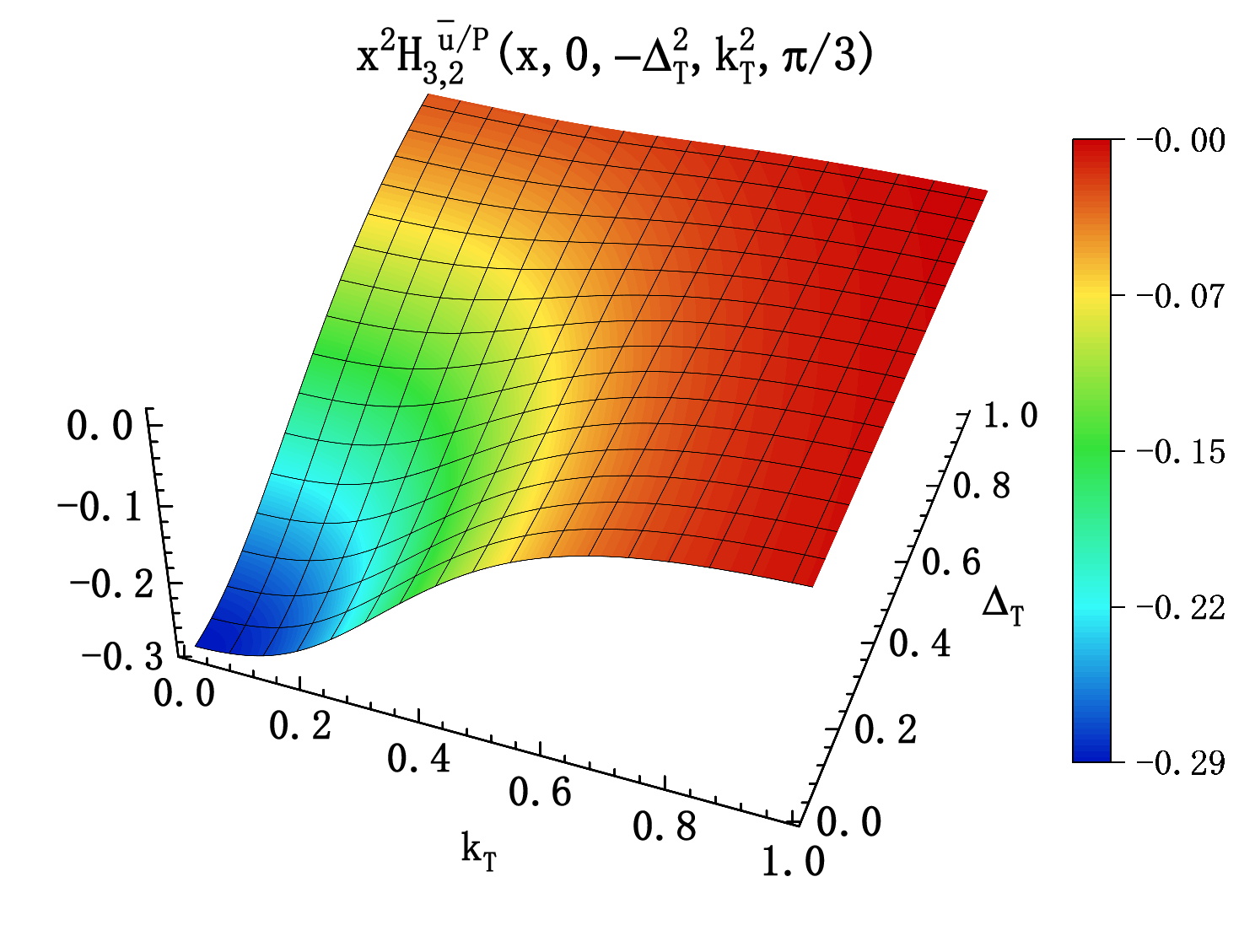}    	
    	\end{minipage}}
    	\subfigure{\begin{minipage}[b]{0.245\linewidth}
    			\centering
    			\includegraphics[width=\linewidth]{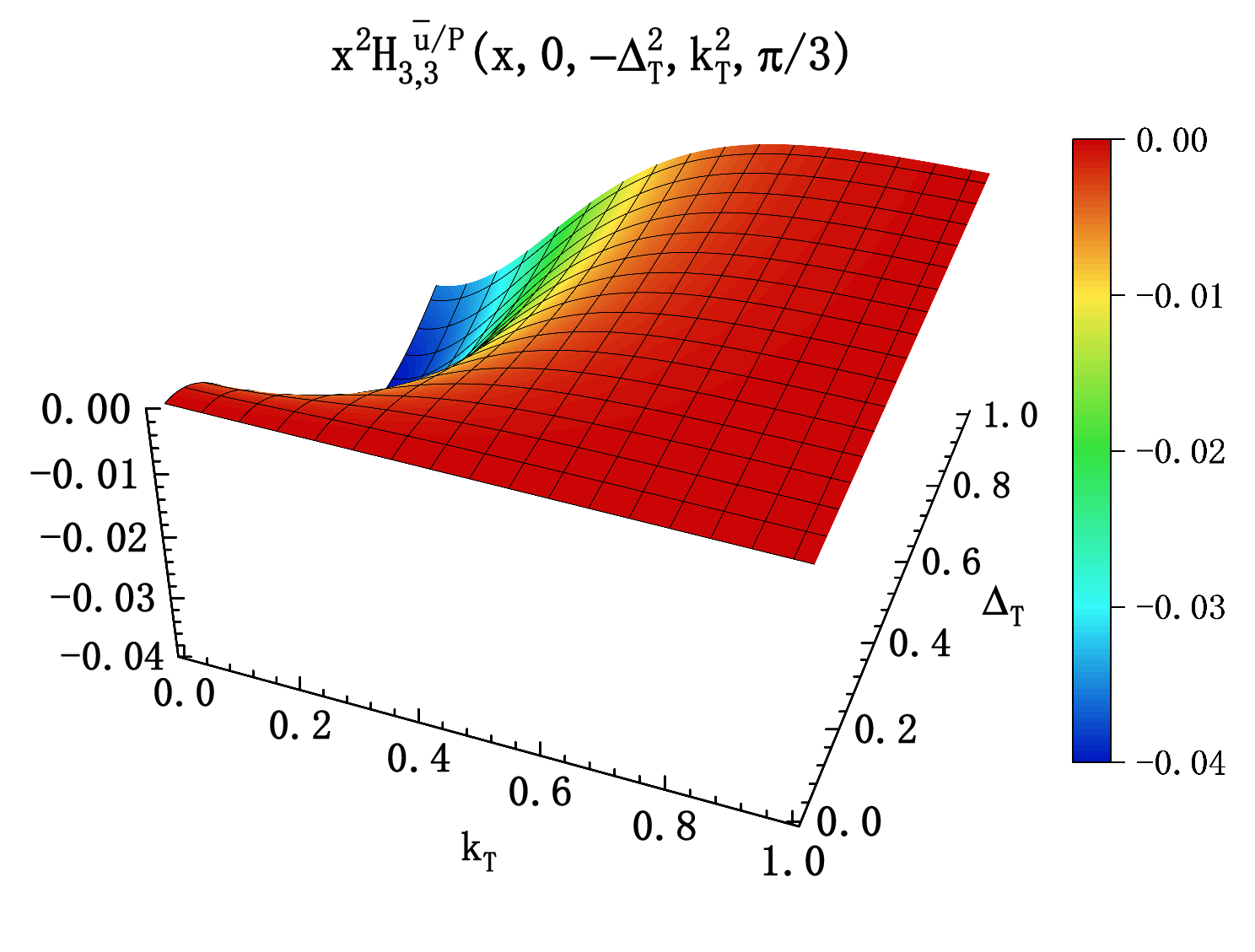}
    	\end{minipage}}
    	\subfigure{\begin{minipage}[b]{0.245\linewidth}
    			\centering
    			\includegraphics[width=\linewidth]{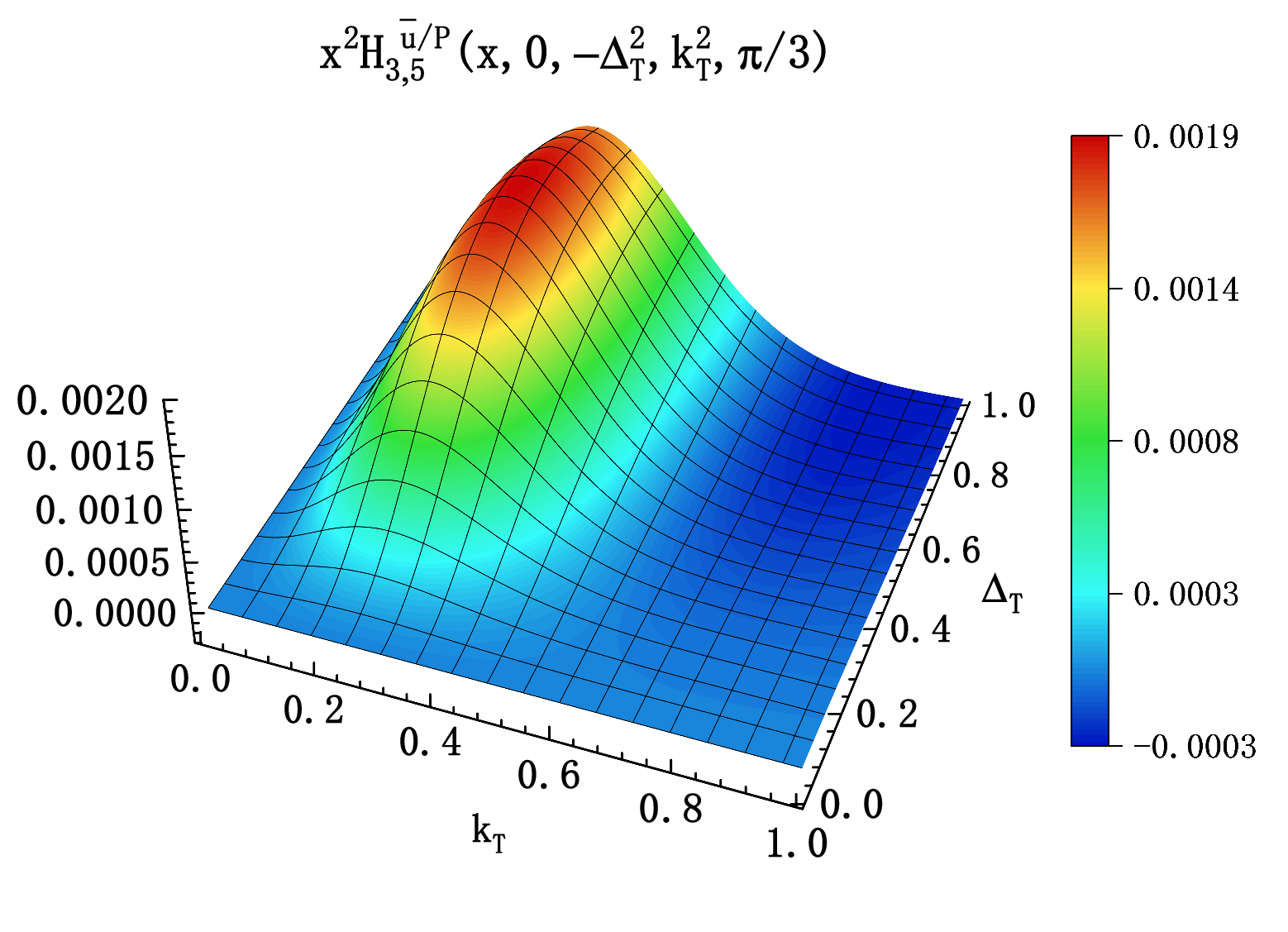}
    	\end{minipage}}
        \subfigure{\begin{minipage}[b]{0.245\linewidth}
        		\centering
        		\includegraphics[width=\linewidth]{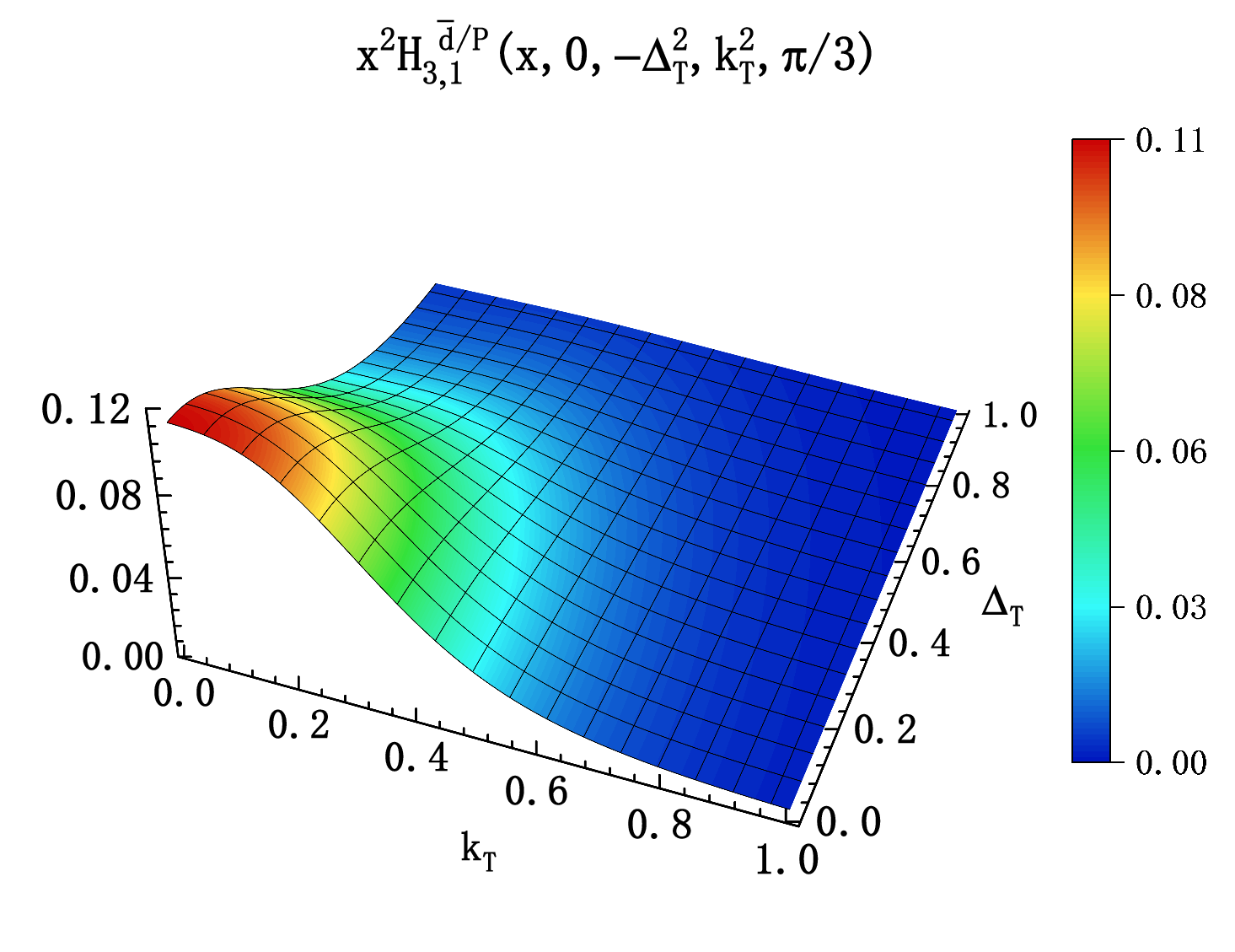}
        \end{minipage}}
        \subfigure{\begin{minipage}[b]{0.245\linewidth}
        		\centering
        		\includegraphics[width=\linewidth]{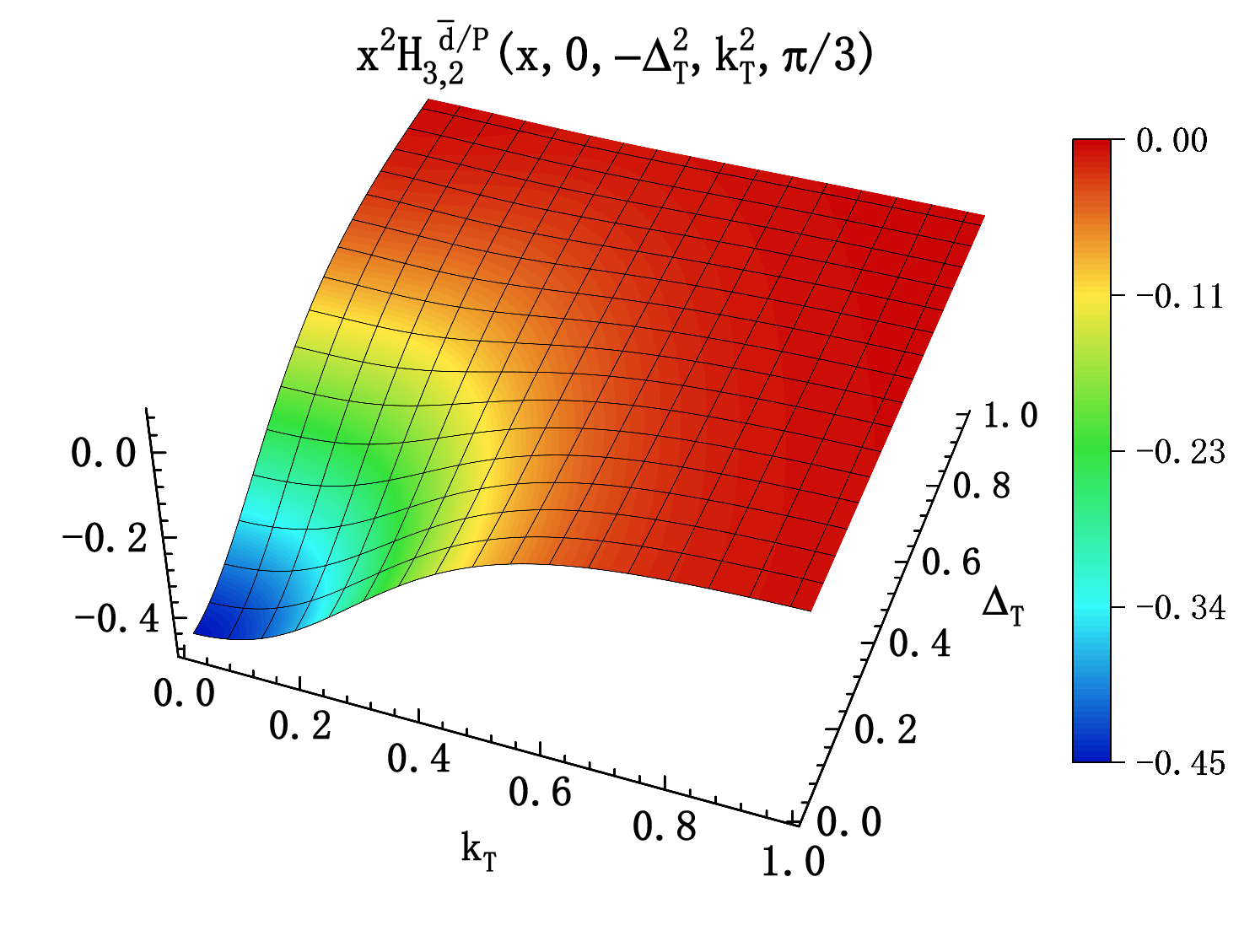}    	
        \end{minipage}}
        \subfigure{\begin{minipage}[b]{0.245\linewidth}
        		\centering
        		\includegraphics[width=\linewidth]{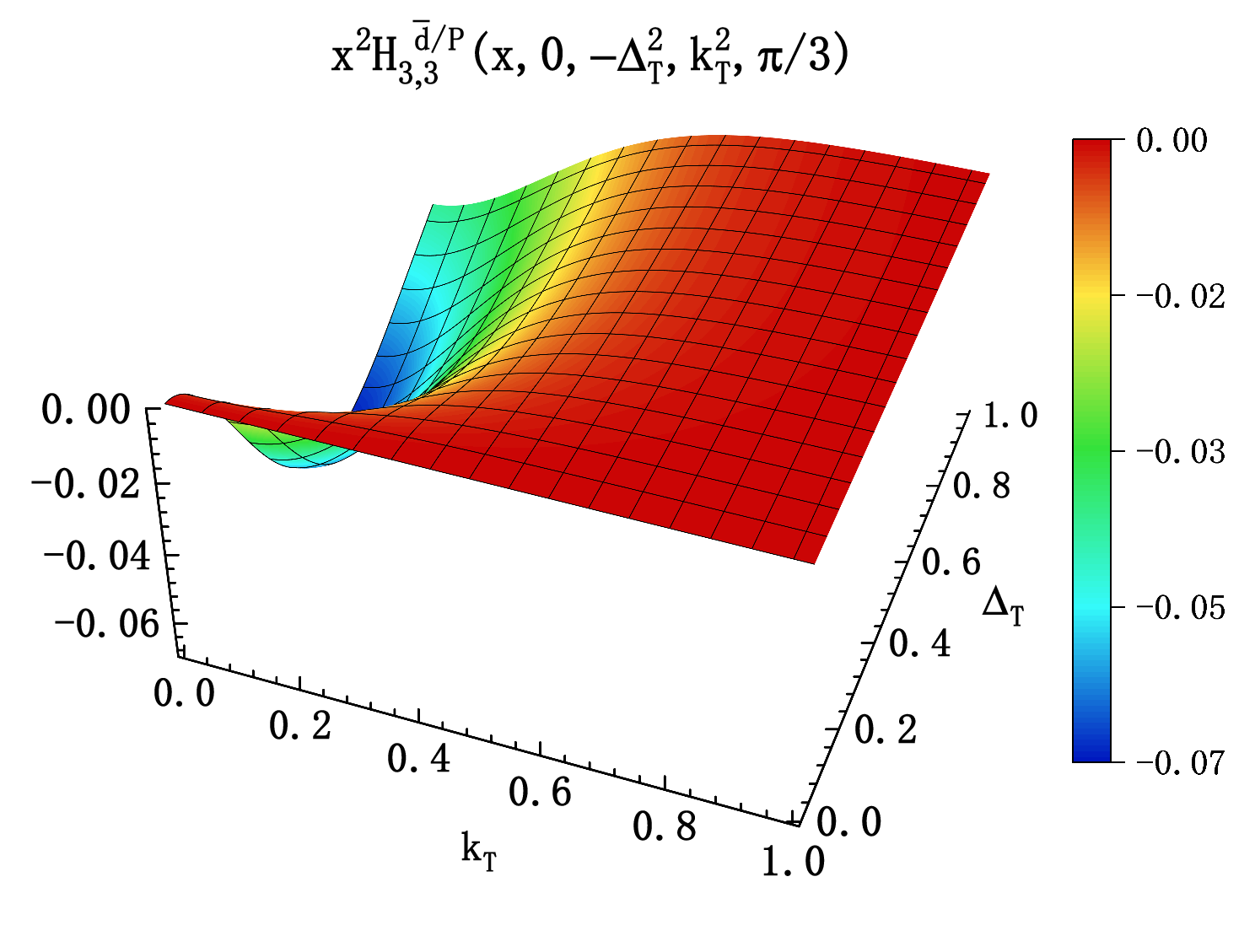}
        \end{minipage}}
        \subfigure{\begin{minipage}[b]{0.245\linewidth}
        		\centering
        		\includegraphics[width=\linewidth]{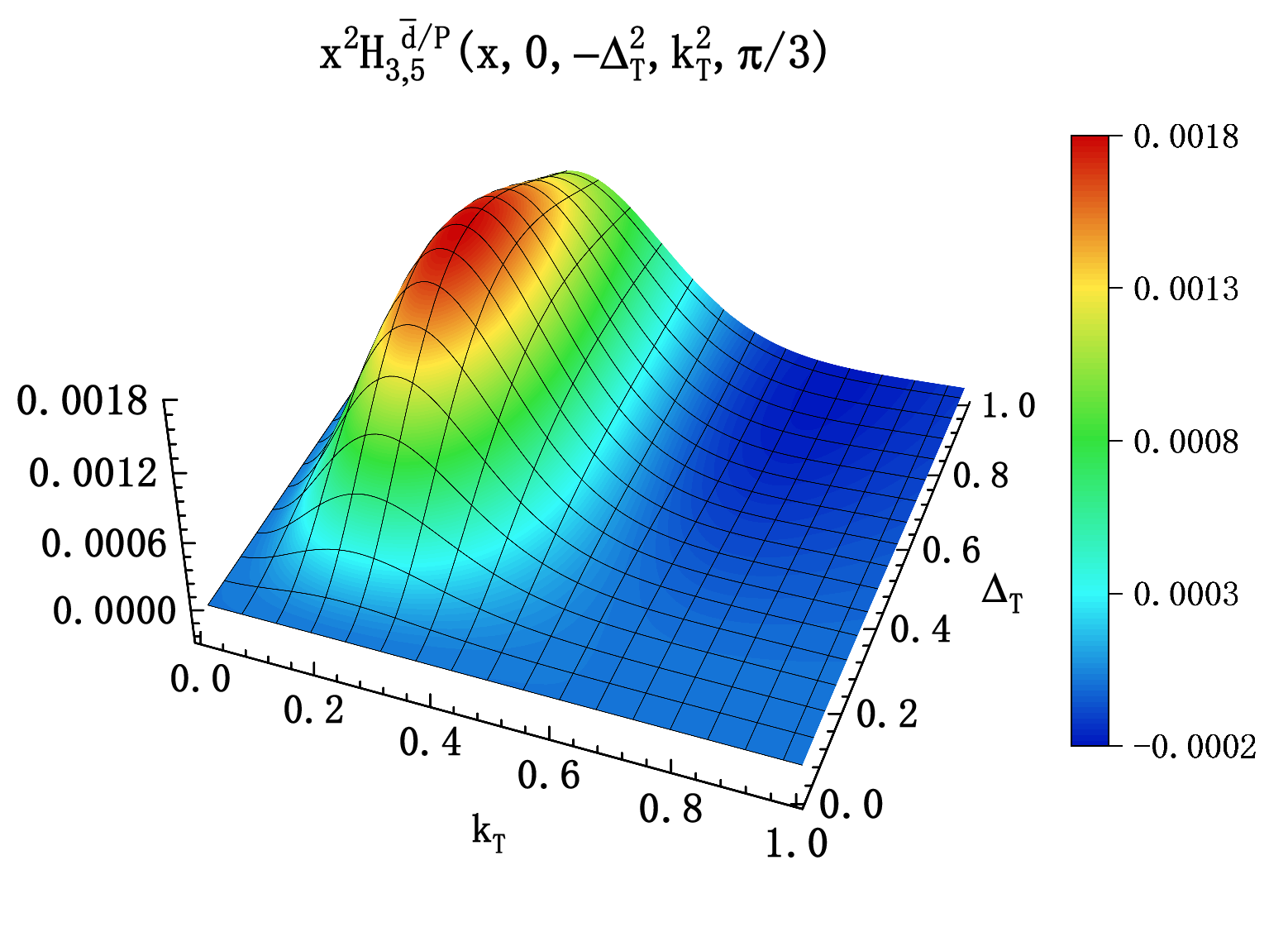}
        \end{minipage}}
    
    	\subfigure{\begin{minipage}[b]{0.245\linewidth}
    			\centering
    			\includegraphics[width=\linewidth]{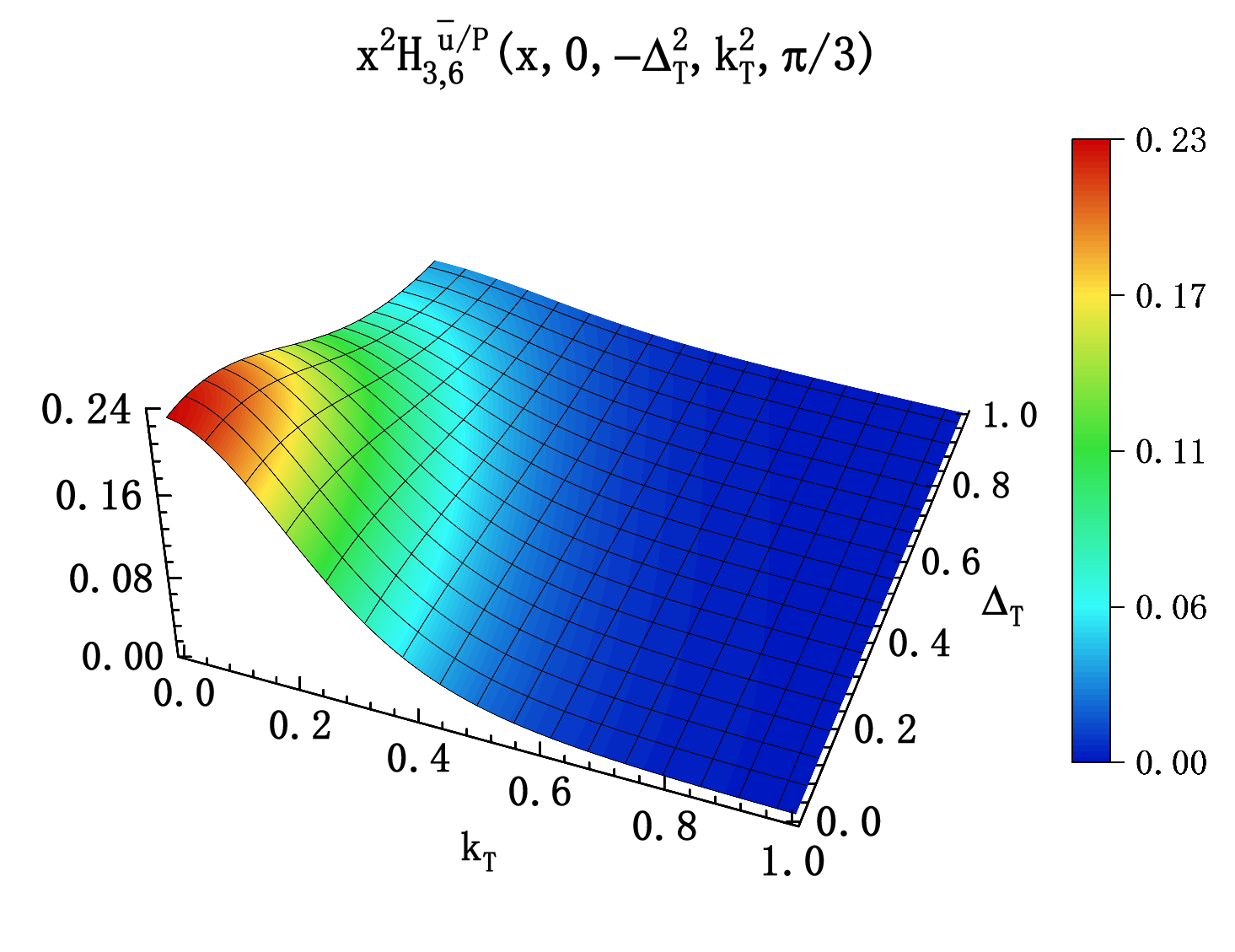}    	
    	\end{minipage}}	
    	\subfigure{\begin{minipage}[b]{0.245\linewidth}
    			\centering
    			\includegraphics[width=\linewidth]{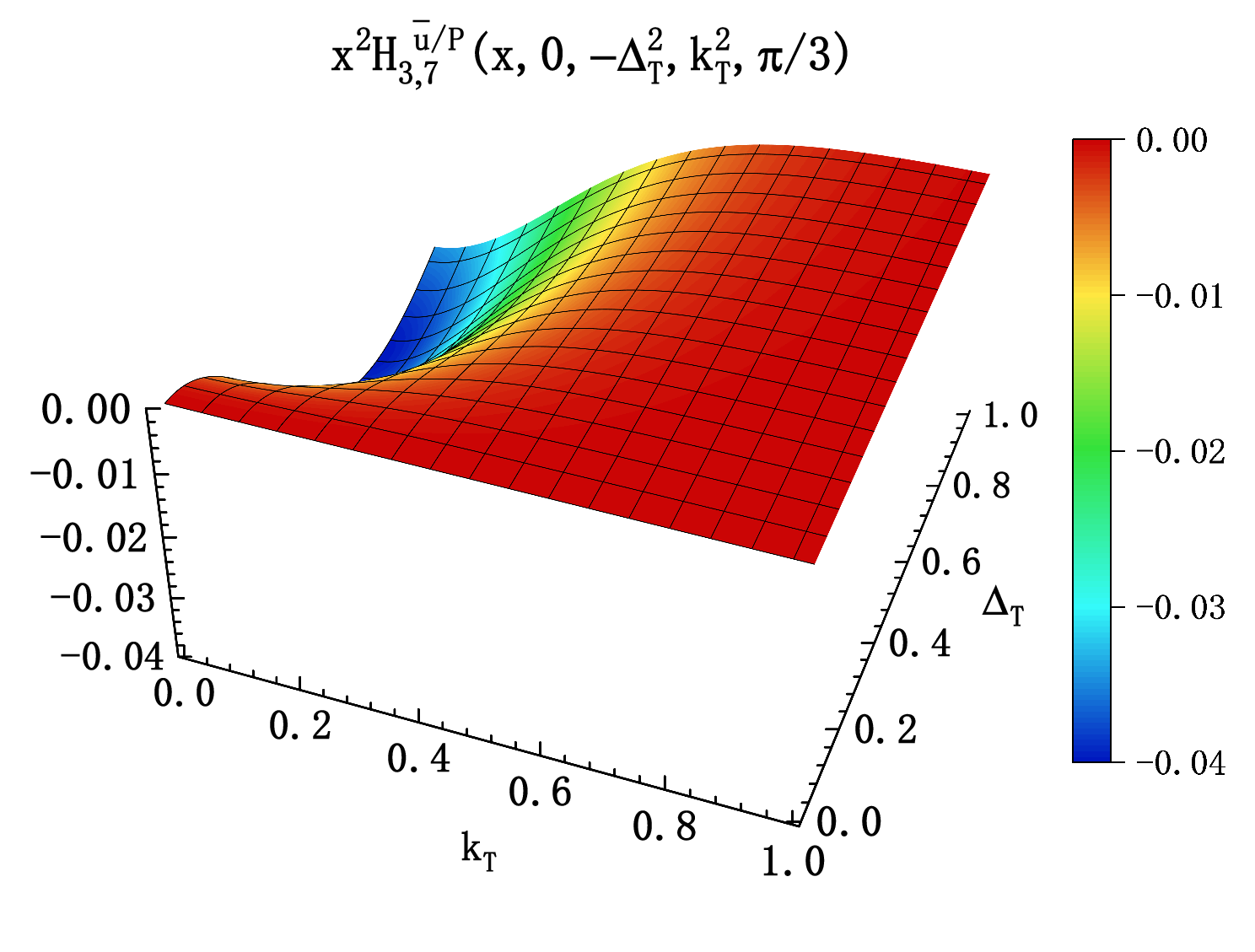}    
    	\end{minipage}}	
    	\subfigure{\begin{minipage}[b]{0.245\linewidth}
    			\centering
    			\includegraphics[width=\linewidth]{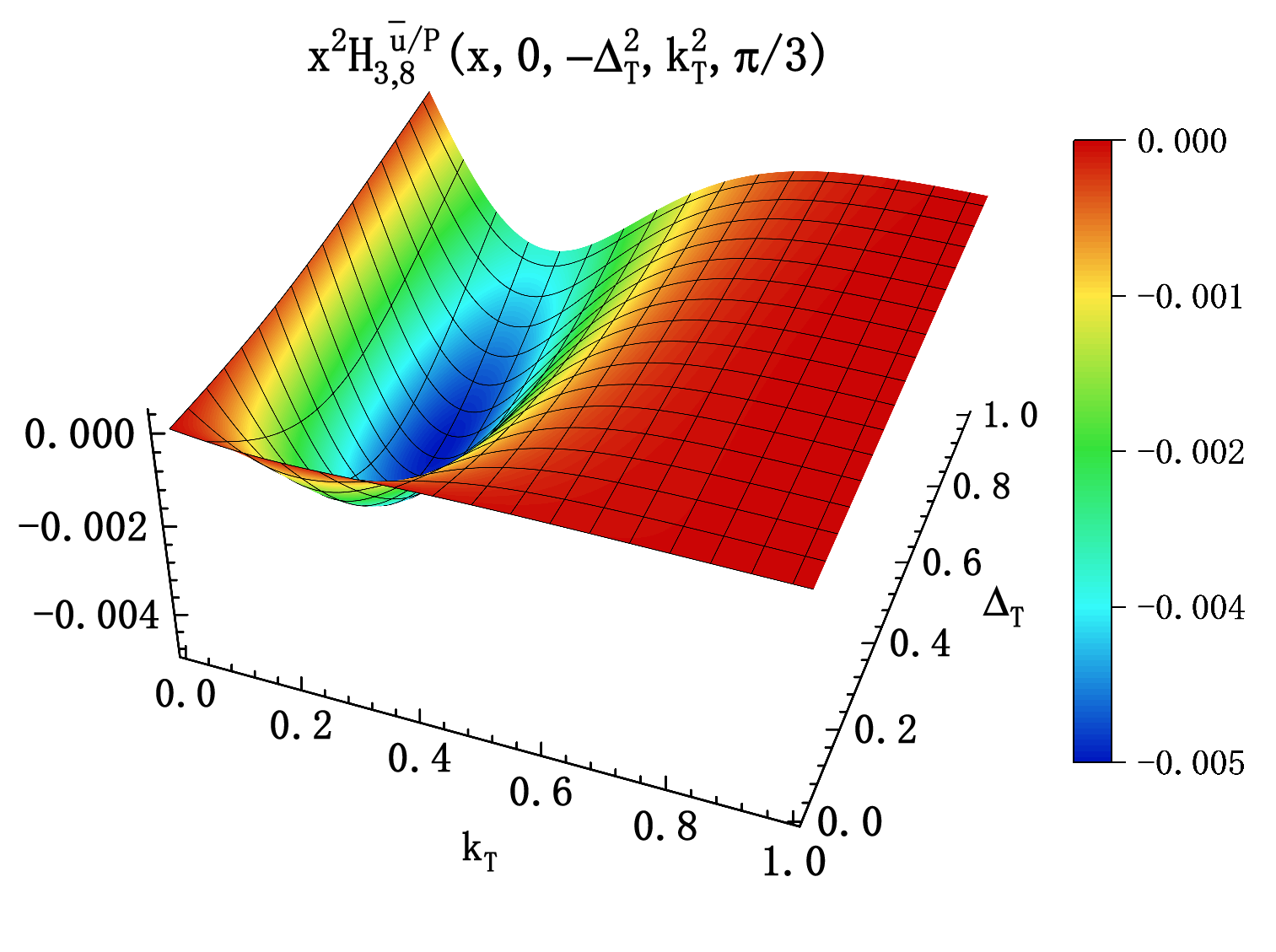}
    	\end{minipage}}
    
    	\subfigure{\begin{minipage}[b]{0.245\linewidth}
    			\centering
    			\includegraphics[width=\linewidth]{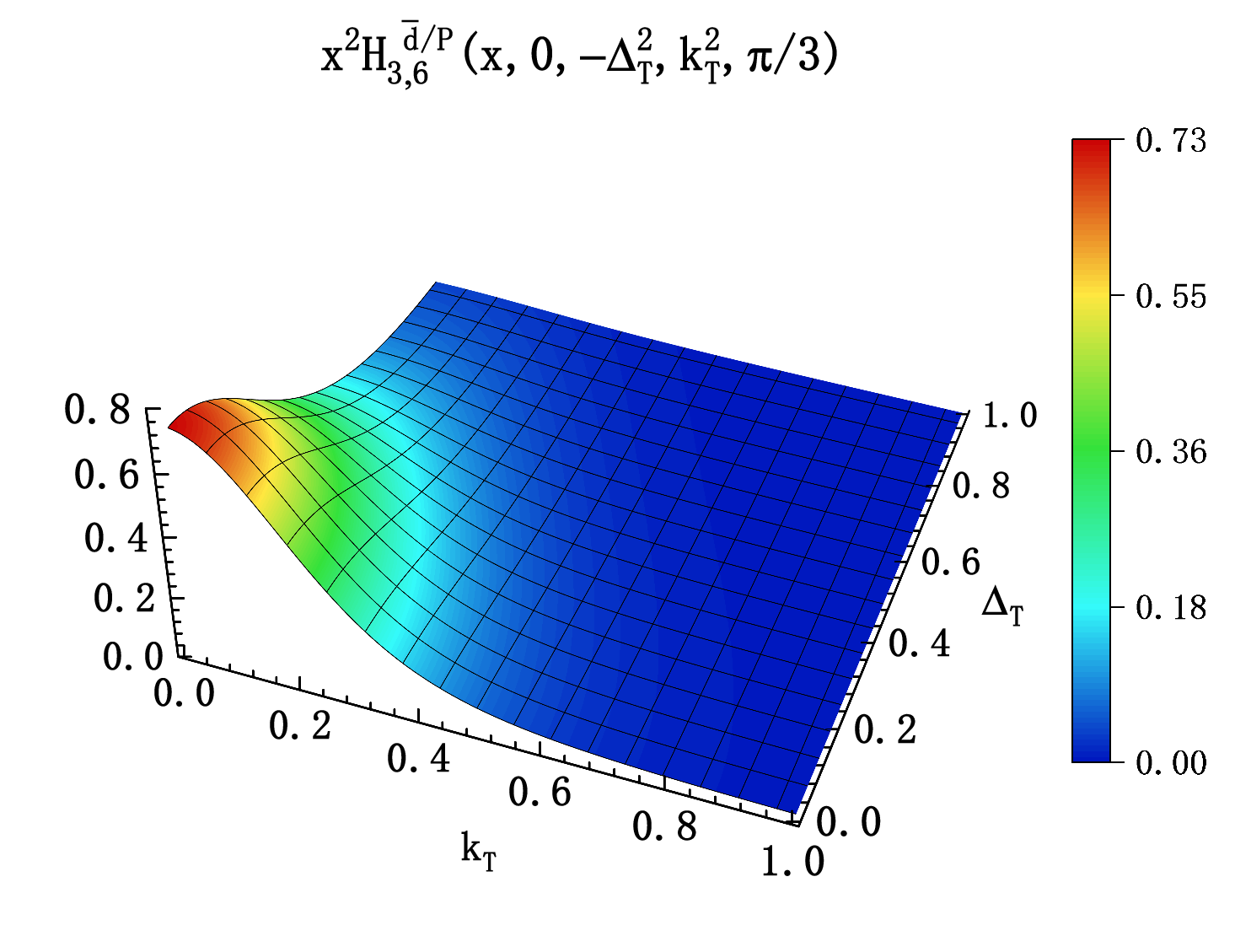}    	
    	\end{minipage}}	
    	\subfigure{\begin{minipage}[b]{0.245\linewidth}
    			\centering
    			\includegraphics[width=\linewidth]{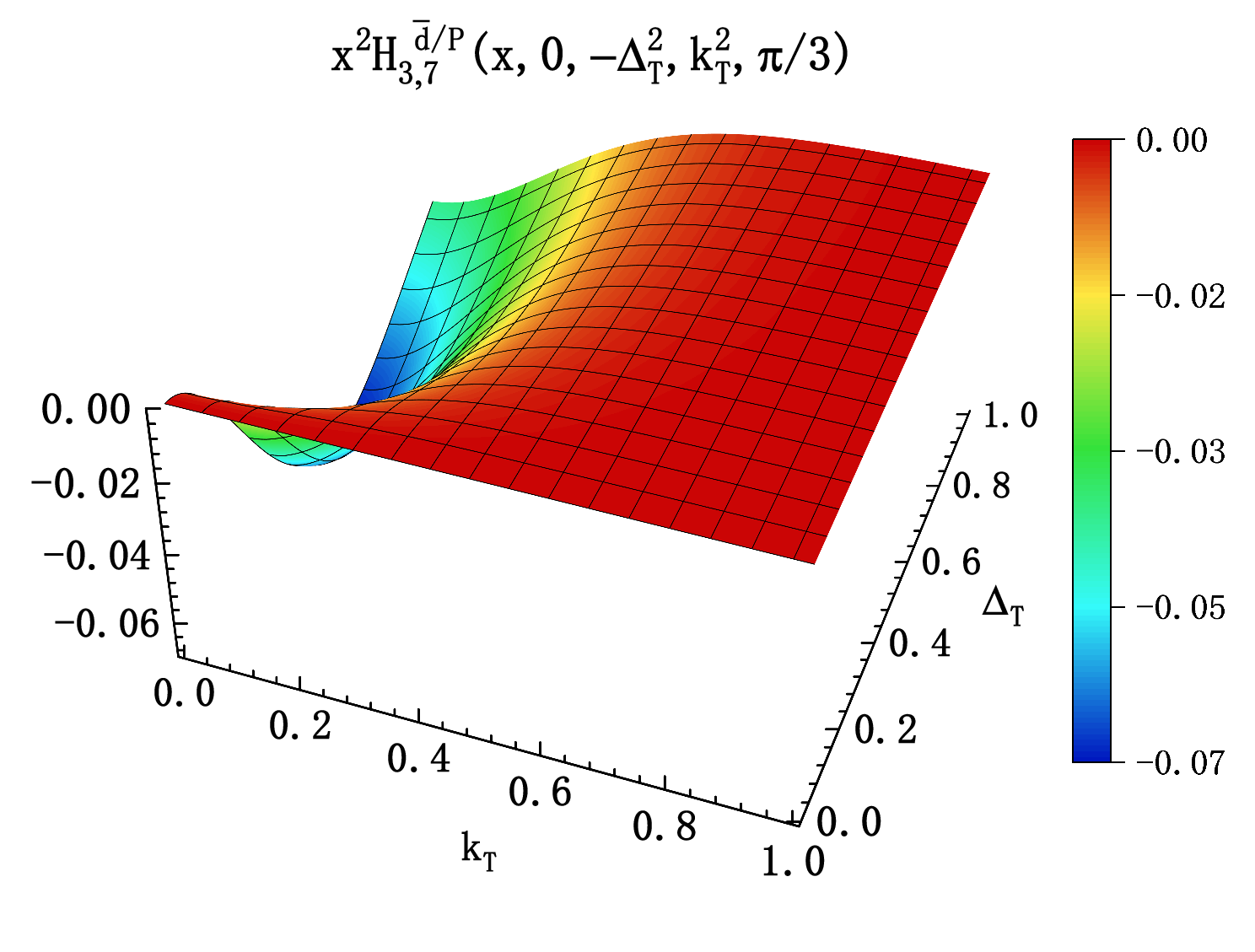}    
    	\end{minipage}}	
    	\centering
    	\subfigure{\begin{minipage}[b]{0.245\linewidth}
    			\centering
    			\includegraphics[width=\linewidth]{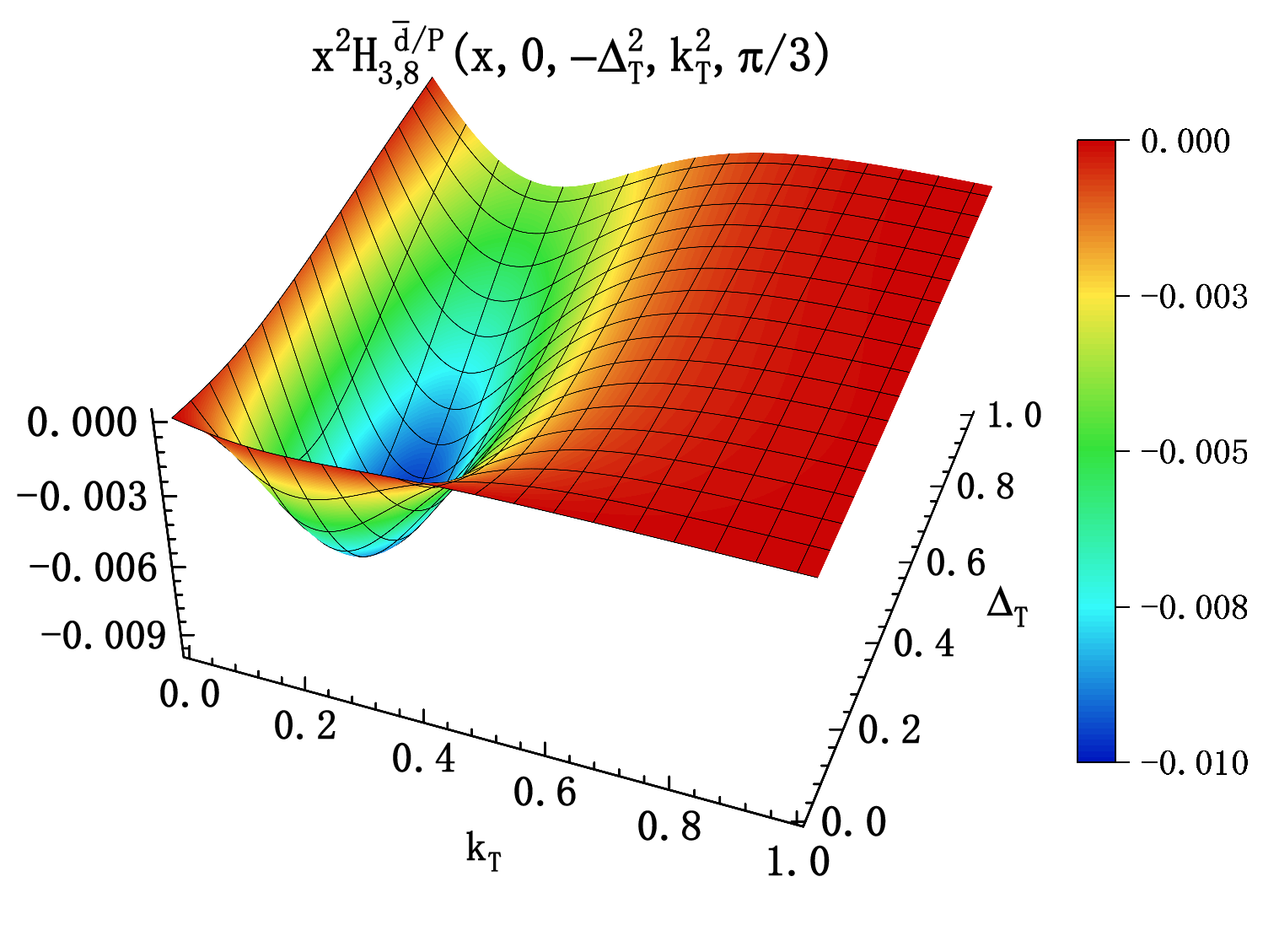}
    	\end{minipage}}
    	\caption{Similar to Fig~\ref{udbarF-2}, but for transversely polarized sea quarks.} \label{udbarH-2}     
    \end{figure*}
    
    \begin{figure*}[htbp]
    	\centering
    	\subfigure{\begin{minipage}[b]{0.245\linewidth}
    			\centering
    			\includegraphics[width=\linewidth]{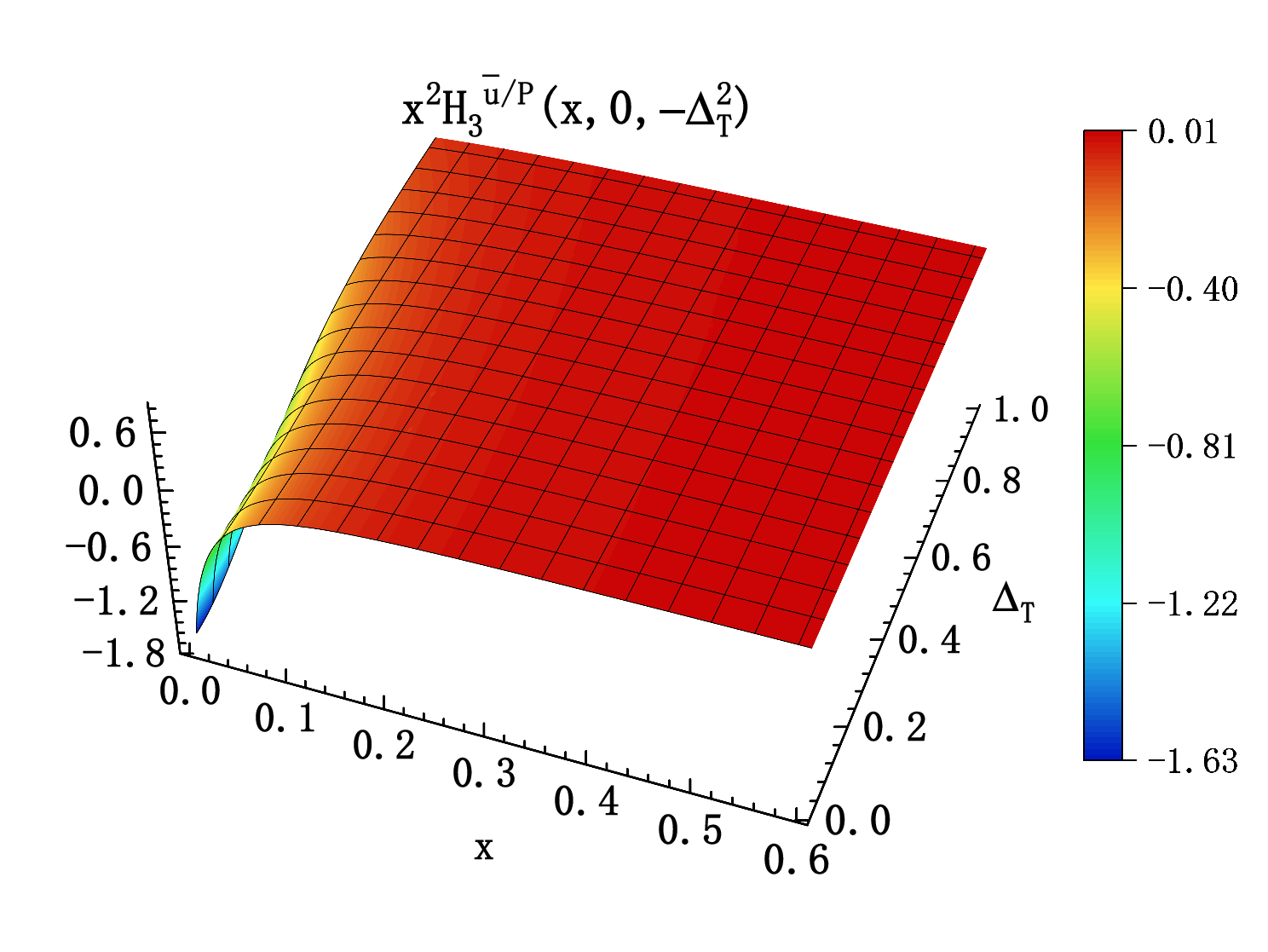}
    	\end{minipage}}
    	\subfigure{\begin{minipage}[b]{0.245\linewidth}
    			\centering
    			\includegraphics[width=\linewidth]{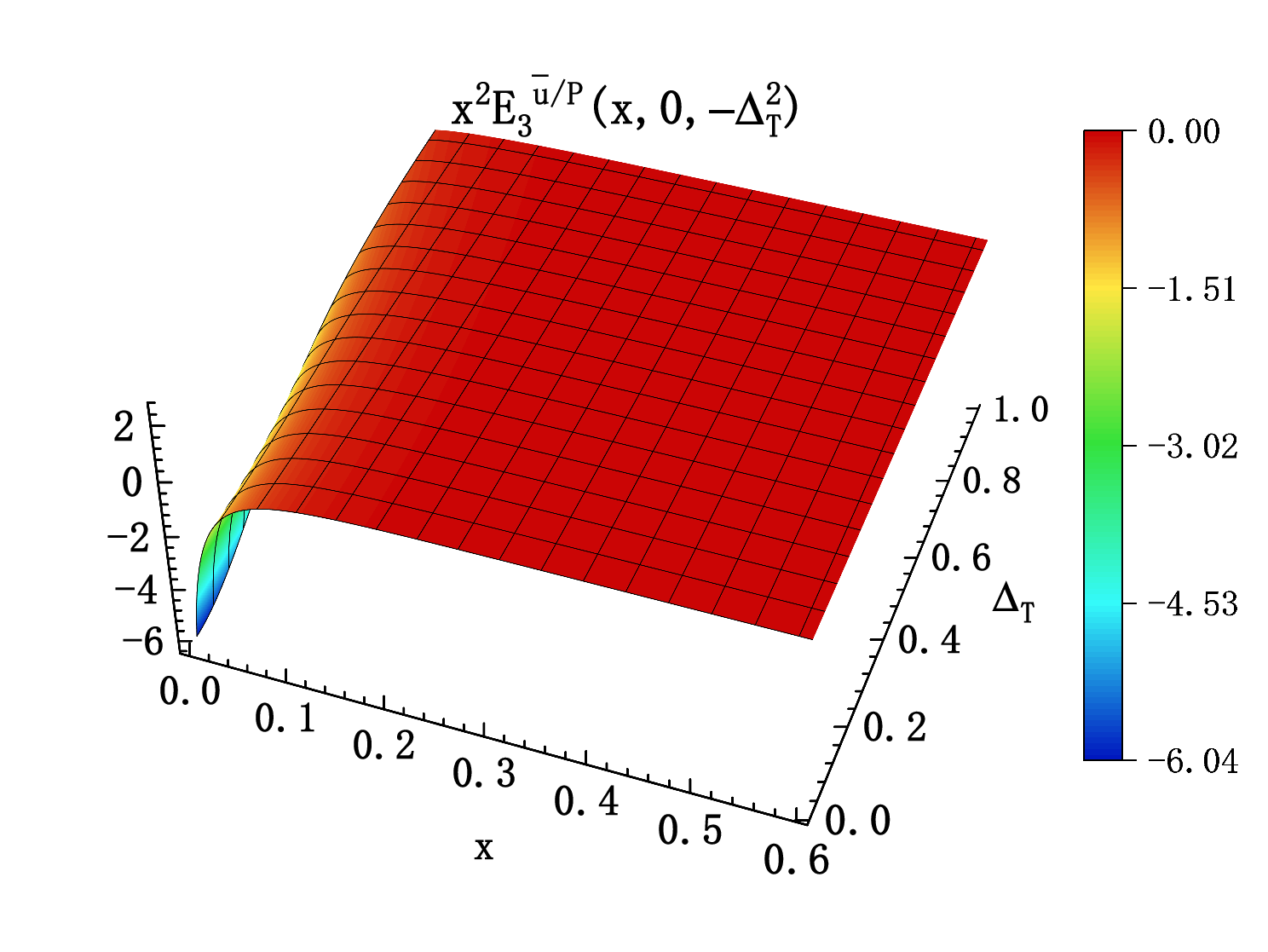}    	
    	\end{minipage}}
    	\subfigure{\begin{minipage}[b]{0.245\linewidth}
    			\centering
    			\includegraphics[width=\linewidth]{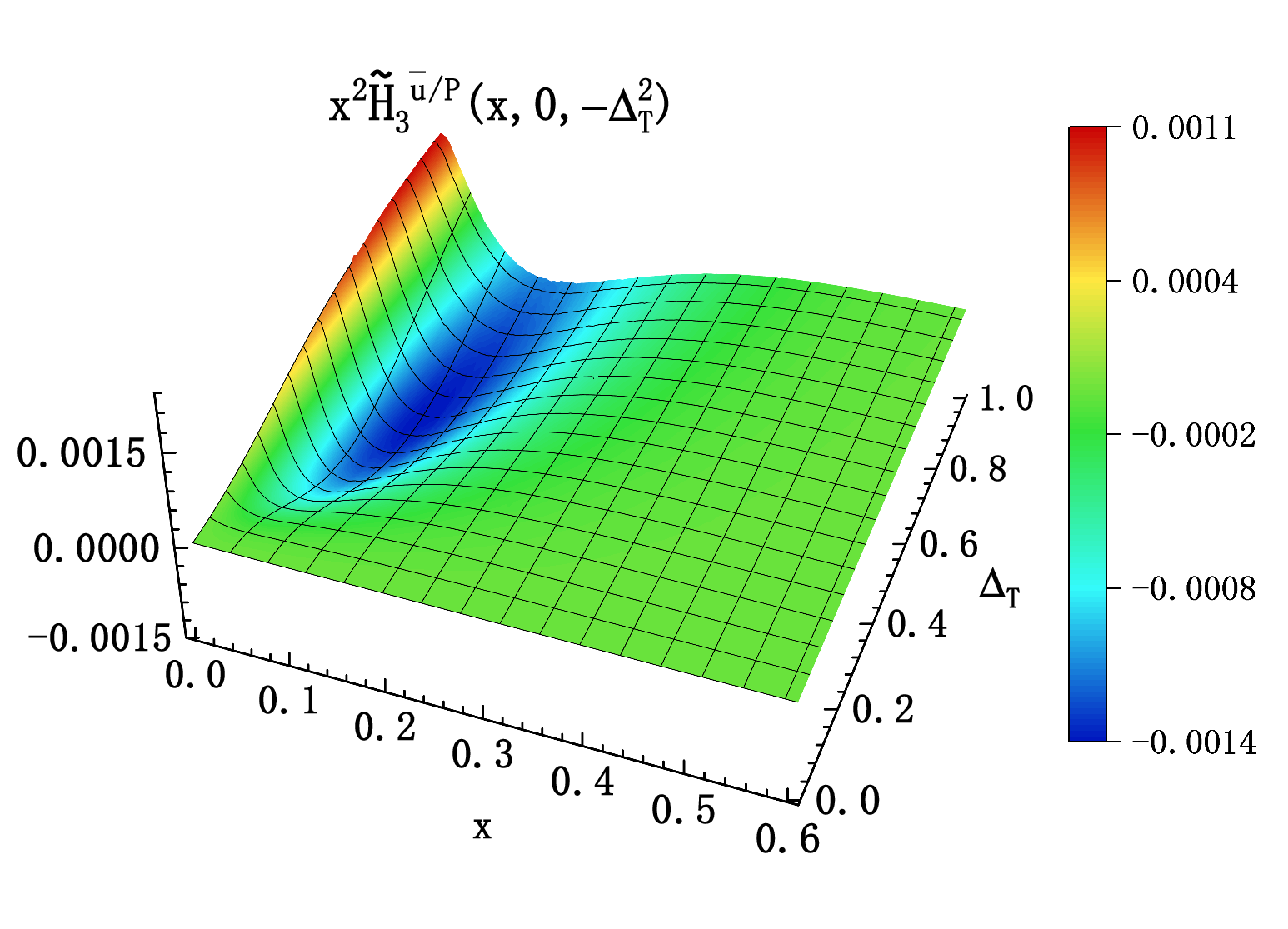}
    	\end{minipage}}
    
    	\subfigure{\begin{minipage}[b]{0.245\linewidth}
    			\centering
    			\includegraphics[width=\linewidth]{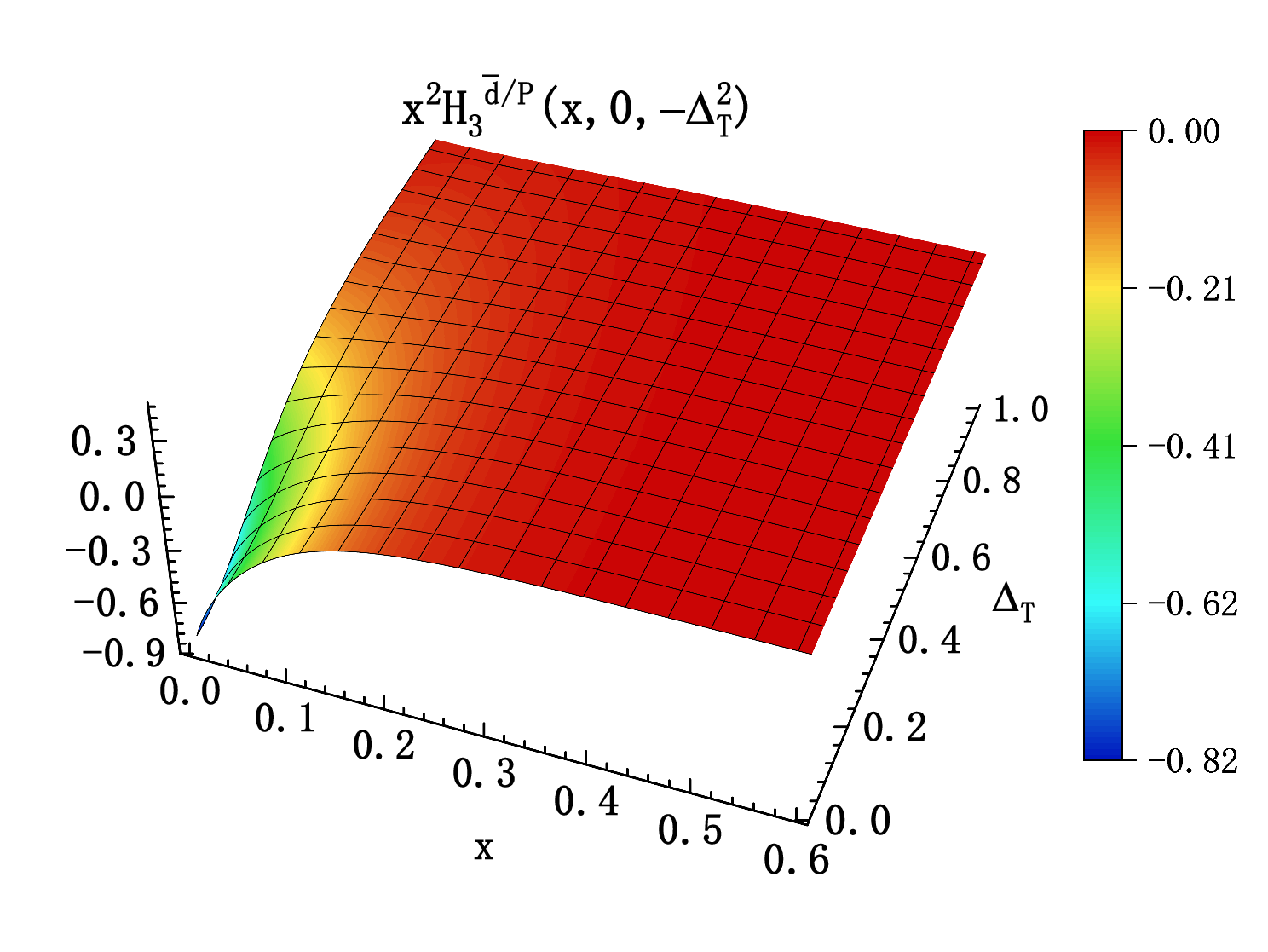}    	
    	\end{minipage}}
       \subfigure{\begin{minipage}[b]{0.245\linewidth}
       		\centering
       		\includegraphics[width=\linewidth]{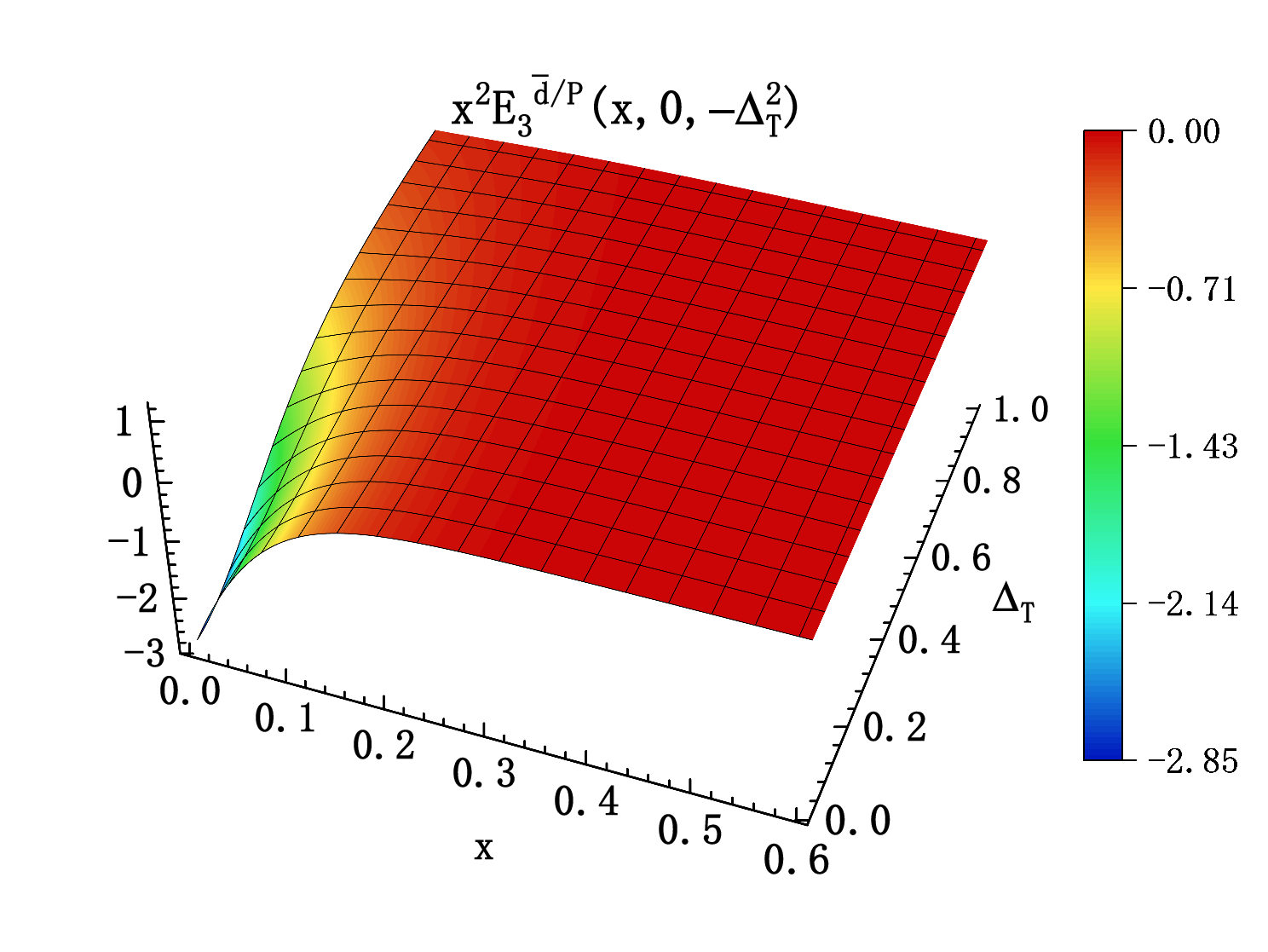}
       \end{minipage}}
       \subfigure{\begin{minipage}[b]{0.245\linewidth}
       		\centering
       		\includegraphics[width=\linewidth]{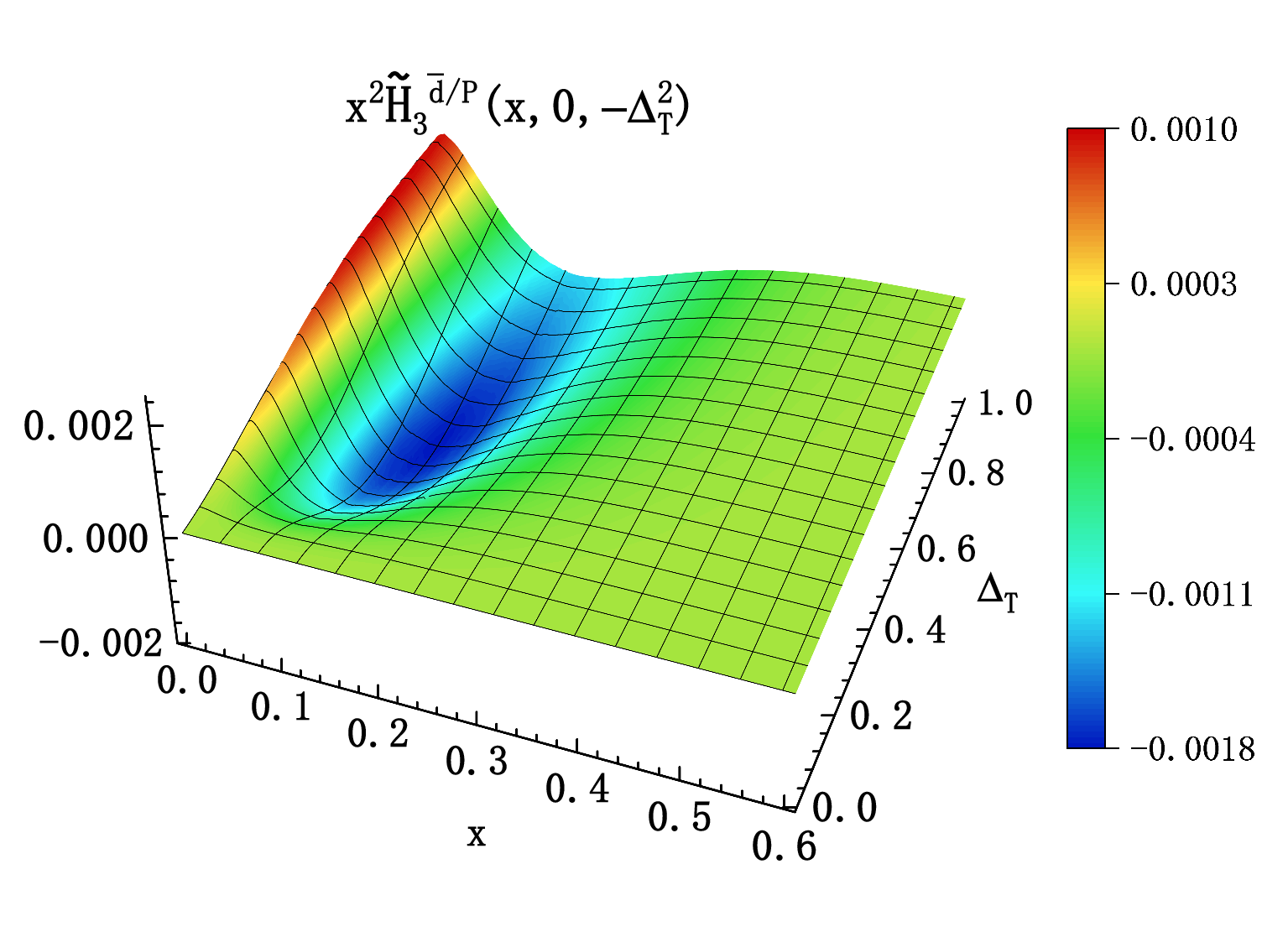}
       \end{minipage}}
   
       \subfigure{\begin{minipage}[b]{0.245\linewidth}
       		\centering
       		\includegraphics[width=\linewidth]{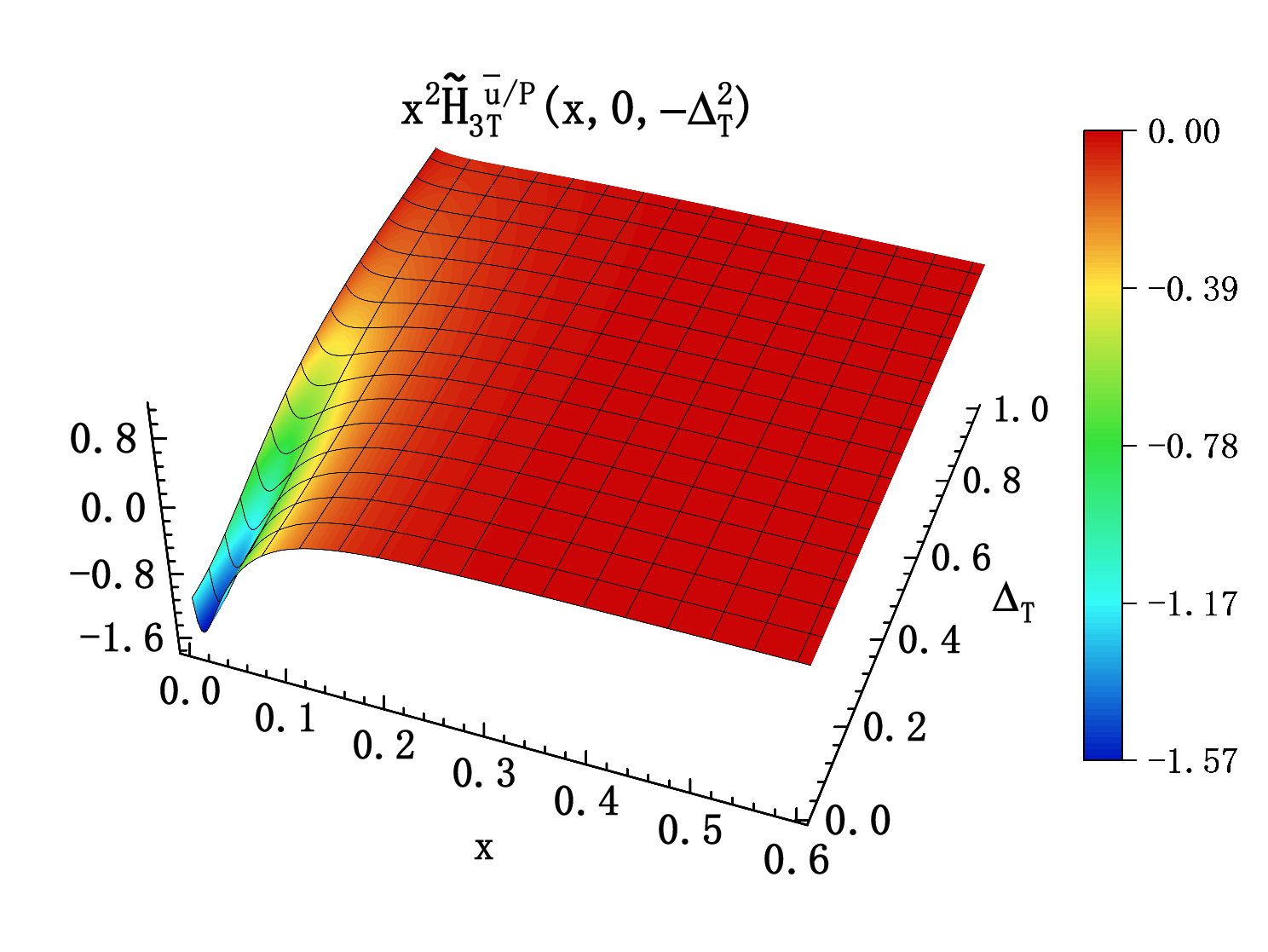}
       \end{minipage}}
       \subfigure{\begin{minipage}[b]{0.245\linewidth}
       		\centering
       		\includegraphics[width=\linewidth]{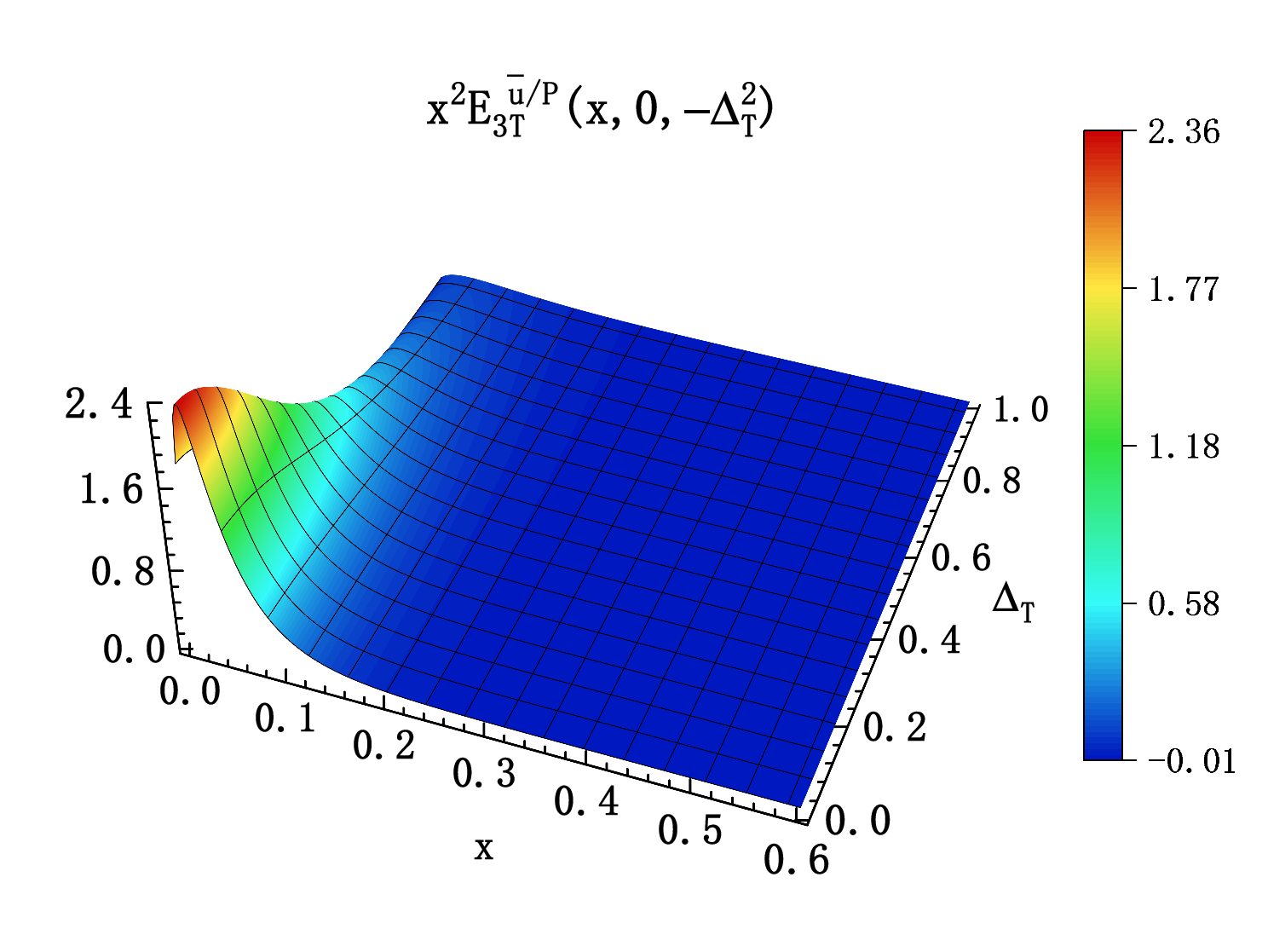}    	
       \end{minipage}}
   
       \subfigure{\begin{minipage}[b]{0.245\linewidth}
       		\centering
       		\includegraphics[width=\linewidth]{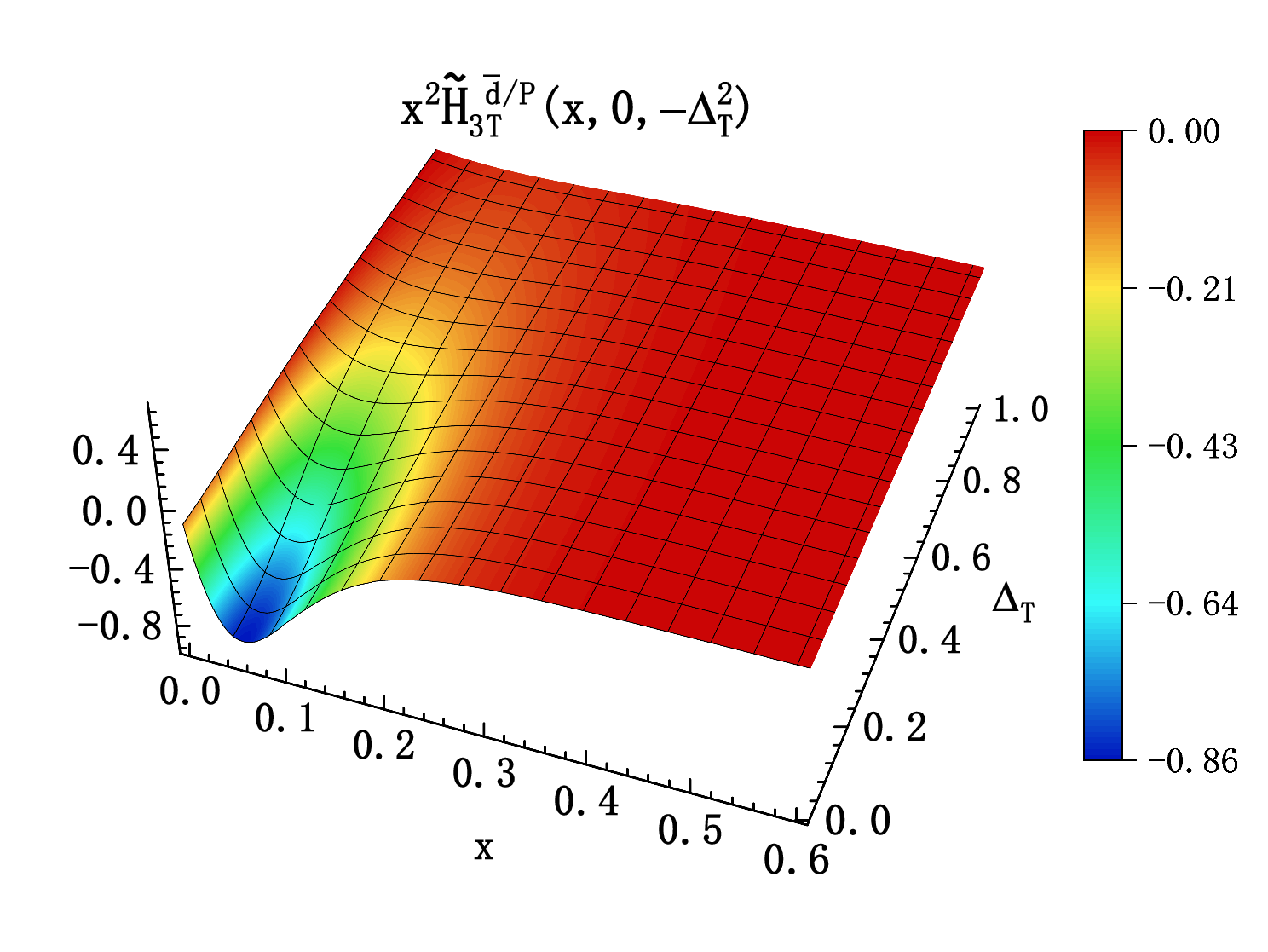}
       \end{minipage}}
       \subfigure{\begin{minipage}[b]{0.245\linewidth}
       		\centering
       		\includegraphics[width=\linewidth]{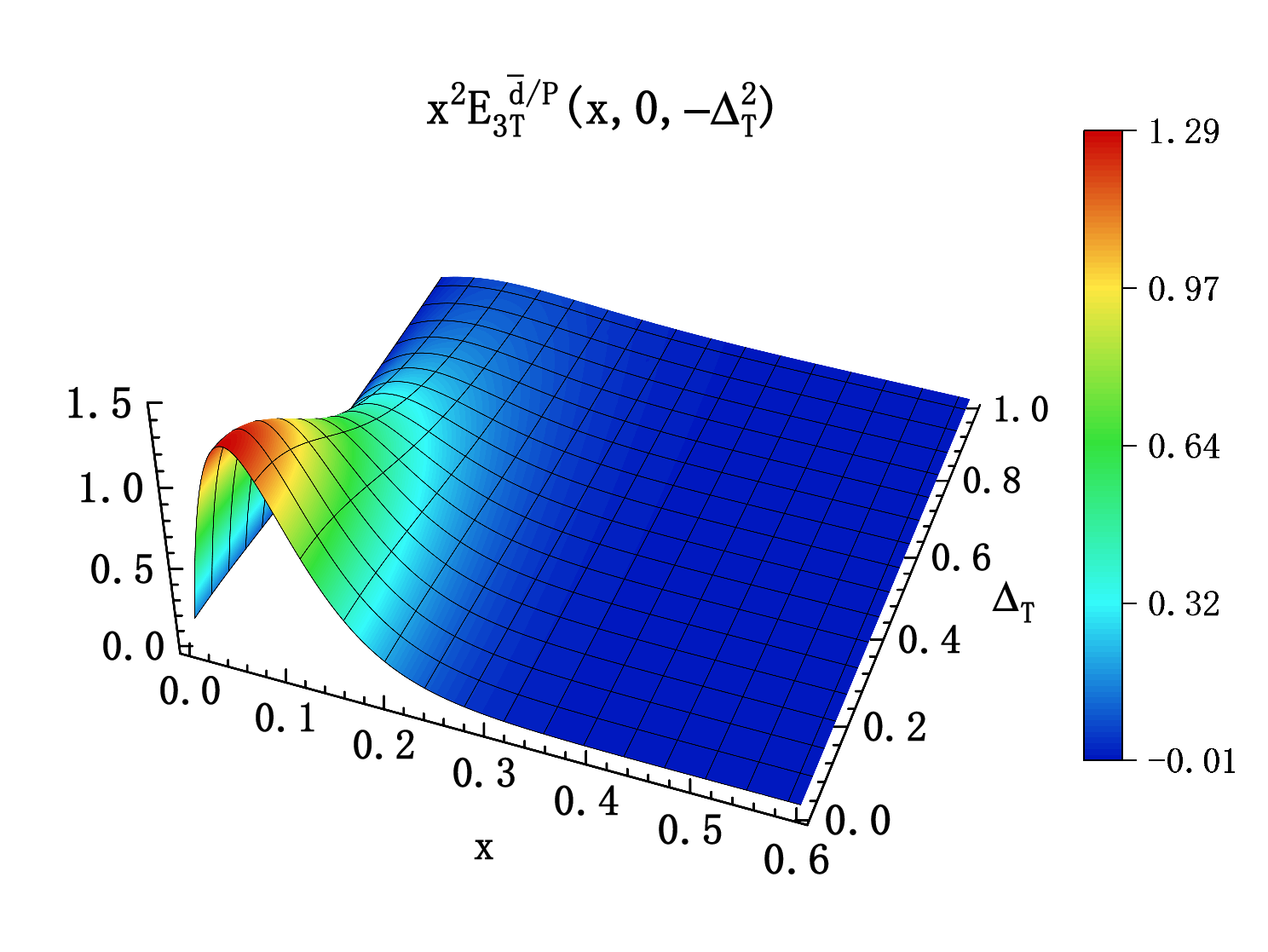}    	
       \end{minipage}}
    	\caption{The twist-4 GPDs (multiplied by $x^2$) $H_3^{\bar{q}/P}$, $E_3^{\bar{q}/P}$, $\widetilde{H}_{3}^{\bar{q}/P}$, $\widetilde{H}_{3T}^{\bar{q}/P}$ and $E_{3T}^{\bar{q}/P}$ ($\bar{q}=\bar{u}$ or $\bar{d}$) as functions of $x$ and $\Delta_T$.} \label{T4GPDs}      
    \end{figure*}
   
In this section, we present numerical results for twist-4 GTMDs and GPDs of the $\bar{u}$ and $\bar{d}$ quarks. For our model parameters $g_1$, $g_2$, $\Lambda_{\bar{q}}$, and $\Lambda_\pi$, we adopt values from Ref.~\cite{Luan:2022fjc}. There, $g_2$ and $\Lambda_\pi$ were determined by fitting $f_1^{\bar{u}/\pi^-}$ (or $f_1^{\bar{d}/\pi^+}(x)$) using the GRV leading-order (LO) parametrization~\cite{Gluck:1991ey}. The MSTW2008 LO parametrization~\cite{Martin:2009iq} is used for $f_1^{\bar{u}/P}$ and $f_1^{\bar{d}/P}$ to extract $g_1$ and $\Lambda_{\bar{q}}$. The parameter values are listed in Table~\ref{tab1}.
At zero skewness ($\xi = 0$), GTMDs depend on four kinematic variables: $x$, $\boldsymbol{k}_T$, $\boldsymbol{\Delta}_T$, and $\theta$. The angle $\theta$ between $\boldsymbol{k}_T$ and $\boldsymbol{\Delta}_T$ spans $0$ to $\pi$, so $\boldsymbol{k}_T\cdot\boldsymbol{\Delta}_T$ ranges from $k_T\Delta_T$ to $-k_T\Delta_T$. 
In this work, we set $\theta=\pi/3$, for which all GTMDs are nonvanishing.

\begin{center}\label{tab1}
    	\setlength{\tabcolsep}{5mm}
    	\renewcommand\arraystretch{1.5}
    	\begin{tabular}{ c | c | c }
    		\hline
    		Parameters & $\bar{u}$ & $\bar{d}$ \\
    		\hline
    		\hline
    		$g_1$ & 9.33 & 5.79 \\
    		\hline
    		$g_2$ & 4.46 & 4.46 \\
    		\hline
    		$\Lambda_\pi(GeV)$ &  0.223 & 0.223 \\
    		\hline
    		$\Lambda_{\bar{q}}(GeV)$ &  0.510 &  0.510 \\
    		\hline
    	\end{tabular}
    	\captionof{table}{Values of the parameters obtained from Ref.~\cite{Luan:2022fjc}.} \label{tab1}
\end{center}

To simultaneously show the $x$ and ${\Delta}_T$ dependences of twist-4 GTMDs for $\bar{u}$ and $\bar{d}$ quarks, we display their three-dimensional plots at fixed ${k}_T=0.1$ GeV and $\theta=\pi/3$ in Figs.~\ref{udbarF},~\ref{udbarG}, and~\ref{udbarH}. Furthermore, the ${k}_T$ and ${\Delta}_T$ dependences of these twist-4 GTMDs at fixed $x = 0.1$ and $\theta=\pi/3$ is presented in Figs.~\ref{udbarF-2},~\ref{udbarG-2}, and~\ref{udbarH-2}.
    
\subsection{$(x, k_T)$-dependences of twist-4 GTMDs}

In the upper and lower panels of Fig.~\ref{udbarF}, we plot the three-dimensional images of twist-4 GTMDs (multiplied by a prefactor $x^2$) for unpolarized sea quarks $F_{3,1}^{\bar{q}/P}$, $F_{3,3}^{\bar{q}/P}$, $F_{3,4}^{\bar{q}/P}$ as functions of $x$ and ${\Delta}_T$ at fixed ${k}_T=0.1$ GeV and $\theta=\pi/3$. 
All the GTMDs are sizable. 
The plots show that $F_{3,1}^{\bar{q}/P}$ and $F_{3,4}^{\bar{q}/P}$ are negative, while $F_{3,3}^{\bar{q}/P}$ is positive.  
Their magnitudes peak at small $x$ ($x<0.05$) and decrease with increasing momentum transfer ${\Delta}_T$. Notably, the peak for the $\bar{u}$ quark occurs at smaller $x$ than for the $\bar{d}$ quark. As the longitudinal momentum fraction $x$ increases beyond $0.05$, sea quarks carry a larger fraction of the nucleon momentum, and the magnitudes of these distributions decrease much faster with $x$ than with $\boldsymbol{\Delta}_T$. This indicates that the dominant contributions to twist-4 F-type sea-quark GTMDs reside in the low-$x$ region, a generic feature of unpolarized sea-quark distributions. For both $\bar{u}$ and $\bar{d}$, $x^2F_{3,1}^{\bar{q}/P}$ and $x^2F_{3,4}^{\bar{q}/P}$ are negative over the entire $(x,{\Delta}_T)$  region, whereas $x^2F_{3,3}^{\bar{q}/P}$ remains positive.

In Fig.~\ref{udbarG}, we present the dependence of the nonvanishing twist-4 GTMDs (multiplied by a prefactor $x^2$) for longitudinally polarized sea quarks in a proton as functions of $x$ and $\Delta_{T}$, with $k_T$ fixed at 0.1 GeV and $\theta = \pi/3$. 
It is found that $G_{3,1}^{\bar{q}/P}$ and $G_{3,2}^{\bar{q}/P}$ for $\bar{q}= \bar{u}$ and $\bar{d}$ are positive, while  $G_{3,3}^{\bar{q}/P}$ and $G_{3,4}^{\bar{q}/P}$ are negative, 
and the magnitude of $G_{3,1}^{\bar{q}/P}$ is sizable.
Fig.~\ref{udbarH} illustrates similar results, but for transversely polarized sea quarks in a proton.
The plots show that $H_{3,1}^{\bar{q}/P}$, $H_{3,2}^{\bar{q}/P}$ and $H_{3,6}^{\bar{q}/P}$ are sizable, while the other GTMDs are negligible.

\subsection{$(\Delta_{T}, k_T)$-dependences of twist-4 GTMDs}

To simultaneously present the $k_T$ and $\Delta_T$ dependences of the F-type twist-4 GTMDs for unpolarized $\bar{u}$ and $\bar{d}$ quarks, we show the corresponding three-dimensional plots at fixed $x=0.1$ and $\theta=\pi/3$ in Fig.~\ref{udbarF-2}. 
The magnitudes of these distributions have maximum values at $k_T=0$ GeV and $\Delta_T=0$ GeV and decrease as $k_T$ and $\Delta_T$ increase. 
For both $\bar{u}$ and $\bar{d}$ quarks, $x^2F_{3,1}^{\bar{q}/P}$ and $x^2F_{3,4}^{\bar{q}/P}$ are negative, whereas $x^2F_{3,3}^{\bar{q}/P}$ is positive throughout the considered $k_T$ and $\Delta_T$ ranges.
     
Fig.~\ref{udbarG-2} presents the results from our model for the G-type twist-4 GTMDs for longitudinally polarized sea quarks in a proton as functions of $k_T$ and $\Delta_{T}$ at a fixed $x = 0.1$ and $\theta=\pi/3$. The distributions $x^2G_{3,1}^{\bar{q}/P}$ and $x^2G_{3,2}^{\bar{q}/P}$ exhibit similar dependence on $k_T$, both increasing as $k_T$ decreases. 
However, their $\Delta_{T}$ dependences are different. 
Specifically, $x^2G_{3,1}^{\bar{q}/P}$ shows a monotonic decrease as $\Delta_{T}$ increases, whereas $x^2G_{3,2}^{\bar{q}/P}$ initially increases and then decreases with increasing $\Delta_{T}$. As for $x^2G_{3,3}^{\bar{q}/P}$, its magnitude first increases and then decreases with the increase of $k_T$ and $\Delta_{T}$. 
Unlike $x^2G_{3,1}^{\bar{q}/P}$, $x^2G_{3,2}^{\bar{q}/P}$, and $x^2G_{3,3}^{\bar{q}/P}$, $x^2G_{3,4}^{\bar{q}/P}$ increases as $\Delta_{T}$ increases. We also find that the signs of $x^2G_{3,1}^{\bar{q}/P}$ and $x^2G_{3,2}^{\bar{q}/P}$ are negative, while $x^2G_{3,3}^{\bar{q}/P}$ is positive. Particularly, as $k_T$ increases, $x^2G_{3,4}^{\bar{q}/P}$ changes from a positive value to a negative one, indicating a sign change at a specific value of $x$ and ${\Delta}_{T}$.
      
In Fig.~\ref{udbarH-2}, we display the dependences of the G-type twist-4 GTMDs for transversely polarized sea quarks as functions of $k_T$ and $\Delta_{T}$ at a fixed $x = 0.1$ and $\theta=\pi/3$. 
We find that the overall sizes of the GTMDs for the transversely polarized sea quarks are smaller than those for the unpolarized quarks and longitudinally polarized quarks.
We find that  $x^2H_{3,1}^{\bar{q}/P}$, $x^2H_{3,2}^{\bar{q}/P}$, $x^2H_{3,6}^{\bar{q}/P}$ are sizable, while the other GTMDs are much smaller.
For $x^2H_{3,1}^{\bar{q}/P}$, $x^2H_{3,2}^{\bar{q}/P}$, $x^2H_{3,6}^{\bar{q}/P}$, the distributions peak in the region where $k_T$ and ${\Delta}_T$ approach zero. On the contrary, the other four GTMDs peak at nonzero $\Delta_T$.

\subsection{Twist-4 GPDs}
 
In Fig.~\ref{T4GPDs}, we present the model results for the twist-4 GPDs (multiplied by $x^2$) $H_3^{\bar{q}/P}$, $E_3^{\bar{q}/P}$, $\widetilde{H}_{3}^{\bar{q}/P}$, $\widetilde{H}_{3T}^{\bar{q}/P}$, and $E_{3T}^{\bar{q}/P}$ for $\bar{q}=\bar{u}$ and $\bar{d}$ as functions of $x$ and $\Delta_T$. 
The distributions $H_3^{\bar{u}/P}$ and $H_3^{\bar{d}/P}$ have identical three-dimensional profiles to those of $E_3^{\bar{u}/P}$ and $E_3^{\bar{d}/P}$, respectively. 
The twist-4 GPDs $x^2\widetilde{H}_{3T}^{\bar{q}/P}$ and $x^2E_{3T}^{\bar{q}/P}$ exhibit similar behaviors with opposite signs. 
All distributions tend toward zero as $x$ increases. Their peaks occur at relatively small $x$ and shift toward larger $x$ as $\Delta_T$ increases. For both $\bar{u}$ and $\bar{d}$ quarks, $x^2H_3^{\bar{q}/P}$, $x^2E_3^{\bar{q}/P}$, and $x^2\widetilde{H}_{3T}^{\bar{q}/P}$ are negative, whereas $x^2E_{3T}^{\bar{q}/P}$ is positive. The distributions $x^2\widetilde{H}_{3}^{\bar{u}/P}$ and $x^2\widetilde{H}_{3}^{\bar{d}/P}$ take both positive and negative values, with positive values appearing at lower $x$ and larger $\Delta_T$.

\section{CONCLUSION}\label{Sec5} 
     
In this work, we have investigated the twist-4 GTMDs of sea quarks in the proton at zero skewness within the LCQM. Using the overlap representation, we have expressed the twist-4 GTMDs in terms of LCWFs. 
The sea-quark degrees of freedom are generated through the meson-baryon fluctuation mechanism, in which the proton is treated as a composite system consisting of a pion and a baryon, with the pion containing a $q\bar{q}$ Fock component.
Based on the LCWFs, we have derived analytic expressions for the F-, G-, and H-type twist-4 GTMDs of $\bar{u}$ and $\bar{d}$ quarks. We have also established the relations between the twist-4 GTMDs and GPDs and obtained the corresponding twist-4 GPDs at $\xi=0$, including $H_3^{\bar{q}/P}$, $E_3^{\bar{q}/P}$, $\widetilde{H}_3^{\bar{q}/P}$, $\widetilde{H}_{3T}^{\bar{q}/P}$, $E_{3T}^{\bar{q}/P}$, and $H_{3T}^{\bar{q}/P}$. Within the present model, $F_{3,2}^{\bar{q}/P}$, $H_{3,4}^{\bar{q}/P}$, and $H_{3T}^{\bar{q}/P}$ vanish at $\xi=0$.
Using the model parameters described above, we have presented numerical results for the twist-4 GTMDs of $\bar{u}$ and $\bar{d}$ quarks as functions of the longitudinal momentum fraction $x$, the transverse momentum $k_T$, and the transverse momentum transfer $\Delta_T$. We have also investigated the corresponding twist-4 GPDs as functions of $x$ and $\Delta_T$. The numerical results reveal characteristic differences in the magnitudes, signs, and kinematic dependences of the various twist-4 distributions for $\bar{u}$ and $\bar{d}$ quarks.
Our results provide a model-based description of the twist-4 sea-quark structure of the proton and may serve as useful input for future studies of higher-twist parton distributions and their phenomenological implications.

\section*{Acknowledgements}
This work is partially supported by the National Natural Science Foundation of China under grant number 12150013.

\end{document}